\documentclass[a4paper,11pt]{article}
\usepackage[dvipsnames]{xcolor}
\usepackage[utf8]{inputenc}
\usepackage[bbgreekl]{mathbbol}
\usepackage{geometry}
\usepackage{multicol}
\usepackage{float} 
\usepackage{longtable}

\usepackage{extarrows}
\usepackage{xcolor}
\usepackage{amsmath, array, amssymb, amsfonts,amsthm}
\usepackage{upgreek}
\usepackage{bbding}
\usepackage{pifont}
\usepackage{xfrac}
\usepackage[inline]{enumitem}
\setlist{nolistsep}
\usepackage[french, italian, english]{babel}
\usepackage[all]{xy}
\usepackage{txfonts}  
\usepackage{sectsty}
\usepackage{booktabs}
\usepackage{caption}
\usepackage{dsfont}
\usepackage{mathtools}
\usepackage{slashed}
\usepackage[makeroom]{cancel}
\usepackage[hidelinks]{hyperref}
\usepackage{textcomp}
\usepackage{multicol}
\usepackage[title]{appendix}
\usepackage[square,numbers,sort,compress,semicolon,merge]{natbib}
\let\cite\citep 
\usepackage{hyperref}
\hypersetup{colorlinks=true, urlcolor=blue, citecolor=blue, linktoc=page}

\usepackage{tikz}
\usepackage{tikz-cd}
\usetikzlibrary{cd}
\tikzcdset{
arrow style=tikz,
diagrams={>={Straight Barb[scale=0.8]}}
}
\usetikzlibrary{matrix,arrows,decorations.pathmorphing}

\DeclareMathAlphabet{\mathpzc}{OT1}{pzc}{m}{it}

\usepackage{fancybox}

\usepackage{mathrsfs}
\usepackage{scrextend}

\makeatletter
\renewcommand*\env@matrix[1][\arraystretch]{%
  \edef\arraystretch{#1}%
  \hskip -\arraycolsep
  \let\@ifnextchar\new@ifnextchar
  \array{*\c@MaxMatrixCols c}}
\makeatother

\newcommand{\defeq}{\vcentcolon=}
\newcommand{\rdefeq}{=\vcentcolon}
\renewcommand\P{\mathcal{P}}
\newcommand\M{\mathcal{M}}
\newcommand\N{\mathcal{N}}
\newcommand\R{\mathcal{R}}
\newcommand\RR{\mathbb{R}}
\newcommand\CC{\mathbb{C}}
\newcommand\EE{\mathbb{E}}
\newcommand\C{\mathcal{C}}

\newcommand\id{\textit{id}}
\newcommand\T{\mathcal{T}}
\newcommand\G{\mathcal{G}}
\renewcommand\H{\mathcal{H}}

\renewcommand\S{\mathcal{S}}

\newcommand\SU{\mathcal{SU}}
\newcommand\U{\mathcal{U}}
\newcommand\SO{\mathcal{SO}}

\newcommand\K{\mathcal{K}}

\renewcommand\O{\mathcal{O}}
\newcommand\GL{\mathcal{GL}}
\newcommand\D{\mathcal{D}}

\newcommand\vphi{\varphi}

\renewcommand\u{\text{\bf u}}

\newcommand\mC{\mathscr{C}}

\renewcommand\epsilon{\varepsilon}

\newcommand\rarrow{\rightarrow}

\newcommand\aut{\mathfrak{aut}}
\newcommand\diff{\mathfrak{diff}}

\newcommand\LieG{\mathfrak{g}}

\newcommand\so{\mathfrak{so}}

\newcommand\ups{{\bs \upupsilon}}

\renewcommand\t{\tilde}

\renewcommand\b{\bar }
\newcommand\w{\wedge}
\renewcommand\d{\partial}
\newcommand\s{\sigma}
\newcommand\bs{\boldsymbol}

\renewcommand\-{^{-1}}
\newcommand\Ad{\text{Ad}}

\renewcommand\id{\text{id}}

\makeatletter

\newcommand{\Rmnum}[1]{\expandafter\@slowromancap\romannumeral #1@}
\makeatother

\makeatletter
\newcommand{\leqnomode}{\tagsleft@true\let\veqno\@@leqno}
\newcommand{\reqnomode}{\tagsleft@false\let\veqno\@@eqno}
\makeatother

\DeclareMathOperator{\Diff}{Diff}
\DeclareMathOperator{\Aut}{Aut}

\DeclareMathOperator{\im}{Im}

\theoremstyle{definition}


\begin{document}


\title{Mechanics as a general-relativistic gauge field theory, \\ and Relational Quantization}

\author{J. T. \textsc{François} $\,{}^{a,\, b,\, 
*}$ \and L. \textsc{Ravera} $\,{}^{c,\,d,\,e,\,\star}$ }

\date{}

\maketitle
\begin{center}
\vskip -0.6cm
\noindent
${}^a$ Department of Philosophy, University of Graz -- Uni Graz. \\
Heinrichstraße 26/5, 8010 Graz, Austria.\\[2mm]
 
${}^b$ Department of Mathematics \& Statistics, Masaryk University -- MUNI.\\
Kotlářská 267/2, Veveří, Brno, Czech Republic.\\[2mm] 
 

${}^c$ DISAT, Politecnico di Torino -- PoliTo. \\
Corso Duca degli Abruzzi 24, 10129 Torino, Italy. \\[2mm]

${}^d$ Istituto Nazionale di Fisica Nucleare, Section of Torino -- INFN. \\
Via P. Giuria 1, 10125 Torino, Italy. \\[2mm]

${}^e$ \emph{Grupo de Investigación en Física Teórica} -- GIFT. \\
Universidad Cat\'{o}lica De La Sant\'{i}sima Concepci\'{o}n, Concepción, Chile. \\[2mm]

\vspace{1mm}

${}^*$ {\small{jordan.francois@uni-graz.at}} \qquad \quad ${}^\star$ {\small{lucrezia.ravera@polito.it, {Corresponding author}}}
\end{center}


\vspace{-3mm}

\begin{abstract}
We treat {the Mechanics of point particles} as a 1-dimensional general-relativistic gauge field theory, {which may be referred to as Mechanical Field Theory (MFT)}, 
exploiting the bundle geometry of
\emph{Mechanical Field Space} (MFS). 
The diffeomorphism covariance of MFT encodes its relational character, arising -- as in all general-relativistic physics -- via the conjunction of a hole and a point-coincidence argument. 
Any putative ``boundary problem", meaning the claim that `spacetime' boundaries break diffeomorphism and gauge symmetries, thereby dissolves.
It is highlighted that the standard path integral (PI) on the MFS, the exact analogue of the PI used in 
gauge field theory, is 
conceptually and technically \emph{distinct} from the standard PI of Quantum Mechanics.

We then use the Dressing Field Method to give a \emph{manifestly invariant and relational reformulation} of MFT, which reproduces the standard textbook formulation when a clock field is chosen as 
a (natural) dressing field.
The dressed, or basic, PI on the MFS, defining \emph{Relational Quantization} -- i.e. the quantization of invariant relational d.o.f. -- is shown to reproduce the standard PI of Quantum Mechanics. This establishes the soundness of Relational Quantization as a general guiding principle: We outline it for general-relativistic gauge field theories.

\end{abstract}

\noindent
\textbf{Keywords}: Gauge field theory, Bundle geometry, Mechanical Field Space, Path integral, Relational Quantization,  Relational Quantum Gravity. 

\vspace{-3mm}

\tableofcontents

\section{Introduction}  
\label{Introduction}  

The paradigmatic insight of general-relativistic gauge field theory (gRGFT), encoded in their  local symmetries, is \emph{relationality}: i.e. the fact that physical spatiotemporal and internal field-theoretical degrees of freedom (d.o.f.) dynamically co-define each other. 
As 
articulated in \cite{JTF-Ravera-Foundation2025}, this fact arises from combining the generalized hole argument and point-coincidence argument. 
The default formalism of gRGFT is therefore manifestly covariant under its local group of transformations, the semi-direct product of diffeomorphisms and internal gauge groups $\Diff(M)\ltimes \H$, and tacitly relational. This makes for many apparent conceptual difficulties that may lead to a priori puzzling and interrelated questions: most notably the determination of the physical d.o.f. and observables, the ``boundary problem" or problem of subsystems in gRGFT, and the ``problem of time".
A \emph{relational} (re)formulation of gRGFT, i.e. a reformulation that is manifestly \emph{invariant} under $\Diff(M)\ltimes \H$, 
{might}
dissolve 
{some of}
these difficulties. 
In  \cite{JTF-Ravera2024gRGFT} we  developed in full detail such a relational formulation based on the Dressing Field Method (DFM), a systematic approach to the reduction of local symmetries.
See 
\cite{JTF-Ravera2024-SUSY, JTF-Ravera-UnconventionalSUSY,JTF-Ravera2025-OffShellSUSY, JTF-Ravera2025-AdP} for applications 
to supersymmetric field theory. 

Our program aims at a relational, formally unified and conceptually clear, rewriting of fundamental physics. 
This must encompass quantum theory.
In  \cite{JTF-Ravera2024gRGFT} we hinted that the next phase of this program was to systematically develop the DFM-based notion of \emph{Relational Quantization} (RQ): i.e. the quantization of a dressed theory. 
It is in essence the claim that a sensible  quantum theory is one arising from the quantization of physical, relational, d.o.f.  
Applied to fields, it would lead to \emph{relational Quantum Field Theory} (rQFT).  
{If the core idea is not new, having a long branching history; One important chapter being quantization of constrained systems, with roots in the work of Bergmann \cite{Bergmann1949} and  Dirac \cite{Dirac1950}, another being the covariant phase space formalism, which in physics is often traced back to \cite{Witten1986, CrnkovicWitten1986, Zuckerman1986, Crnkovic1987} (but see \cite{Gieres2021} for a more complete history). 
Our approach is closer in spirit to the latter insofar as it starts with the field space of a gRGFT. 
It is  distinct in that it systematically leverages  the \emph{bundle geometry} of field space -- the DFM being a systematic way to realise its  \emph{basic cohomology}.}
The goals of this paper are twofold: {First to establish a DFM-based  \emph{Relational Quantization} a fruitful working hypothesis, 
then to outline it for gRGFT.}

The first goal is aimed at by considering Classical Mechanics 
{of point particles}
as a formal and conceptual \mbox{laboratory}. 
Mechanics {is often seen} 
as a mere stepping stone towards the more ``sophisticated" topic of Field Theory (FT), especially 
gRGFT. 
Its well-established standard formulations,  Lagrangian and Hamiltonian/symplectic,
did provide inspiration for analogue treatments of FT. 
The~local symmetries of gRGFT introduce further conceptual and technical difficulties,  necessitating (apparently) modifications or refinements of the formalisms: gauge-fixings (GF), constraints, etc.
It would thus seem that Mechanics  provides little insight into such more advanced subjects. 

Likewise, standard Quantum Mechanics (QM) is certainly a stepping stone towards Quantum Field Theory (QFT), 
its standard formulations providing motivations for analogue quantization schemes of FT -- notably the  Hamiltonian/canonical approach and the Dirac-Feynman path integral (PI), less often the Schrödinger approach.  
Here again, the local symmetries of gRGFT introduced difficulties that call for new notions and special tools: GF, ghosts, BRST cohomology, anomaly cancellations, etc.\footnote{
This appears to be enough to tame  the internal gauge groups of Gauge Field Theory (GFT) and delivers quantum GFT, or gauge QFT, but it comes short for general-relativistic FT. So, no quantum theory of gravity is established, let alone a full quantum gRGFT framework.}
\enlargethispage{1.5\baselineskip}
So, aside from its singular interpretational challenges and the occasionally made observation that QM is a 1D QFT \cite{Zee2010}, it would appear that QM  sheds little light onto the technical and conceptual structure needed for the quantization of gRGFT.

{A few authors have argued that even simple non-relativistic Classical Mechanics of point particles, revisited, still has something to teach us; this is notably the case via so-called ``parametrized" formulations \cite{Lanczos2012, Struckmeier2005, Rovelli-Vidotto2014}, sometimes couched into the language of ``partial \emph{vs} complete observables", cf. e.g. \cite{Rovelli2002, Dittrich2006, Dittrich2007, Thiemann2006}.}\footnote{
{Fewer still have argued that a better grasp of the radical paradigmatic shift of QM is key to make progress towards Quantum Gravity (QG). Famously so for the Everett-DeWitt many world interpretation, or Rovelli's relational interpretation  \cite{Rovelli1996, Rovelli2004, Rovelli2014}.}}
{We happened to \mbox{find ourselves} directionally in  agreement, and we shall here argue the case in the terms of the project mentioned above.}
First, in sections \ref{Geometry of Mechanical Field Space: Kinematics} and \ref{Mechanical Field Theory on Phi} we will provide a detailed treatment of  Mechanics as a 1D gRGFT; 
{for ease of reference, we shall
call this formulation}  ``Mechanical Field Theory" (MFT). 
For this we will rely on the framework of  \cite{JTF-Ravera2024gRGFT}; 
{calling}
\emph{Mechanical Field Space} (MFS)
{the space of parametrized mechanical coordinates (``mechanical fields"),}
we {systematically} {exploit} its bundle geometry, {in particular to handle adequately its  automorphism  groups and to \mbox{introduce} the notion of ``associated bundle of regions".}
This  generalizes the  ``parametrized" approaches {mentioned above},
giving a complete mathematical treatment paralleling exactly that of gRGFT. 
We shall stress that, as part of the gRGFT framework, the relationality of Mechanics  arises in exactly the same way it does in general-relativistic physics. 
Analogues of the problem of time and of the boundary problem arise in MFT, but are shown to dissolve when its relational structure is understood -- as they {ought to} do in gRGFT. 
We will spell out the links to the standard approach to Mechanics, also known as (a.k.a.) the ``unparametrized" approach, showing it to relate to the invariant relational d.o.f. encoded in the moduli base space $\M$ of the MFS bundle. 
Insights drawn may export to gRGFT.
%
{Considering next}
the issue of quantization of MFT, 
we write the PI on MFS in exact analogy with the standard PI for GFT, {stressing} that it \emph{is not} the standard PI approach to QM. 
{The latter is} instead the functional integral on $\M$; a first  hint at RQ. 


{This general framework lays the ground to describe, in}
 section \ref{Dressing Field Method and Relational Quantization}, 
 how the DFM allows to realise the basic cohomology of MFS, and how it therefore relates to the physical relational d.o.f. of MFT.  
We then apply the DFM to give an explicit invariant relational reformulation of MFT: it is shown to reproduce exactly the standard, unparametrized,  formulation of Mechanics when the dressing field is the clock field -- 
{which echoes and recovers  
the partial/complete observables notions alluded to above}.
This also shows in a technically clear way how apparent conceptual issues raised by the local symmetries of MFT (e.g. the boundary problem) are dissolved by a relational, dressed, reformulation. 
{We also highlight how the DFM accounts for possible change of dressing field, i.e. of clock in MFT: This naturally leads to a relational definition of a network of ``good clocks", consistent with metrological practice.} 
{To then showcase RQ,}
we write the dressed PI on MFS, which is basic (invariant), showing it to reproduce (to \emph{be}) the textbook standard PI  formulation of QM.
In other words, {we indicate that QM is a 1D rQFT,
giving}
a proof of concept that RQ, i.e. the quantization of the relational invariant d.o.f., is a natural viable scheme. 
In section \ref{Relational Quantization for general-relativistic gauge field theories}, we outline the basics of RQ for gRGFT. 
We show in particular how the relational formulation automatically implements anomalies cancellation mechanism, encompassing Wess-Zumino counter-terms as a special case. 

Section \ref{Conclusion} contains closing remarks and hints at forthcoming and future developments, 
notably regarding models of \emph{Relational Quantum Gravity} {(RQG)}. 
{For completeness, appendix \ref{Covariant phase space formalism for MFT} also presents the covariant phase space formulation of MFT and its relation to the standard symplectic treatment of (non-relativistic) mechanics, thereby making explicit the connection of the present framework with covariant phase space methods, gauge reduction, and edge-mode constructions commonly employed in gauge theories and generally covariant systems.}

{For the reader's convenience, we summarize the abbreviations
used throughout the paper in Table \ref{Table00}. 
The principal mathematical notations we adopt are collected in Table \ref{Table01} of Appendix \ref{Principal mathematical notations}.}

\begin{center}
{\small{
\begin{tabular}{c||c|c|} 
          &\textbf{Acronym}	\\[.2em] \hline\hline\\[-1em]
Degrees of freedom &  d.o.f.  \\[.2em] \hline\\[-1em]   
Dressing Field Method &  DFM  \\[.2em] \hline\\[-1em]   
Field theory &  FT  \\[.2em] \hline\\[-1em]  
Gauge Field Theory &  GFT  \\[.2em] \hline\\[-1em]             
Gauge fixing &  GF  \\[.2em] \hline\\[-1em]   
General-relativistic gauge field theory &  gRGFT  \\[.2em] \hline\\[-1em]
General Relativity &  GR \\[.2em] \hline\\[-1em]
Mechanical Field Space  &  MFS  \\[.2em] \hline\\[-1em]   
Mechanical Field
Theory  &  MFT \\[.1em] \hline
Non-relativistic  &  NR \\[.1em] \hline
Path integral &  PI \\[.1em] \hline
Quantum Field Theory &  QFT \\[.1em] \hline
Quantum Mechanics &  QM \\[.1em] \hline
Quantum Gravity &  QG \\[.1em] \hline
Relational Quantization &  RQ \\[.1em] \hline
Relational Quantum Field Theory &  rQFT \\[.1em] \hline
Relational Quantum Gravity &  RQG \\[.1em] \hline
\end{tabular}
\captionof{table}{{List of abbreviations.}}\label{Table00}}} 
\end{center}

\section{Geometry of Mechanical Field Space $\Phi$: Kinematics}  
\label{Geometry of Mechanical Field Space: Kinematics}  

We  specialize the formalism detailed in  \cite{JTF-Ravera2024gRGFT}  (see also \cite{Francois2023-a}) 
developing the view of Mechanics as a  model of the general-relativistic framework, thus generalizing  the so-called ``parametric approach" as discussed e.g. in \cite{Rovelli2004, Rovelli-Vidotto2014} which emphasizes conceptual aspects, and in the nice classic by Lanczos \cite{Lanczos2012} which offers a more systematic technical presentation.
To highlight the key conceptual structures, we  focus here on Mechanics of structureless point-particles, i.e. no rotation and no spin, but the generalization to particles with spin and to extended bodies with {d.o.f.} poses no special challenges.

\medskip
 
We shall consider the Mechanics of $N$ structureless point particles.  
It is the theory of 
fields 
 on a 1-dimensional manifold $I$, 
 $\phi=(x, t): I \rarrow \sf T$, $\tau \mapsto \phi(\tau)=\big(x(\tau), t(\tau) \big)$, 
 where $\sf T$ is a target space representing spatio-temporal d.o.f., and the \emph{clock field} $t$ is  s.t.
 $\dot{t}=\tfrac{dt}{d\tau} >0$.
 The fields $\phi=(x, t)$ represent a ``parametrization" of a kinematical history of the set of $N$ particles.
 In standard non-relativistic (NR) mechanics, for $N=1$ we have ${\sf T} =\EE^{3} \times \T$ where $\EE^3$ is the affine Euclidean space and $\T$  is the affine ``clock" timeline; for $N>1$, typically but not necessarily, one uses a single clock field $t$ for all $N$ \emph{spatial fields} $x=x_N$, so ${\sf T} =\RR^{3N} \times \T^{(N)}$ and 
 $(x, t)=\big(x_1, \ldots, x_N; t_{(N)} \big)$.
 This~naturally generalizes to ${\sf T}=M^{3N} \times \T^{(N)}$ where $M^3$ is a 3-manifold; that is the description of  classical particles in curved 3-space (which accommodate e.g. Newton-Cartan theory).
 The framework to be presented may generalize to special relativistic mechanics, 
 but we shall limit ourselves here to non-relativistic mechanics for simplicity. 
 


From the viewpoint of the manifold $I$, the target space $\sf T$ of the fields $\phi=(x, t)$ is ``internal", 
so that a natural transformation group $H$ of $\sf T$ plays the role of rigid ``internal group", while maps $\upxi: I \rarrow H$  form a ``gauge group" $\H$ acting on fields as $\phi\mapsto \phi^\upxi =\big(x^\upxi, t^\upxi\big)$.\footnote{Fields then may be seen as carrying an ``internal" index $\phi^\alpha=\big(x^a, t\big)= \big(x_i^a, t_{(i)} \big)$
with $\alpha=(a, 0)$ and $a=\{1,2,3\}$,  
and $i\in \{1, \ldots N\}$,
which are representation indices for the action of $H$ and $\H$. 
}
For non-relativistic mechanics, one may e.g. consider the group of spatial translations $H=\RR^{3N}$, so $\big(x^\upxi, t^\upxi\big) = \big(x+\upxi\,, t\big)$ -- see e.g.
 \cite{JTF-Ravera2025NRrelQM}.
 The fields $\phi$ are thus \emph{gauge fields} on $I$. 
 In~addition, they naturally support the action of $\Diff(I)$, and so do elements of $\H$, so that the full group of transformations acting on $\phi$ is 
 $\Diff(I) \ltimes \H$, as befitting a general-relativistic gauge field theory as described in \cite{JTF-Ravera2024gRGFT}.
 The difference with Mechanics is that not all $\H$ induces unphysical transformations. 
 For this reason, but more importantly to focus on key conceptual aspects, in the following we restrict our attention to the sole action of $\Diff(I)$, thus considering Mechanics as a case of ``general-relativistic" field theory: we may call it \emph{Mechanical Field Theory} (MFT).

\subsection{The Mechanical Field Space and its natural transformation groups}
\label{The Mechanical Field Space and its natural transformation groups}

 The field space of MFT, we call it the \emph{Mechanical Field Space} (MFS) $\Phi$, supports a natural right action by $\Diff(I)$
\begin{equation}
\label{right-action-Diff}
\begin{aligned}
\Phi \times \Diff(I) &\rarrow \Phi, \\
\left(\phi, \psi \right) &\mapsto R_{\psi}\,\phi 
\defeq \psi^*\phi
\rdefeq 
\phi^\psi, \\
\left((x, t), \psi \right) &\mapsto R_{\psi}\,(x,t) 
 \defeq \big(\psi^*x, \psi^* t\big)
 \rdefeq 
 \big( x^{\,\psi}, t^{\,\psi} \big).
\end{aligned}
\end{equation}
Indeed it satisfies, $R_{\psi'} R_\psi \phi:= {\psi'}^*\psi^*\phi= (\psi \circ \psi')^*\phi=:R_{\psi \circ \psi'} \phi$. 
The MFS is fibered by this action, the fiber through a point $\phi$ being its $\Diff(I)$-orbit $\mathcal O(\phi)$. 
The set of orbits, or \emph{moduli space} of fields, we denote by $\M \defeq \Phi/\Diff(I)$. 
{The latter is closely related to the so-called extended configuration (phase) space \cite{Rovelli2002, Rovelli2004, Rovelli2014}, see the end of section \ref{Models of MFT}.}
So,~the MFS is an infinite-dimensional principal bundle with structure group  $\Diff(I)$,
 \begin{equation}
  \begin{aligned}
  \Phi &\xrightarrow{\pi} \M,  \\
   \phi=(x, t) &\mapsto \pi(\phi)=\pi(x, t)=: [\phi]=[x, t].
\end{aligned}
 \end{equation}     
The projection $\pi$ is s.t. $\pi \circ R_\psi= \pi$. 
The fiber over a point $[\phi] \in \M$, $\pi\-([\phi])=\mathcal O(\phi)$,  is diffeomorphic to the structure group $\Diff(I)$ as a manifold.
A local section of the MFS over a subset $\U \subset \M$ is a smooth map $\sigma: \U  \rarrow \Phi_{|\U}$, $[\phi] \mapsto \sigma([\phi])\defeq \phi^\sigma=(x^\sigma, t^\sigma)$, s.t. $\pi \circ \sigma = \id_\U$. 
Within the region $\Phi_{|\U}$ over $\U$, the image of the local section $\im(\sigma)$ intersects fibers once, thus selecting a single representative $\phi^\sigma$ in each $\Diff(I)$-orbit $\mathcal O(\phi)$: i.e. a choice of  local section $\sigma$ is a \emph{gauge-fixing}.
Distinct local sections are related by $\sigma' = R_{\bs\uppsi} \circ \sigma= \bs\uppsi^* \sigma$ where $\bs\uppsi: \U \rarrow \Diff(I)$, $[\phi]\mapsto \bs\uppsi([\phi])$,  is a transition function of $\Phi$; meaning that gauge-fixing is not a $\Diff(I)$-invariant  operation. 

As a manifold, $\Phi$ has a diffeomorphisms group $\bs{\Diff}(\Phi)$, but \emph{as a principal bundle} its maximal transformation group is its group of \emph{automorphisms}
\begin{align}
    \bs{\Aut}(\Phi) := \big\{ \, \Xi \in \bs{\Diff}(\Phi) \, |\, \Xi \circ R_\psi = R_\psi \circ \Xi  \, \big\},
\end{align}
whose elements  preserve the fibration structure and thus  naturally induce diffeomorphisms of the base, $\bs{\Diff}(\M)$. 
It~contains a normal subgroup, the group of \emph{vertical automorphisms}
\begin{align}
    \bs{\Aut}_v(\Phi) := \big\{ \,\Xi \in \bs{\Aut}(\Phi) \ |\ \pi \circ \Xi = \pi \, \big\},
\end{align}
which is isomorphic to the \emph{gauge group} of $\Phi$
\begin{align}
 \label{GaugeGroup}
 \bs{\Diff}(I) := \big\{ \,\bs\psi : \Phi \rarrow  \Diff(I) \ |\  \bs\psi(\phi^\psi) =\psi^{-1} \circ \bs\psi(\phi) \circ \psi \,  \big\}.
\end{align}
The isomorphism being given by $\Xi(\phi)=R_{\bs\psi(\phi)}\, \phi$ still, i.e. $\bs\psi\in \bs{\Diff}(I)$ induces $\Xi \in \bs{\Aut}_v(\Phi)$.
The equivariance of elements $\bs\psi$ of $\bs\Diff(I)$ implies that $ \Xi' \bs \circ\,\! \Xi \in  \bs{\Aut}_v(\Phi)$ is induced by $\bs\psi' \circ  \bs\psi \in\bs\Diff(I)$: i.e. the composition operation $\bs\circ$ in $ \bs{\Aut}_v(\Phi)$ translates to  the   composition operation $\circ$ of   $\Diff(I)$. 
Since 
$ \bs\Aut_v(\Phi)$ is a normal subgroup of $\bs\Aut(\Phi)$, 
we have the short exact sequence (SES)
 \begin{align}
 \label{SESgroup}
\id_\Phi\rarrow \bs\Diff(I) \simeq \bs{\Aut}_v(\Phi)  \xlongrightarrow{\triangleleft}   \bs{\Aut}(\Phi) \longrightarrow \bs{\Diff}(\M) \rarrow \id_\M.
 \end{align}
The group $\bs{\Aut}_v(\Phi) \simeq \bs{\Diff}(I)$ embodies what may be called ``\emph{field-dependent}" gauge transformations, in this case \emph{field-dependent} diffeomorphisms of $I$.\footnote{A generalization of $\bs{\Aut}_v(\Phi) \simeq \bs{\Diff}(I)$ may be considered to also embody this notion: the group of vertical diffeomorphisms $\bs{\Diff}_v(\Phi) \defeq \left\{ \, \Xi \in \bs{\Diff}(\Phi)\, |\, \pi \circ \Xi = \pi \, \right\}$ 
isomorphic to, or generated by, $C^\infty\big(\Phi,  \Diff(I) \big)$. The latter can be shown to be the group of bisections of the Lie groupoid $\b{\bs\Gamma} =  \Phi \rtimes \Diff(I) \rightrightarrows \Phi$ associated with the right action of $\Diff(I)$ on $\Phi$ \cite{Mackenzie2005, Schmeding-Wockel2015, Maujouy2022}. This is treated in detail in section 3 of \cite{JTF-Ravera2024gRGFT},  and \cite{Francois2023-b}.
}
The structure group $\Diff(I)$ then supplies the notion of  ``\emph{field-independent}"  gauge transformations, i.e. \emph{field-independent} diffeomorphisms of $I$.
The linear version of \eqref{SESgroup} gives a SES of Lie algebras defining the Atiyah Lie algebroid of the MFS $\Phi$. 
 
\subsection{Differential structures}
\label{Differential structures}
As a manifold, $\Phi$ has a tangent bundle $T\Phi$, a cotangent bundle $T^\star\Phi$, and more generally a space of forms $\Omega^\bullet(\Phi)$. 
In considering these structures, it is  important to distinguish  the pushforward and pullback on $I$ and $\Phi$: we  use $*$ to denote these operations on $I$ (as we did in \eqref{right-action-Diff}), and $\star$ for their counterparts on $\Phi$.

\paragraph{Tangent bundle and subbundles} 

Vector fields $\mathfrak X : \Phi \rarrow T\Phi$ are sections of the tangent bundle, $\mathfrak X \in \Gamma(T\Phi)$,  and a Lie algebra under the bracket $[\ , \, ]_{\text{\tiny{$\Gamma(T\Phi)$}}}: \Gamma(T\Phi) \times \Gamma(T\Phi) \rarrow \Gamma(T\Phi)$. 
We write a vector field at $\phi \in \Phi$ as $\mathfrak X_{|\phi} =\tfrac{d}{ds}  \Psi_s(\phi) \, \big|_{s=0}$, with flow $\Psi_s \in \bs\Diff(\Phi)$  s.t.  $\Psi_{s=0}(\phi)=\phi$.
As derivations of the algebra of functions $C^\infty(\Phi)$ we may write: $\mathfrak X = \mathfrak X(\phi) \tfrac{\delta}{\delta \phi}$, where $\tfrac{\delta}{\delta \phi}$ is  the functional differentiation w.r.t. $\phi$, and $\mathfrak X(\phi)$ are the functional components. 

 The pushforward by the projection is $\pi_\star: T_\phi\Phi \rarrow T_{\pi(\phi)}\M=T_{[\phi]}\M$. 
The pushforward by the right action of $\psi \in \Diff(I)$ is  $R_{\psi\star}: T_\phi\Phi \rarrow T_{\psi^*\phi}\Phi$.  
 In general $R_{\psi\star} \mathfrak X_{|\phi} \neq \mathfrak X_{|\psi^*\phi}$, 
 i.e. a generic vector field ``rotates" as it is pushed vertically along fibers, 
so that $\pi_\star \mathfrak X$ is not a well-defined vector field on the moduli space $\M$. 
This is not so for the Lie subalgebra of \emph{right-invariant} vector fields
 \begin{align}
 \label{Inv-vector-fields}
\Gamma_{\text{\!\tiny{inv}}}(T\Phi):=\left\{\mathfrak X \in \Gamma(T\Phi)\, |\, R_{\psi\star} \mathfrak X_{|\phi}=  \mathfrak X_{|\psi^*\phi}  \right\},
 \end{align}
 which have well-defined projections on $\M$
as they do not rotate as they are pushed vertically. Indeed, for $\mathfrak X \in \Gamma_{\text{\!\tiny{inv}}}(T\Phi)$, we have $\pi_\star \mathfrak X_{|\psi^*\phi}=\pi_\star R_{\psi\star} \mathfrak X_{|\phi} = (\pi \circ R_\psi)_\star \mathfrak X_{|\phi} = \pi_\star \mathfrak X_{|\phi} =: \mathfrak Y_{|[\phi]}  \in T_{[\phi]}\M$.  Then, $\pi_\star \mathfrak X =: \mathfrak Y \in \Gamma(T\M)$. 
 The~property of invariant vector fields implies that their flows are automorphisms of $\Phi$, 
the Lie subalgebra $\Gamma_{\text{\!\tiny{inv}}}(T\Phi)$ is thus the Lie algebra of  $\bs\Aut(\Phi)$:
  \begin{align}
  \label{LieAlg-Aut}
\bs\aut(\Phi)=\Big(\Gamma_{\text{\!\tiny{inv}}}(T\Phi); \, [\mathfrak X, \mathfrak X']_{\text{{\tiny $\aut$}}}  \defeq -[\mathfrak X, \mathfrak X']_{\text{\tiny{$\Gamma(T\Phi)$}}}  \Big).
  \end{align}

The \emph{vertical tangent bundle} $V\Phi := \ker \pi_\star$ is a canonical  subbundle of the tangent bundle $T\Phi$.
Vertical vector fields are elements of $\Gamma(V\Phi):=\left\{  \mathfrak X \in \Gamma(T\Phi)\, |\, \pi_\star \mathfrak X=0 \right\}$. 
Since $V\Phi$ is a subbundle,  $\Gamma(V\Phi)$ is a Lie ideal of $\Gamma(T\Phi)$. 
 \emph{Fundamental} vertical vector fields are generated by the action of the Lie algebra $\diff(I)$ of  the structure group:
   \begin{align}
 \diff(I)=\Big(\Gamma(TI); \, [X,Y]_{\text{{\tiny $\diff(I)$}}}\defeq -[X , Y ]_{\text{{\tiny $\Gamma(TI)$}}}\Big). 
   \end{align}
For $X=\tfrac{d}{ds} \psi_s \big|_{s=0} \in \diff(I)$ with flow $\psi_s \in \Diff(I)$, the corresponding fundamental vector at $\phi\in \Phi$ is: 
 \begin{align}
 \label{Fund-vect-field}
 X^v_{|\phi} := \tfrac{d}{ds}\, R_{\psi_s}\, \phi\, \big|_{s=0} = \tfrac{d}{ds}\,  \psi^*_s \,\phi \, \big|_{s=0} =: \mathfrak L _X \phi =\big(\mathfrak L _X x, \mathfrak L _X t \big).
 \end{align} 
The Lie derivative on $I$ is as usual also given by the Cartan formula $ \mathfrak L _X=[\iota_x, d]=\iota_Xd + d\iota_X$, with $d=d\tau \tfrac{\d}{\d\tau}$ the de Rham exterior derivative on $I$. It is a degree 0 derivation of the algebra $\Omega^\bullet(I)$ of forms on $I$, since $\iota_X$ is of degree $-1$ and $d$ is of degree $1$. 
Also, since $I$ is 1-dimensional, $\Omega^\bullet(I)= \Omega^0(I)\oplus \Omega^1(I)$. 
As $\phi \in \Omega^0(I)$ and $X=X(\tau)\tfrac{\d}{\d\tau}$, we have 
$X^v_{|\phi}= \iota_X d \phi = X[\phi\big]=X(\tau)\dot \phi$, i.e.  $X^v_{|\phi}= \big(\iota_X dx, \iota_X dt \big) = \big(X[x], X[t] \big)= \big(X(\tau) \dot x, X(\tau)\dot t \,\big)$.

Manifestly, fundamental vector fields satisfy $\pi_\star X^v\equiv 0$, since 
$\pi_\star X^v_{|\phi} =\tfrac{d}{ds} \pi \circ R_{\psi_s} \phi\, \big|_{s=0} = \tfrac{d}{ds} \pi(\phi)\, \big|_{s=0}$. 
One shows  that the map $|^v :\diff(I) \rarrow \Gamma(V\Phi)$, $X \mapsto X^v$, is a Lie algebra morphism: i.e. $([X,Y]_{\text{{\tiny $\diff(I)$}}})^v=(-[X, Y]_{\text{{\tiny $\Gamma(TP)$}}})^v=[X^v, Y^v]$.
The pushforward by the right-action of $\Diff(I)$ of  a fundamental vertical vector field is:
\begin{equation}
\begin{aligned}
\label{Pushforward-fund-vect}
 R_{\psi\star} X^v_{|\phi} :=&\, \tfrac{d}{ds} R_\psi \circ R_{\psi_s} \phi \,  \big|_{s=0} 
     =  \tfrac{d}{ds} R_{\psi_s \circ \psi} \phi \, \big|_{s=0} =  \tfrac{d}{ds} R_{\psi_s \circ \psi} \, R_{\psi^{-1} \circ \psi} \, \phi \, \big|_{s=0}  \\
    =&\,  \tfrac{d}{ds} R_{(\psi\- \circ \psi_s \circ \psi)}  \, R_{\psi} \, \phi \, \big|_{s=0} =   \tfrac{d}{ds} R_{(\psi\- \circ \psi_s \circ \psi)}  \, \psi^*\phi \, \big|_{s=0}   \\
        =&\!: \left(  (\psi\-)_* \, X \circ \psi \right)^v_{|\psi^*\phi}. 
\end{aligned}
\end{equation}
Therefore,  fundamental vector fields generated by $\diff(I)$ are  not right-invariant. 

On the other hand,  the vertical vector fields induced by $\bs\diff(I)$, the Lie algebra of the gauge group $\bs\Diff(I)$, are right-invariant. To $\bs\psi_s \in \bs\Diff(I)$ corresponds  $\bs X =\tfrac{d}{ds}\, \bs\psi_s\, \big|_{s=0} \in \bs\diff(I)$. By the definition  \eqref{GaugeGroup} of the gauge group, whose elements transform as  $R^\star_\psi \bs\psi = \psi\- \circ \bs\psi \circ \psi $, 
we have
 \begin{align}
 \label{LieAlg-GaugeGroup}
 \bs\diff(I):=\left\{\, \bs X: \Phi \rarrow \diff(I)\  |\ R^\star_\psi \bs X = (\psi\-)_*\, \bs X \circ \psi \,  \right\}.
 \end{align}
 This transformation property, which is also  written as $\bs X(\phi^\psi)= \bs X(\psi^*\phi) =  (\psi\-)_*\, \bs X(\phi) \circ \psi $, has infinitesimal version  given by the Lie derivative on $\Phi$ along the  fundamental vector field corresponding to $\psi$:
 \begin{align}
 \label{inf-equiv-gauge-Lie-alg}
\bs L_{X^v} \bs X = X^v(\bs X) = \tfrac{d}{ds} \, R^\star_{\psi_s} \bs X \, \big|_{s=0} = \tfrac{d}{ds} \, (\psi\-_s)_*\, \bs X \circ \psi_s  \, \big|_{s=0} 
				=: \mathfrak L_X \bs X =[X, \bs X]_{\text{{\tiny $\Gamma(TI)$}}} =[\bs X, X]_{\text{{\tiny $\diff(I)$}}}. 
 \end{align}
A fundamental vector field generated by $\bs X \in \bs\diff(I)$ is
  \begin{align}
  \label{phi-dep-vert-vect-field}
 \bs X^v_{|\phi} := \tfrac{d}{ds} R_{\bs\psi_s(\phi)} \phi\, \big|_{s=0} = \tfrac{d}{ds}  (\bs\psi_s(\phi))^* \phi \, \big|_{s=0} =: \mathfrak L _{\bs X} \phi.
 \end{align}
 Its pushforward by the right-action of $\Diff(I)$ is
  \begin{align}
\label{R-inv-phi-dep-vvf}
 R_{\psi\star} \bs X^v_{|\phi} :=&\, \tfrac{d}{ds} R_\psi \circ R_{\bs\psi_s(\phi)} \phi \,  \big|_{s=0} 
        =  \tfrac{d}{ds} R_{(\psi\- \circ \bs\psi_s(\phi) \circ \psi)}  \, R_{\psi} \, \phi \, \big|_{s=0} =   \tfrac{d}{ds} R_{\bs\psi_s(\psi^*\phi)}  \, \psi^*\phi \, \big|_{s=0}
                         \rdefeq \bs X^v_{|\psi^*\phi}. 
 \end{align}
Furthermore, one shows that the ``verticality map" $|^v : \bs\diff(I) \rarrow \Gamma_{\text{\!\tiny{inv}}}(V\Phi)$, $\bs X \mapsto \bs X^v$, is a Lie algebra \emph{anti}-morphism: i.e. $([\bs X, \bs Y]_{\text{{\tiny $\diff(I)$}}})^v=(-[\bs X, \bs Y]_{\text{{\tiny $\Gamma(TP)$}}})^v=-[\bs X^v, \bs Y^v]$.
 Therefore, since the Lie subalgebra of right-invariant vertical vector fields is the Lie algebra of the group $\bs\Aut_v(\Phi)$, we have 
  \begin{align}
  \label{Gauge-Lie-alg-morph}
 \bs\diff(I)  \simeq \bs\aut_v(\Phi)=\big(\Gamma_{\text{\!\tiny{inv}}}(V\Phi); -[\ , \ ]_{\text{\tiny{$\Gamma(T\Phi)$}}}  \big).
  \end{align}
From the above we obtain the  infinitesimal version of \eqref{SESgroup}, which is the SES describing the Atiyah Lie algebroid~of~$\Phi$,
\begin{align}
 \label{Atiyah-Algebroid}
0\rarrow \bs\diff(I)  \simeq \bs\aut_v(\Phi)  \xlongrightarrow{|^v}   \bs{\mathfrak{aut}}(\Phi) \xlongrightarrow{\pi_\star} \bs{\diff}(\M) \rarrow 0.
\end{align}
A splitting of this SES, i.e. a map $ \bs{\mathfrak{aut}}(\Phi) \rarrow \bs\diff(I)$ or equivalently  a map $\bs{\diff}(\M) \rarrow  \bs{\mathfrak{aut}}(\Phi)$, allows to decompose a (right-invariant) vector field on $\Phi$ as a sum of a gauge element and a vector field on $\M$. Such a splitting is supplied by a choice of Ehresmann connection $1$-form on $\Phi$, the definition of which we remind in section \ref{Connections on Mechanical Field Space}.

\bigskip

Finally, we state a key result for the geometric definition of gauge transformations on $\Phi$. The pushforward by a vertical automorphism $\Xi \in \bs\Aut_v(\Phi)$, induced by  an  element of the gauge group $\bs \psi \in \bs\Diff(I)$,
is  the map $\Xi_\star: T_\phi\Phi \rarrow T_{\Xi(\phi)}\Phi=T_{\bs\psi^*\phi}\Phi$. 
For a generic vector field $\mathfrak X \in \Gamma(T\Phi)$ it is  
\begin{equation}
\begin{aligned}
\label{pushforward-X}
 \Xi_\star \mathfrak X_{|\phi} &= R_{\bs\psi(\phi) \star} \mathfrak X_{|\phi} + \left\{   \bs\psi(\phi)\-_* \bs d \bs\psi_{|\phi}(\mathfrak X_{|\phi})   \right\}^v_{|\,\Xi(\phi)}  \\
 					    &=R_{\bs\psi(\phi) \star} \left( \mathfrak X_{|\phi} +  \left\{  \bs d \bs\psi_{|\phi}(\mathfrak X_{|\phi})  \circ \bs\psi(\phi)\-  \right\}^v_{|\phi} \right). 
\end{aligned}
\end{equation}
This~relation can be used to obtain the result  for repeated pushforwards, i.e. iterated gauge transformations:  e.g. to get the result for $(\Xi' \bs \circ \Xi)_\star \mathfrak X_{|\phi}$,  one only needs to substitute  $\bs\psi \rarrow \bs\psi'  \circ  \bs\psi$ above.

\paragraph{Differential forms} 

The de Rham complex of $\Phi$ is $\big( \Omega^\bullet(\Phi); \bs d  \big)$ with $\bs d$ the de Rham (exterior) derivative on field space, which is s.t. $\bs d ^2 =0$ and defined via the Koszul formula.  The exterior product $\w$ is defined on scalar-valued forms as usual, so that  
$\big( \Omega^\bullet(\Phi, \mathbb K), \w, \bs d \big)$ is a differential graded algebra. The exterior product is also defined on the space $\Omega^\bullet(\Phi, \sf A)$ of forms with values in an algebra $(\sf A, \cdot)$, using the product in $\sf A$ instead of the product in $\mathbb K$. 
So~$\big( \Omega^\bullet(\Phi, \sf A), \w, \bs d \big)$ is a again a differential graded algebra.\footnote{ On the other hand, an exterior product cannot be defined on $\Omega^\bullet(\Phi, \bs V)$ where $\bs V$ is merely a vector space.}
The case of immediate interest to us is ${\sf A}=\big(\Omega^\bullet(I), \w \big)$ the algebra of forms on $I$. 
We write a  form  $\bs \alpha \in \Omega^\bullet(\Phi)$ evaluated at $\phi\in \Phi$ as
\begin{align}
\bs\alpha_{|\phi} = \alpha\big(\! \w^\bullet \!\bs d\phi_{|\phi}; \phi \big),
\end{align}
where $\bs d\phi=\big(\bs dx, \bs d t \big) \in \Omega^1(\Phi)$ is the basis 1-form on $\Phi$ 
 and $\alpha(\ ;\ )$ is the functional expression of $\bs\alpha$, alternating multilinear in the first arguments and with arbitrary $\phi$-dependence  in the second argument. 
 For example, given a 0-form $\bs f$ we have 
 $\bs d f= d\phi \tfrac{\delta}{\delta \phi} \bs f = \bs d x \tfrac{\delta}{\delta x} \bs f + \bs d t \tfrac{\delta}{\delta t} \bs f$.

The action by pullback of   $\Diff(I)$ on a form $\bs \alpha \in \Omega^\bullet(\Phi)$ defines its \mbox{\emph{equivariance}}, $R^\star_\psi \alpha$. 
\mbox{The~action} by pullback of   $\bs\Aut_v(\Phi)\simeq \bs\Diff(I)$  defines  \emph{gauge transformations}, which we write $\bs\alpha^{\bs\psi} \defeq \Xi^\star \bs\alpha  $. 
We have
 \begin{equation}
\begin{aligned}
R^\star_\psi \bs\alpha_{|\phi^\psi} &= \alpha\big(\! \w^\bullet \! R^\star_\psi \bs d\phi_{|\phi^\psi};\,  R_\psi \phi \big) = \alpha\big(\! \w^\bullet \! R^\star_\psi \bs d\phi_{|\phi^\psi}; \, \phi^\psi \big), 
\quad  \text{for } \ \psi \in \Diff(I), \\
\bs\alpha^{\bs\psi}_{|\phi} \defeq \Xi^\star \bs\alpha^{}_{|\Xi(\phi)} &= \alpha\big(\! \w^\bullet \! \Xi^\star \bs d\phi^{}_{|\Xi(\phi)};\  \Xi( \phi) \big) = \alpha\big(\! \w^\bullet \! \Xi^\star \bs d\phi_{|\phi^{\bs\psi}};\,  \phi^{\bs\psi} \big),
\quad \text{for } \  \Xi \in \bs\Aut_v(\Phi)  \sim \bs\psi \in \bs\Diff(I).    \label{GT-general}
\end{aligned}
\end{equation}
The infinitesimal equivariance and  vertical transformations of $\bs\alpha$ are given by its Lie derivative along the elements of $\Gamma(V\Phi)$ generated  respectively by $\diff(I)$ and $\bs\diff(I)$:\footnote{Actually, more generally they are given by the Nijenhuis-Lie derivative along vector-valued forms, as described in section 3.2.2 of \cite{JTF-Ravera2024gRGFT}.
See also \cite{Francois2023-b}. 
Indeed, vertical vector fields generated by $\diff(I)$ and $\bs\diff(I)$ can be understood to be $V\Phi$-valued $0$-forms. See \cite{Kolar-Michor-Slovak}.}
\begin{align}
\bs L_{X^v}\bs\alpha = \tfrac{d}{ds}  R^\star_{\psi_s} \bs\alpha \big|_{s=0}  \quad  \text{ with } \ X \in \diff(I),  
 \qquad \quad  \bs L_{\bs X^v}\bs\alpha = \tfrac{d}{ds}  \Xi_s^\star \bs\alpha \big|_{s=0}\quad  \text{ with } \ \bs X \in \bs\diff(I). \label{GT-inf-general}
\end{align}
Notably for our purpose, the equivariance of a $\Omega^\bullet(I)$-valued form $\bs\alpha$ is $R^\star_\psi \bs\alpha =\psi^*\bs \alpha$,
and infinitesimally $L_{X^v} \bs\alpha = \mathfrak L_X \bs\alpha$.
There are forms of particular interest whose gauge transformations are read directly from their  defining properties.

\medskip

First, \emph{equivariant} forms are those whose equivariance assumes a simple form.
\emph{Standard equivariant} forms are valued in representations $(\rho, \bs V)$ of the structure group $\Diff(I)$ and s.t.
\begin{align}
\Omega_\text{eq}^\bullet(\Phi, \rho) \defeq \left\{ \, \bs\alpha \in \Omega^\bullet(\Phi, \bs V)\,|\, R^\star_\psi\bs\alpha_{|\phi^\psi}=\rho(\psi)\-\bs\alpha_{|\phi}\, \right\}. 
 \end{align} 
Their infinitesimal equivariance property is $\bs L_{X^v}\bs\alpha=-\rho_*(X) \bs\alpha$ for $X \in \diff(I)$.
\emph{Cocyclic equivariant} forms \cite{Francois2019_II, JTF-Ravera2024gRGFT} have equivariance  controlled by a 1-cocycle for the action of $\Diff(I)$ on $\Phi$, 
 i.e. a map
 \begin{equation}
\begin{aligned}
\label{cocycle}
C: \Phi \times \Diff(I) &\rarrow G, \quad \text{$G$ some Lie group,}   \\
	(\phi, \psi) &\mapsto C(\phi; \psi) \qquad \text{s.t.} \quad C(\phi; \psi'\circ \psi) =C(\phi; \psi') \cdot C(\phi^{\psi'}; \psi). 
 \end{aligned} 
  \end{equation}
In case $G=U(1)$, we have  $C(\phi; \psi)=\exp{i\, c(\phi; \psi)}$ and 
   \begin{equation}
\begin{aligned}
\label{cocycle-bis}
c: \Phi \times \Diff(I) &\rarrow \RR, \\ 
	(\phi, \psi) &\mapsto c(\phi; \psi) \qquad \text{s.t.} \quad c(\phi; \psi'\circ \psi) =c(\phi; \psi') +  c(\phi^{\psi'}; \psi). 
 \end{aligned} 
  \end{equation}
 Manifestly,   $\phi$-independent 1-cocycles are  group morphisms,  i.e. 1-cocycles are generalisations of  representations. 
Given a $G$-space $\bs V$, one defines cocyclic equivariant forms as 
 \begin{align}
 \label{C-eq-form}
 \Omega^\bullet_\text{eq}(\Phi, C) \defeq \left\{ \bs\alpha \in \Omega^\bullet(\Phi, \bs V)\, | \, R^\star_\psi \bs\alpha_{|\phi^\psi} = C(\phi  ; \psi)\- \bs\alpha_{|\phi}\,\right\}.
 \end{align} 
 The property \eqref{cocycle}-\eqref{cocycle-bis} ensures compatibility with the right action: $R_{\psi'}^\star R^\star_{\psi} = R^\star_{\psi' \circ\,\psi}$. The infinitesimal equivariance  is $\bs L_{X^v}\bs\alpha=-a(X; \phi) \bs\alpha$, where $a(X, \phi):= \tfrac{d}{ds}\, C(\phi, \psi_s) |_{s=0}$ is a 1-cocycle for the action of $\diff(I)$~on~$\Phi$:
  \begin{equation}
 \begin{aligned}
 \label{inf-cocycle}
a: \Phi \times \diff(I) &\rarrow \mathfrak g, \quad \text{$\mathfrak g$ the Lie algebra of $G$,}   \\
	(\phi, X) &\mapsto a(X; \phi) \qquad \text{s.t.} \quad X^v\! \cdot a(Y; \phi)- Y^v\!\cdot a(X; \phi)+ \big[a(X; \phi)\,,\ a(Y; \phi)\big]_{\mathfrak g}=a([X,Y]_{\text{{\tiny $\diff(I)$}}}; \phi). 
 \end{aligned} 
   \end{equation}
 The infinitesimal relation \eqref{inf-cocycle} ensures compatibility with the right action: $[\bs L_{X^v}, \bs L_{X^v}]=\bs L_{[X^v, Y^v]}=\bs L_{([X, Y]_{\text{{\tiny $\diff(I)$}}})^v}$. 
 Notice this is a non-Abelian generalisation of the Wess-Zumino (WZ) consistency condition for  anomalies $a(X; \phi)$.
 The WZ consistency condition being reproduced for Abelian $G$.
 
 The subspace of \emph{invariant} forms are those whose  equivariance is trivial, and \emph{horizontal} forms are those vanishing on vertical vector field:
 \begin{equation}
  \begin{aligned}
 \Omega_\text{inv}^\bullet(\Phi)&=\left\{ \, \bs\alpha \in \Omega^\bullet(\Phi)\,|\, R^\star_\psi \bs\alpha=\bs\alpha \,\right\}, \  \text{infinitesimally }\  \bs L_{X^v}\bs\alpha=0, \\
 \Omega_\text{hor}^\bullet(\Phi)&=\left\{ \, \bs\alpha \in \Omega^\bullet(\Phi)\,|\, \iota_{X^v}\bs\alpha=0 \right\}.
  \end{aligned}
   \end{equation}
 A form which is both equivariant and horizontal is  \emph{tensorial}. \emph{Standard} tensorial forms  are
 \begin{align}
 \label{tens-forms}
\Omega_\text{tens}^\bullet(\Phi, \rho)\defeq \left\{ \, \bs\alpha \in \Omega^\bullet(\Phi, \bs V)\,|\, R^\star_\psi\bs\alpha=\rho(\psi)\-\bs\alpha,\, \text{ \& }\ \iota_{X^v}\bs\alpha=0\, \right\}. 
\end{align} 
Similarly, \emph{cocyclic tensorial} forms are
\begin{equation}
 \begin{aligned}
 \label{twisted-tens-forms}
\Omega_\text{tens}^\bullet(\Phi, C)\defeq&\, \left\{ \, \bs\alpha \in \Omega^\bullet(\Phi, \bs V)\,|\, R^\star_\psi\bs\alpha=C(\phi; \psi)\-\bs\alpha,\, \text{ \& }\ \iota_{X^v}\bs\alpha=0\, \right\}.
\end{aligned} 
\end{equation}
In both standard and cocyclic cases, we have  $\Omega_\text{tens}^0(\Phi)=\Omega_\text{eq}^0(\Phi)$. 
It is  well-known  that the de Rham derivative $\bs d$ does not preserve $\Omega_\text{tens}^\bullet(\Phi)$ -- horizontality being lost -- which is why  \emph{connections} on $\Phi$ are needed  to define \emph{covariant derivatives} on the spaces of tensorial forms. We come to this in section \ref{Connections on Mechanical Field Space} next.

Finally, forms that  are both  invariant and horizontal are called  \emph{basic}: 
 \begin{align}
\Omega_\text{basic}^\bullet(\Phi)\defeq \left\{ \, \bs\alpha \in \Omega^\bullet(\Phi)\,|\, R^\star_\psi\bs\alpha=\bs\alpha\, \text{ \& }\ \iota_{X^v}\bs\alpha=0\, \right\}. 
\end{align} 
This space is preserved by $\bs d$, so 
$\big( \Omega^\bullet_\text{basic}(\Phi), \bs d \big)$ is a subcomplex of the de Rham complex of $\Phi$: the \emph{basic subcomplex}.  
The cohomology of $\big( \Omega^\bullet_\text{basic}(\Phi), \bs d \big)$ defines  the \emph{equivariant cohomology} of $\Phi$ \cite{Guillemin-Sternberg1999}. Since  $[\bs d, \pi^\star]=0$, it is isomorphic to the cohomology  $\big(\Omega^\bullet(\M), \bs d \big)$ of the base moduli space.
Therefore, basic forms can also be defined as
 \begin{align}
\Omega_\text{basic}^\bullet(\Phi)\defeq\left\{ \, \bs\alpha \in \Omega^\bullet(\Phi)\,|\, \exists\, \bs\beta \in \Omega^\bullet(\M) \text{ s.t. } \bs\alpha=\pi^\star\bs\beta \,  \right\}
= \text{Im}(\pi^\star). 
\end{align} 
The basic cohomology is especially important when 
it is unpractical (or impossible) to work concretely on $\M$, as is generically the case in gRGFT. 
Observe that the analogue of $\Gamma_{\text{\!\tiny{inv}}}(T\Phi)$ for forms 
is not $\Omega^\bullet_\text{inv}(\Phi)$ 
but $\Omega_\text{basic}^\bullet(\Phi)$. Only  basic forms induce forms in $\Omega^\bullet(\M)$, thus contain only physical d.o.f.
In section \ref{Building basic forms via dressing} we review  a systematic method to build the basic version $\bs\alpha^\u \in \Omega_\text{basic}^\bullet(\Phi)$ of a form $\bs\alpha \in \Omega^\bullet(\Phi)$: the Dressing Field Method (DFM) .

\paragraph{Gauge transformations}

As seen above, the  gauge transformation of a form $\bs\alpha \in \Omega^\bullet(\Phi)$ is its  pullback by $\bs\Aut_v(\Phi)$ and expressible in terms of the elements of $\bs\Diff(I)$,  hence the notation: $\bs\alpha^{\bs\psi} \defeq \Xi^\star \bs\alpha$.
Concrete expressions are obtained by using the  pullback/pushforward duality, together with  \eqref{pushforward-X}:  For any $(\mathfrak X, \ldots) \in \Gamma(T\Phi)$   one has 
\begin{equation}
\label{GT-geometric}
\begin{aligned}
\bs\alpha^{\bs\psi}_{|\phi} (\mathfrak X_{|\phi}, \ldots) =  \Xi^\star \bs\alpha_{|\Xi(\phi)} (\mathfrak X_{|\phi}, \ldots)= \bs\alpha^{}_{|\Xi(\phi)} (\Xi_\star \mathfrak X_{|\phi}, \ldots) &= \bs\alpha^{}_{|\phi^{\bs\psi(\phi)}}\left( R_{\bs\psi(\phi) \star} \left( \mathfrak X_{|\phi} +  \left\{  \bs d \bs\psi_{|\phi}(\mathfrak X_{|\phi})  \circ \bs\psi(\phi)\-  \right\}^v_{|\phi} \right), \ldots\right)  \\
&=R_{\bs\psi(\phi)}^\star\, \bs\alpha_{|\phi^{\bs\psi(\phi)}} \left(  \mathfrak X_{|\phi} +  \left\{  \bs d \bs\psi_{|\phi}(\mathfrak X_{|\phi})  \circ \bs\psi(\phi)\-  \right\}^v_{|\phi}, \ldots \right).
\end{aligned}
\end{equation}
From this, it is manifest that the gauge transformation of a form is controlled by its equivariance and verticality properties. 
In particular,  the vertical transformation of a tensorial form is controlled simply by its equivariance: 
\begin{equation}
\label{GT-tensorial}
\begin{split}
\text{For }\ \bs\alpha \in \Omega_\text{tens}^\bullet(\Phi, \rho), \quad \bs\alpha^{\bs\psi} = \rho(\bs\psi)\- \bs\alpha; 
\qquad \quad &\text{  for }\  
 \bs\alpha \in \Omega_\text{tens}^\bullet(\Phi, C), \quad \bs\alpha^{\bs\psi} = C(\bs\psi)\- \bs\alpha,\\
\end{split}
\end{equation}
where we introduce the simplified notation $[C(\bs\psi)](\phi) \defeq C\big(\phi; \bs\psi(\phi)\big)$ for the 1-cocycle.\footnote{If $G=U(1)$, then $[C(\bs\psi)](\phi)=\exp{i\, [c(\bs\psi)](\phi)}$ with $[c(\bs\psi)](\phi) \defeq c\big(\phi; \bs\psi(\phi)\big)$.}
The infinitesimal gauge transformations, under $\bs\aut_v(\Phi) \simeq \bs\diff(I)$, are given by the linearisation  of \eqref{GT-geometric}, 
\begin{align}
\label{Inf-GT-general}
\bs L_{\bs X^v} \bs\alpha = \tfrac{d}{ds} \, R_{\bs\psi_s}^\star \bs\alpha\, \big|_{s=0} + \iota_{\{\bs{dX}\}^v} \bs\alpha,
\end{align}
with $\bs X=\tfrac{d}{ds}\bs\psi_s \big|_{s=0}$.\footnote{ 
Remark that $\{\bs{dX}\}^v$ can be seen as an element of $\Omega^1(\Phi, V\Phi)$, so $\iota_{\{\bs{dX}\}^v}$ is an algebraic derivation (of degree 0) as defined in \cite{Kolar-Michor-Slovak}.}
For example, for a $\Omega^\bullet(I)$-valued form $\bs\alpha$, this specialises to
$\bs L_{\bs X^v} \bs\alpha = \mathfrak L_{\bs X} \bs\alpha + \iota_{\{\bs{dX}\}^v} \bs\alpha$,\footnote{In some works on the covariant phase space literature of gRGFT \cite{Hopfmuller-Freidel2018, Chandrasekaran_Speranza2021, Freidel-et-al2021, Freidel-et-al2021bis, Speziale-et-al2023}, a so-called ``anomaly operator" $\Delta_{\bs X}$ was introduced and defined as: 
 $\Delta_{\bs X}\defeq \bs L_{\bs X^v} - \mathfrak L_{\bs X} - \iota_{\{\bs{dX}\}^v}$. See how the above clarifies the geometric meaning of this operator, here appearing in the context of MFT.
We observe that $\Delta_{\bs X}$ can only be non-zero on the field space  $\Phi$ of  theories admitting background non-dynamical structures or fields ``breaking"  gauge  and/or diffeomorphism covariance, i.e. theories that  fundamentally fail to comply with the core physical (symmetry) principles  of gRGFT. }
while for tensorial forms it gives the infinitesimal versions of \eqref{GT-tensorial},
\begin{align}
\label{GT-inf-tensorial}
\bs L_{\bs X^v} \bs \alpha = -\rho_*(\bs X) \bs \alpha, \qquad \bs L_{\bs X^v} \bs \alpha = -a(\bs X) \bs \alpha, 
\end{align}
where we  introduce the notation $[a(\bs X)](\phi)\defeq a\big(\bs X(\phi); \phi\big)$ for the linearised 1-cocycle.
Given their definition, basic forms are  gauge invariant: 
\begin{align}
\label{GT-basic}
\text{For }\ \bs\alpha \in \Omega_\text{basic}^\bullet(\Phi), \quad \bs\alpha^{\bs\psi} = \bs\alpha \quad \text{so}\quad  L_{\bs X^v} \bs \alpha=0,
\end{align}
as expected for forms containing only physical, $\bs\Diff(I)$-invariant, d.o.f. and inducing forms on the moduli space~$\M$.
 

To illustrate, let us consider the  example of the basis 1-form $\bs d\phi = \big(\bs d x, \bs d t \big)  \in \Omega^1(\Phi)$, which is important 
since its gauge transformation $\bs d\phi^{\bs\psi}\defeq \Xi^\star \bs d \phi$ features in the general formulae \eqref{GT-general}  for the gauge transformation of any form. 
The equivariance and verticality properties of $\bs d\phi$ are given by definition: 
 \begin{align}
 \label{basis-1-form}
R^\star_\psi \bs d\phi \defeq \psi^* \bs d \phi, \qquad \text{and} \qquad \bs \iota_{X^v}\bs d\phi \defeq \mathfrak L_X \phi, 
 \end{align}
the verticality property reproducing the $\diff(I)$-transformation of the field $\phi=(x, t)$.
 It is then immediate that
 \begin{align}
  \label{GT-basis-1-form}
  \bs d\phi^{\bs\psi}   \defeq \Xi^\star \bs d \phi = \bs\psi^*   \big(  \bs d\phi   +   \mathfrak L_{ \bs d \bs\psi  \circ \bs\psi\-} \phi  \big).   
 \end{align}
Similar results are arrived at in a heuristic way in the literature on covariant phase space of gRGFT (see e.g. \cite{DonnellyFreidel2016}). 
In \cite{JTF-Ravera2024gRGFT}, it is derived from geometric first principles as we did here.
The linear version~is 
 $\bs L_{\bs X^v} \bs d\phi   =  \mathfrak L_{\bs X} \bs d \phi +   \mathfrak L_{\bs d \bs X} \phi$,
as is clear also from \eqref{Inf-GT-general}.

\subsection{Connections on Mechanical Field Space}  
\label{Connections on Mechanical Field Space}  

As reminded above, connections on the MFS $\Phi$ are necessary to define covariant derivatives on $\Omega_\text{tens}^\bullet(\Phi)$:
  For standard tensorial forms one needs an Ehresmann connection 1-form, while for cocyclic tensorial forms one needs a generalisation called \emph{cocyclic connection} 1-form \cite{Francois2019_II, JTF-Ravera2024gRGFT, JTF-Ravera2025NRrelQM}. 
  We briefly review these two notions.
\smallskip

{An} Ehresmann connection $\bs \omega \in \Omega^1_{\text{eq}}\big(\Phi, \diff(I) \big)$ on the Mechanical Field Space $\Phi$ is defined by  two properties:
\begin{equation}
\label{Variational-connection}
\begin{aligned}
\bs\omega_{|\phi} \big( X^v_{|\phi}\big)&=X, \quad \text{for } X\in \diff(I), \\
R^\star_\psi \bs\omega_{|\phi^\psi} &= \psi\-_* \, \bs\omega_{|\phi}  \circ\, \psi. 
\end{aligned}
\end{equation}
The infinitesimal equivariance, under $\diff(I)$, is
$\bs L_{X^v} \bs\omega = \tfrac{d}{ds}\, R^\star_{\psi_s} \bs\omega\, \big|_{s=0} = \tfrac{d}{ds}\,  {\psi_s\-}_* \, \bs\omega  \circ\, \psi_s \,\big|_{s=0} 
		= [X, \bs \omega]_{\text{{\tiny $\Gamma(TP)$}}} = [\bs\omega, X]_{\text{{\tiny $\diff(I)$}}}$.
The space of connection $\C$ is an affine space modelled on the vector space $\Omega^1_{\text{tens}}\big(\Phi, \diff(I)\big)$:  
Given $\bs\omega \in \C$ and $\bs\beta \in \Omega^1_{\text{tens}}\big(\Phi, \diff(I)\big)$, we have that $\bs\omega'=\bs\omega + \bs \beta \in \C$. 
As foretold, a $\bs\omega$ induces a covariant derivative on standard tensorial forms via  $\bs D :  \Omega^\bullet_\text{tens} \big(\Phi, \rho) \rarrow   \Omega^{\bullet+1}_\text{tens} \big(\Phi, \rho)$, $\bs\alpha \mapsto \bs{ D\alpha}=\bs{d\alpha} +\rho_*(\bs\omega) \bs\alpha$. 
The curvature of $\bs\omega$ is $\bs \Omega \in \Omega^2_\text{tens} \big(\Phi, \diff(I)\big)$ and   given by Cartan structure equation
\begin{align}
\label{Curvature-Cartan-eq}
\bs\Omega = \bs{d \omega} +\tfrac{1}{2}[\bs \omega, \bs \omega]_{\text{{\tiny $\diff(I)$}}},
\end{align}
and thus satisfies the Bianchi identity $\bs{D\Omega}=\bs{d\Omega}+[\bs\omega, \bs \Omega]_{\text{{\tiny $\diff(I)$}}} \equiv0$. 
On tensorial forms, one shows that $\bs D \circ \bs D =\rho_*(\bs \Omega)$.

Given the defining properties \eqref{Variational-connection}, using \eqref{pushforward-X}/\eqref{GT-geometric} one finds that a connection gauge transforms as 
\begin{align}
\label{Vert-trsf-connection}
\bs\omega^{\bs\psi }\defeq \Xi^\star \bs\omega =\bs\psi\-_* \, \bs\omega  \circ \bs\psi + \bs\psi\-_*\bs{d\psi}. 
\end{align}
Correspondingly, its $\aut_v(\Phi)\simeq \bs\diff(I)$-transformation is
$\bs L_{\bs X^v} \bs\omega =\bs{dX}+ [\bs\omega, \bs X]_{\text{{\tiny $\diff(I)$}}}$. 
Similarly, the finite  gauge transformations of the curvature is $\bs\Omega^{\bs\psi }\defeq \Xi^\star \bs\Omega =\bs\psi\-_* \, \bs\Omega  \circ \bs\psi$, and is as special cases of \eqref{GT-tensorial}. Linearly it is $\bs L_{\bs X^v} \bs\Omega =[\bs\Omega, \bs X]_{\text{{\tiny $\diff(I)$}}}$, as a special case of \eqref{GT-inf-tensorial}.
\smallskip

A \emph{cocyclic connection} 1-form $\bs \varpi \in \Omega^1_{\text{eq}}\big(\Phi, \LieG \big)$ is defined by the two following properties:
\begin{equation}
\label{Variational-twisted-connection}
\begin{aligned}
\bs\varpi_{|\phi} \big( X^v_{|\phi}\big)&= \tfrac{d}{ds} C(\phi; \psi_s) \,\big|_{s=0}=a(X, \phi) \ \in \LieG, \quad \text{for } X\in \diff(I), \\[2mm]
R^\star_\psi \bs\varpi_{|\phi^\psi} &= \Ad_{C(\phi; \psi)\-} \, \bs\varpi_{|\phi} + C(\phi; \psi)\-\bs d C(\ \, ; \psi)_{|\phi}. 
\end{aligned}
\end{equation}
The infinitesimal equivariance, under $\diff(I)$, is thus given by
$\bs L_{X^v} \bs\varpi = \tfrac{d}{d\tau}\, R^\star_{\psi_\tau} \bs\varpi\, \big|_{\tau=0} = \bs d a(X; \ ) + [\bs\varpi, a(X;\ )]_{\text{{\tiny $\LieG$}}}$.
The~space of twisted connections $\b\C$ is an affine space modelled on the vector space $\Omega^1_{\text{tens}}\big(\Phi, \LieG\big)$: 
Given $\bs\omega \in \b\C$ and $\bs\beta \in \Omega^1_{\text{tens}}\big(\Phi, \LieG\big)$, we have that $\bs\omega'=\bs\omega + \bs \beta \in \b\C$. 
A \emph{cocyclic covariant derivative} is  defined as the first order linear operator $\b{\bs D} :  \Omega^\bullet_\text{tens} \big(\Phi, C) \rarrow   \Omega^{\bullet+1}_\text{tens} \big(\Phi, C)$, $\bs\alpha \mapsto \b{\bs D} \bs\alpha \defeq \bs{d\alpha} +\bs\varpi \bs\alpha$. 
The curvature 2-form of $\bs\varpi$ is  defined by 
\begin{align}
\label{Twisted-curvature}
\b{\bs\Omega} \defeq \bs{d \varpi} +\tfrac{1}{2}[\bs \varpi, \bs \varpi]_{\text{{\tiny $\LieG$}}} \ \  \in  \Omega^2_\text{tens} \big(\Phi, \LieG \big).
\end{align}
It thus satisfies the Bianchi identity, $\b{\bs D} \b{\bs\Omega}=\bs d\b{\bs\Omega}+[\bs\varpi, \b{\bs \Omega}]_{\text{{\tiny $\LieG$}}}=0$. 
On cocyclic tensorial forms, we have $\b{\bs D} \circ \b{\bs D} =\rho_*(\b{\bs \Omega})$.

The gauge transformation of a twisted connection is, by \eqref{Variational-twisted-connection} and \eqref{pushforward-X}/\eqref{GT-geometric}, 
\begin{align}
\label{Vert-trsf-twisted-connection}
\bs\varpi^{\bs\psi }\defeq \Xi^\star \bs\varpi =\Ad_{C(\bs\psi)\- } \, \bs\varpi  + C(\bs\psi)\-\bs{d} C(\bs\psi),
\end{align}
See \cite{Francois2019_II} for a  proof.
Correspondingly, its transformation under $\bs\aut_v(\Phi)\simeq \bs\diff(I)$  is
$\bs L_{\bs X^v} \bs\varpi =\bs{d} a(\bs X)+ [\bs\varpi, a(\bs X)]_{\text{{\tiny $\LieG$}}}$. 
Finite and infinitesimal general gauge transformations of the curvature are given by 
$\b{\bs\Omega}^{\bs\psi }\defeq \Xi^\star \b{\bs\Omega} =\Ad_{C(\bs\psi)\-} \, \b{\bs\Omega}$ 
and $\bs L_{\bs X^v}\b{\bs\Omega} =[\b{\bs\Omega}, a(\bs X)]_{\text{{\tiny $\LieG$}}}$. 
These illustrate \eqref{GT-tensorial} and \eqref{GT-inf-tensorial}.

\subsection{Associated bundle of regions and integration}  
\label{Associated bundle of region and integration}  
 
We aim to define integration on $I$ geometrically, as an operation  on what we call the \emph{associated bundle of regions}. 
Bundles associated to a principal bundle are built from representations, or spaces supporting the action, of its structure group. 
The associated bundle of region is built from the ``defining representation" of $\Diff(I)$:
the $\sigma$-algebra of closed sets (intervals) of $I$, 
$\bs i(I)\defeq \big\{\, i \subset I\ | \ i \text{ closed set } \big\}$.\footnote{It is actually the defining representation of $\Diff(I)$ as a (Lie) pseudo-group \cite{Kobayashi1972}.} 
As usual, one first defines the right action of $\Diff(I)$ on the direct product space $\Phi \times \bs i(I) $:
 \begin{equation}
 \label{right-action-diff}
\begin{aligned}
\big( \Phi \times \bs i(I) \big) \times \Diff(I) &\rarrow  \Phi \times \bs i(I), \\
\big( (\phi, i), \psi \big) &\mapsto \b R_\psi (\phi,  i) \defeq \big( \psi^*\phi, \psi\-(i) \big). 
\end{aligned}
\end{equation}
The  \emph{associated bundle of regions} is then defined as the quotient of the product space by this right action:
 \begin{equation}
 \label{Assoc-bundle-regions}
\begin{aligned}
\mathcal R(I)
= \Phi \times_\text{{\tiny $\Diff(I)$}} \bs i(I) \defeq   \Phi \times  \bs i(I) /\!\sim.
\end{aligned}
\end{equation}
It is a standard result of bundle theory that its space of sections $\Gamma\big(\mathcal R(I)\big)\defeq \big\{ \b{\bs r}: \M \rarrow \mathcal R(I) \big\}$ is isomorphic to
 \begin{equation}
 \label{equiv-regions}
\begin{aligned}
\Omega^0_\text{tens}\big(\Phi,  \bs i(I)  \big)\defeq \big\{\, \bs r: \Phi \rarrow \bs i(I), \phi \rarrow \bs r(\phi) \ |\  R^\star_\psi {\bs r} = \psi\-({\bs r}) \, \big\}.
\end{aligned}
\end{equation}
A 0-form   $\bs r(\phi)=\bs r(x, t)$ may be understood as a ``field-dependent"  set  of $I$, i.e. a region of $I$ defined in a ``$\phi$-\emph{relative}" and $\Diff(I)$-equivariant way.
By \eqref{GT-tensorial}-\eqref{GT-inf-tensorial}, its gauge transformations, under $\bs\Aut_v(\Phi) \simeq \bs\Diff(I)$ and $\bs\aut_v(\Phi) \simeq \bs\diff(I)$, are respectively: 
\begin{align}
\bs r^{\bs \psi} = \bs\psi\-(\bs r), \quad \text{ and } \quad \bs L_{\bs X^v} \bs r = -\bs X(\bs r). 
\end{align} 

 \paragraph{Integration map} 

Integration on $I$ may provide such an example of equivariant functional.
For the detailed elaboration of the viewpoint on integration on a manifold as an operation on the associated bundle of region on field space, see section 3.5.2 of \cite{JTF-Ravera2024gRGFT}. Here we go straight to the point. 

We consider the ``dual representation" $\Omega^\bullet(I)$ of $\Diff(I)$, i.e. dual to $\bs i(I)$ w.r.t. to the $\Diff(I)$-invariant 
\emph{integration pairing}:
\begin{equation}
\begin{aligned}
\langle\ , \ \rangle: \bs \Omega^\text{top}(i) \times \bs i(I) &\rarrow \RR, \\[-2mm]                  (\omega, i) &\mapsto \langle \omega , i \rangle \defeq \int_i \omega. 
\end{aligned}
\end{equation}
 The invariance property is 
 \begin{align}
 \label{invariance-action}
\langle\, \psi^*\omega ,   \psi\-(i) \rangle =  \langle \omega , i \,\rangle \quad   \rarrow  \quad  \int_{\psi\-(i)} \psi^*\omega = \int_i \omega.
 \end{align}
 A familiar identity
 which upon linearisation gives
 \begin{align}
 \label{inf-invariance-action}
\langle\, \mathfrak L_X \omega ,   i \,\rangle + \langle\,  \omega ,   -X(i) \rangle =0  \quad
\rarrow \quad
  \int_i   \mathfrak L_X \omega  + \int_{-X(i)} \hspace{-2mm}\omega =0.
 \end{align}
 This may be understood as a  \emph{continuity equation} for the action of $\diff(I)$. 
 By Stokes theorem,
the de Rham derivative $d$ on $\Omega^\bullet(I)$ and the boundary operator $\d$ on $\bs i(I)$ are  adjoint operators w.r.t. to the integration pairing:
 \begin{align}
 \label{Stokes}
 \langle\, d \omega ,  i \,\rangle = \langle\,  \omega ,   \d i \,\rangle \quad
 \rarrow \quad
  \int_i   d \omega  = \int_{\d i} \hspace{-2mm}\omega. 
\end{align} 
Observe that integration of a $0$-form, say $\phi$, on $0$-dimensional submanifolds $\tau \subset I$ is the \emph{evaluation operation}:
\begin{align}
\label{ev-op}
{\sf ev}(\phi, \tau) \defeq \int_\tau \phi 
=\phi(\tau).
\end{align}
The banal notation $\phi(\tau)$ should not obscur the fact that this operation is $\Diff(I)$-invariant by definition; ${\sf ev} \circ \b R_\psi = {\sf ev}$.


Considering  now $\bs \alpha \in \Omega^\bullet_\text{eq}\big( \Phi,  \Omega^\text{top}(i) \big)$, the field-dependent top forms on $I$, we define  integration  on $\Phi \times \bs i(I)$:
\begin{align}
\mathcal I(\bs\alpha_{|\phi}, i) = \langle \bs\alpha_{|\phi}, i \,\rangle \defeq \int_i \bs \alpha_{|\phi}.
\end{align}
We may use the shorthand notation $\bs\alpha_i$ for the integral $\mathcal I(\bs\alpha, i)$ when no confusion is likely. 
We have naturally $\bs d \mathcal I(\bs\alpha, i) = \mathcal I(\bs{d\alpha}, i)$ and $\bs\iota_{\mathfrak X}  \mathcal I(\bs\alpha, i ) = \mathcal I( \bs\iota_{\mathfrak X} \bs\alpha, i)$ for $\mathfrak X \in \Gamma(T\Phi)$.
The induced actions of $\Diff(I)$ 
and $\bs\Diff(I)  \simeq \bs\Aut_v(\Phi)$ on such integrals (their ``equivariance" and ``gauge transformations") are:
\begin{equation}
\label{Equiv-GT-Int}
\begin{aligned}
 &\t R^\star_\psi \mathcal I(\bs\alpha,\   )_{|\big(\psi^*\phi,\ \psi\-(i) \big)} \defeq  \langle R^\star_\psi \bs\alpha_{|\psi^*\phi},\, \psi\-(i) \rangle = \int_{ \psi\-(i)} R^\star_\psi \bs\alpha_{|\psi^*\phi}
 = \int_{ \psi\-(i)} \psi^* \bs\alpha_{|\psi^*\phi}
=\int_i \bs\alpha_{|\phi} 
= \mathcal I(\bs\alpha,\   )_{|(\phi,\ i )},\\[2mm]
& \t \Xi^\star \mathcal I(\bs\alpha,\   )_{|\big(\Xi (\phi),\  \bs\psi\-(i) \big)} \defeq  \langle \Xi^\star \bs\alpha_{|\Xi(\phi)},\, \bs\psi\-(i)  \rangle 
= \int_{ \bs\psi\-(i)}  \Xi^\star \bs\alpha_{|\Xi(\phi)}. 
 \end{aligned}
\end{equation}
We may write the above as ${\bs\alpha_i}^\psi =\langle \bs\alpha, i \rangle^{\psi}=\bs\alpha_i $ and ${\bs\alpha_i}^{\bs\psi}=\langle \bs\alpha, i \rangle^{\bs\psi}$, respectively. 
Remark that, in the latter case,  acting with $\bs d$ will affect the transformed region $\bs\psi\-(i)$ due to the $\phi$-dependence of $\bs\psi$. 
We have indeed:
\begin{align}
\label{commut-bXi-d-int}
\bs d \left( \b\Xi^\star \langle \bs\alpha, i \rangle \right)
&=
\langle \bs d \Xi^\star \bs\alpha, \bs\psi\-(i) \rangle 
+
\langle \Xi^\star \bs\alpha, \bs d\bs\psi\-(i)\rangle \notag\\
&=
\langle \Xi^\star \bs d\bs\alpha, \bs\psi\-(i) \rangle 
+
\langle \Xi^\star \bs\alpha, - \bs\psi
\-_* \bs d\bs\psi\circ \bs\psi\-
(i) \rangle \notag \\
&=
\b \Xi^\star \langle  \bs d\bs\alpha, i \rangle 
-
\langle \mathfrak L_{ \bs\uppsi
\-_* \bs d\bs\uppsi}\,\Xi^\star \bs\alpha, 
\bs\uppsi\-(i) \rangle, \notag\\[.5mm]
\text{or}\quad
\bs d\big({\bs\alpha_i}^{\bs\psi} \big)
&=
\left( \bs d\bs\alpha_i\right)^{\bs\psi}
- \langle \mathfrak L_{ \bs\psi
\-_* \bs d\bs\psi}\,\Xi^\star \bs\alpha, 
\bs\psi\-(i)  \rangle.
\end{align}
The identity \eqref{inf-invariance-action} has been used to conclude.
Also, while we used the standard relation $[{\bs \Xi}^\star, \bs d] =  0$  holding on $\Phi$ -- which means that $\bs d$ is, as is well-known, a \emph{natural operator} on $\Phi$ -- this result implies that on $\Phi \times \bs i(I)$ we have $[\b{\bs \Xi}^\star, \bs d]\neq 0$ -- meaning that $\bs d$ is not a natural operator on $\Phi \times \bs i(I)$.
Fortunately, the commutator is a boundary term, $\langle \iota_{ \bs\psi
\-_* \bs d\bs\psi}\,\Xi^\star \bs\alpha , \d\big( \bs\psi\-(i)\big) \rangle$, since $\bs\alpha$ is a top form  on $i$ and by \eqref{Stokes}; a fact that will bear on our discussion of the relational variational principle for MFT in section \ref{Basic Classical Mechanics}.

By linearisation of \eqref{Equiv-GT-Int}, the actions of $\diff(I)$ and $\bs\diff(I)\simeq \bs\aut_v(\Phi)$ are found to be,
\begin{equation}
\label{linear-versions-integral}
\begin{aligned}
&\tfrac{d}{ds}\, \t R_{\psi_s}^\star \mathcal I (\bs\alpha,  i) \,\big|_{s=0} 
= \langle \bs L_{X^v} \bs\alpha,\,  i \rangle 
				   + \langle \bs\alpha,\, -X(i) \rangle
=\int_i  \mathfrak L_{X} \bs\alpha + \int_{-X(i)} \bs\alpha = 0, \\
&\tfrac{d}{ds}\, \t \Xi_s^\star \mathcal I  (\bs\alpha,  i)\, \big|_{s=0} 
= \langle \bs L_{\bs X^v} \bs\alpha,\,  i \rangle 
		 + \langle \bs\alpha,\, -\bs X(i)   \rangle
=\int_i  \bs L_{\bs X^v} \bs\alpha + \int_{-\bs X(i)} \bs\alpha.  
\end{aligned}
\end{equation}
We may write the above, suggestively, as $\delta_X\ \!{\bs\alpha_i} =\delta_X\ \!  \langle \bs\alpha, i \rangle=0$ and $\delta_{\bs X}\ \!{\bs\alpha_i}=\delta_{\bs X}\ \!  \langle \bs\alpha, i \rangle$ respectively. 

For $\bs\alpha \in \Omega_\text{tens}^\bullet\big(\Phi, \Omega^{\text{top}}(i) \big)$  
we have that $\bs\alpha^{\bs \psi} = \Xi^\star \bs\alpha = \bs\psi^* \bs\alpha$, 
so that  its integral is $\bs\Diff(I)$-invariant,  ${\bs\alpha_i}^{\bs \psi} =\bs\alpha_i$. 
Linearly, the $\bs\diff(I)$-invariance of its integral implies
\begin{equation}
\label{Cont-eq-general-case}
\begin{aligned}
\delta_{\bs X} \bs\alpha_i =0 \quad \Rightarrow \quad \langle \bs L_{\bs X^v} \bs\alpha,\,  i \rangle + \langle \bs\alpha,\, -\bs X(i) \rangle &= 0, \qquad \bs X \in \bs\diff(I), \\[.5mm]
  \langle  \mathfrak L_{\bs X} \alpha,\,  i \rangle + \langle \bs\alpha,\, -\bs X(i)   \rangle &= 0 \quad \rarrow \quad \int_i \mathfrak L_{\bs X} \alpha \ + \int_{-\bs X(i)} \bs\alpha =0,
\end{aligned}
\end{equation}
which again can be interpreted as a continuity equation. 
Therefore, when $\bs\alpha$ is tensorial, its integral $\bs\alpha_i=\mathcal I(\bs\alpha, i)$ is  a well-defined object on the bundle of regions $\mathcal R(I)=\Phi \times \bs i(I)/\!\sim$. 
Since $\bs\alpha_i$ is constant along $\Diff(I)$-orbits in  $\Phi \times \bs i(I)$, one may  define a $\Diff(I)$-equivariant $\bs i(I)$-valued function $\bs r_\mathcal{I}$ on $\Phi$, using $\pi_{\bs i(I)}: \Phi \times \bs i(I) \rarrow \bs i(I)$, as: 
 \begin{equation}
 \label{Induced-equiv-fct}
\begin{aligned}
\bs r_{\mathcal I}(\phi)&\defeq \pi_{\bs i(I)}(\phi,  i)_{|\mathcal I(\bs\alpha_{|\phi}, i) =\text{cst}} \equiv i, \\
\bs r_{\mathcal I}(\psi^*\phi)& \defeq \pi_{\bs i(I)}\big(\psi^*\phi, \psi\- (i)\big)_{|\mathcal I(\bs\alpha_{|\phi}, i) =\text{cst}} \equiv \psi\- (i).
 \end{aligned}
\end{equation}
This map outputs
intervals of $I$ defined in a ``$\phi$-\emph{relative}" and $\Diff(I)$-equivariant way, i.e. $\bs r_\mathcal{I} \in\Omega^0_\text{tens}\big(\Phi,  \bs i(I)  \big)$.
By the isomorphism $\Omega^0_\text{tens}\big(\Phi,  \bs i(I)  \big) \simeq \Gamma\big(\mathcal R(I) \big)$, $\bs r_\mathcal{I}$ is is equivalent to a section of the bundle of regions, $\b{\bs r}_{\mathcal I} :\M \rarrow \mathcal R(I)$. 
The $\bs\Diff(I)$-invariance of $\bs\alpha_i$
 also implies that $\bs d({\bs\alpha_i}^{\bs\psi}) = \bs d \bs\alpha_i$, i.e. 
\begin{align}
\label{Invariance-tens-int}
\bs d \, \langle \bs\psi^* \bs\alpha, \bs\psi\-(i)\rangle 
= \bs d  \langle  \bs\alpha, i\,\rangle 
= \langle \bs d \bs\alpha, i\,\rangle 
\quad \rarrow \quad
\bs d \int_{\bs\psi\-(i)}  \bs\psi^* \bs\alpha  = \bs d \int_i \bs \alpha
= \int_i \bs{d\alpha}.
\end{align}
Finally, one proves the following identity:
\begin{equation}
\label{Vert-trsf-int-dalpha}
\begin{aligned}
 \langle  \bs{d\alpha}, i \rangle ^{\bs \psi} 
 = \langle\bs{d\alpha}, i \rangle + \langle \mathfrak L_{\bs{d\psi}\circ \bs\psi\-} \bs\alpha,  i \,\rangle, 
\end{aligned}
\end{equation}
which will be important when discussing the variational principle in MFT. 


\section{Mechanical Field Theory on $\Phi$: Dynamics} 
\label{Mechanical Field Theory on Phi}  

We develop {MFT} on $\Phi$ as a ``general-relativistic" field theory. 
We  illustrate the abstract framework via the simplest, non-relativistic, case.
Its $\Diff(I)$-covariance is shown to encode its relational character. 



The classical (i.e. non-quantum) dynamics is specified by a Lagrangian $L \in \Omega_\text{tens}^0\big(\Phi, \Omega^{\text{top}}(i) \big)$, with equivariance
\begin{equation}
\begin{aligned}
\label{Equiv-L}
R^\star_\psi L &= \psi^* L, \quad \psi\in \Diff(I), \\
\bs L_{X^v} L &= \mathfrak L_X L = d\iota_X L, \quad X\in \diff(I).
\end{aligned}
\end{equation}
We may observe that this can be rewritten in terms of the (trivial Abelian) 1-cocycle $c(\phi, \psi)\defeq \psi^* L- L$,  easily shown to satisfy \eqref{cocycle-bis}, as
\begin{equation}
\begin{aligned}
\label{Equiv-L-bis}
R^\star_\psi L &= L + c(\,\_, \psi), \quad \psi\in \Diff(I), \\
\bs L_{X^v} L &= a(X;\_ \,), \quad X\in \diff(I),
\end{aligned}
\end{equation}
where $a(X; \phi)=\mathfrak L_X L$ is a (trivial Abelian) $\diff(I)$-1-cocycle: the classical $\diff(I)$-anomaly. It satisfies the Abelian version of \eqref{inf-cocycle}, which reproduces the Wess-Zumino consistency condition: $X^v\! \cdot a(Y; \phi)- Y^v\!\cdot a(X; \phi)=a([X,Y]_{\text{{\tiny $\diff(I)$}}}; \phi)$.
As observed earlier, the gauge transformation of tensorial forms is controlled by their equivariance. So, as a special case of \eqref{GT-tensorial}-\eqref{GT-inf-tensorial}, 
the gauge transformation of the Lagrangian is immediately found to be
\begin{equation}
\begin{aligned}
\label{GT-L}
L^{\bs \psi} &= L + c(\,\_, \bs\psi) =\bs\psi^* L, \quad \bs\psi\in \bs\Diff(I), \\
\bs L_{\bs X^v} L &= a(\bs X;\_ \,)=\mathfrak L_{\bs X} L, \quad \bs X\in \bs\diff(I).
\end{aligned}
\end{equation}
Such a transformation (under $\bs\Diff(I)$)  is known in the literature as the ``\emph{field-dependent gauge transformation}" of $L$, while the equivariance (under $\Diff(I)$) is then called ``\emph{field-independent gauge transformation}".
The integral of the Lagrangian $L$ over a region $i\subset I$ is the action functional 
\begin{equation}
\label{Action-functional}
\begin{aligned}
 S : \Phi \times \bs i(I) &\rarrow \RR, \\
  (\phi, i) & \mapsto S\defeq \langle L, i \rangle = \int_i L(\phi). 
\end{aligned}
\end{equation}
It is naturally invariant under the action of $\Diff(I)$, as a special case of \eqref{Equiv-GT-Int}, and since the Lagrangian $L$ is tensorial, it is furthermore $\bs\Diff(I)$-invariant, $S^{\bs\psi} = S$. 
This is the first indication that Mechanical Field Theory has a much larger covariance group than initially conceived, 
as we shall shortly confirm in detail.
The linear invariance of $S$ is
\begin{equation}
\label{Inf-inv-S}
\begin{aligned}
\delta_{\bs X} S = 0 \quad \Rightarrow \quad \langle  \mathfrak L_{\bs X} L,\,  i \rangle + \langle L,\, -\bs X(i)   \rangle = 0,
 \quad \text{or} \quad \int_i \mathfrak L_{\bs X} L\ = \int_{\bs X(i)} \!\!\!\! L,
\quad \text{for }\ \bs X \in \bs\diff(I), 
\end{aligned}
\end{equation}
as a special case of \eqref{Cont-eq-general-case}. The same  holds for $X\in \diff(I)$.
One may see it as a continuity equation for the actions of $\Diff(I)$ and $\bs\Diff(I)$.
 Since $S$ is constant along $\Diff(I)$-orbits in  $\Phi \times \bs i(I)$, it is a well-defined section of the associated bundle of regions $\mathcal R(I)=\Phi \times \bs i(I)/\!\sim$, and induces
 a $\Diff(I)$-equivariant $\bs i(I)$-valued function $\bs r_S$ on $\Phi$ as in
\eqref{Induced-equiv-fct}.

\subsection{The variational principle for MFT, and its covariance group} 
\label{The variational principle for MFT, and its covariance group} 

The relevant object for the variational principle, over a bounded region $i\subset I$ with boundary $\d i=\{i_0, i_1 \}$ (the codimension 1, dim $=0$, two points boundary), is $\bs dS = \langle \bs d L, i \rangle$, with  $\bs d L = d\phi \tfrac{\delta}{\delta \phi}L = \bs d x \tfrac{\delta}{\delta x} L + \bs d t \tfrac{\delta}{\delta t} L \ \in \Omega^1\big(\Phi, \Omega^1(i) \big)$. 
From general considerations laid out in \cite{JTF-Ravera2024gRGFT} (section 5.1.3.),
we can write the variational principle as
\begin{equation}
\label{dS-dL}
\begin{aligned}
\bs dS_{|\phi}= \int_i \bs d L_{|\phi}= 0 \quad
\text{ with } \quad
\bs d L_{|\phi} = \bs E_{|\phi} + d\bs \theta_{|\phi} 
= E(\bs d\phi; \phi) + d \theta(\bs d\phi; \phi),
\end{aligned}
\end{equation}
with $\bs E \in \Omega^1\big(\Phi, \Omega^1(i) \big)$ the field equations 1-form, and $\bs \theta \in \Omega^1\big(\Phi, \Omega^{0}(i) \big)$ is referred to as the presymplectic potential current in the ``covariant phase space" approach to gRGFT (see Appendix \ref{Covariant phase space formalism for MFT}). 
The space of solutions is 
\begin{align}
\S\defeq \big\{\,\phi \in \Phi \ |\ \bs E_{|\phi}=0 \, \big\}.
\end{align}
A fundamental goal is to assess its stability under the action of the structure and gauge groups of $\Phi$. 
To do so, we shall consider the equivariance and gauge transformations of all objects involved in \eqref{dS-dL}. 


First, observe that while $L$ is tensorial, $\bs dL \in \Omega^1\big(\Phi, \Omega^1(I) \big)$ is not since $\bs d$ does not preserve tensoriality (as stressed in section \ref{Connections on Mechanical Field Space}). The equivariance and verticality properties of $\bs d L$ are obtained from the naturality of $\bs d$, $[R^\star_\psi, \bs d]=0$, and \eqref{Equiv-L}:
\begin{equation}
\label{dL-eq-vert}
\begin{aligned}
R^\star \bs dL &=\psi^* \bs dL , \\
 \iota_{X^v} \bs dL &= a(X, \phi)= \mathfrak L_X L= d\iota_X L. 
\end{aligned}
\end{equation}
So, its gauge transformation is easily deduced geometrically, via \eqref{GT-geometric}, to be
\begin{equation}
 \label{GT-dL}  
 \begin{aligned}
(\bs dL)^{\bs \psi} 
&=
\bs\psi^*\big( \bs d L + \mathfrak L_{\bs{d\psi}\circ \bs\psi\-} L \big) \\
&=
\bs\psi^*\big( \bs d L +  d\iota_{\bs{d\psi}\circ \bs\psi\-} L \big).
\end{aligned}
\end{equation}
{This} is crosschecked by $[\Xi^\star, d]=0$ and \eqref{GT-L}. 
This immediately implies that
\begin{equation}
 \label{GT-dS}  
 \begin{aligned}
(\bs dS)^{\bs \psi} 
&= \int_{\bs\psi\-(i)} (\bs dL)^{\bs \psi}
=\int_i \bs d L + d\iota_{\bs{d\psi}\circ \bs\psi\-} L
=\bs d S + \langle\, \iota_{\bs{d\psi}\circ \bs\psi\-} L, \d i \, \rangle,
\end{aligned}
\end{equation}
which is indeed a special case of 
\eqref{Vert-trsf-int-dalpha}. 
This result  indicates that $\S$ is stable under the action of the gauge group $\bs\Diff(I)$.
This we now establish explicitly by finding the gauge transformation of $\bs E$. And since it comes with no additional work, we also give that of the presymplectic potential $\bs\theta$. 
For this we first write their equivariance and verticality properties. 
The former are immediate, 
\begin{align}
\label{equiv-E-theta} 
R^\star_\psi \bs E = \psi^* \bs E, 
\quad \text{and} \quad 
R^\star_\psi \bs \theta = \psi^* \bs \theta.
\end{align}
To find their verticality properties, we first use  \eqref{dS-dL} and \eqref{dL-eq-vert}, from which follows the identity
\begin{align}
 \label{id-dL-E-theta}   
 \iota_{X^v} \bs E = d\left(\iota_X L - \iota_{X^v}\bs\theta \right).
\end{align}
Then, using the identity derived in \cite{JTF-Ravera2024gRGFT}, eq. (318), according to  which in gRGFT one has
\begin{align}
  \label{key-identity}
  \iota_X L - \iota_{X^v}\bs\theta 
  =
  E( \iota_X\phi; \phi)
- d\theta( \iota_X\phi; \phi), 
\end{align}
and observing that in MFT $\phi \in \Omega^0(I)$ so that $\iota_X \phi =0$, we get 
\begin{align}
\label{vertic-E-theta} 
\iota_{X^v} \bs E = 0, 
\quad \text{and} \quad 
\iota_{X^v} \bs \theta = \iota_X L.
\end{align}
Therefore,  the field equations are tensorial, $\bs E \in \Omega^1_\text{tens}\big(\Phi, \Omega^1(I) \big)$,  
not the presymplectic potential, $\bs \theta \in \Omega^1_\text{eq}\big(\Phi, \Omega^1(I) \big)$. 
Their respective gauge transformations are thus  immediately found to be,
\begin{align}
\label{GT-E-theta} 
 \bs E^{\bs\psi} = \bs\psi^* \bs E, 
\quad \text{and} \quad 
 \bs \theta^{\bs\psi} = \bs\psi^* \big(  \bs \theta + \iota_{\bs{d\psi}\circ \bs\psi\-} L \big).
\end{align}
These are  special cases of the gauge transformations of $\bs E$, \cite{JTF-Ravera2024gRGFT} eqs. (319)-(324),  and $\bs\theta$, \cite{Francois2023-a} eq. (238), in gRGFT. 

The tensoriality of the field equations 1-form $\bs E$ on the MFS $\Phi$, and its gauge transformation above, imply that $\S$ is indeed stable under the action of $\bs\Diff(I)$. 
More precisely, \eqref{equiv-E-theta} and \eqref{GT-E-theta} establish that the space of solution is a principal $\Diff(I)$-subbundle of the MFS,
\begin{align}
  \S  &\xrightarrow{\pi} \M_\S,
\end{align}
 with base $\M_\S$ the moduli space of solutions (its point $[\phi]=[(x, t)]$ are $\Diff(I)$-classes), and 
with gauge group $\bs\Diff(I)_{|\S} \simeq \bs\Aut_v(\S)$. 
This means that the
\emph{covariance group} of MFT is not just the (structure) group $\Diff(I)$ (of field-independent diffeomorphisms) but the much bigger (gauge) group $\bs\Diff(I)$ (of field-dependent diffeomorphisms). 
This is typical of gRGFT, as established in full generality in \cite{JTF-Ravera2024gRGFT}.
To the best of our knowledge, this fact was first noticed in the case of General Relativity by \cite{Bergmann1961, Bergmann-Komar1972}, and later by \cite{Salisbury-Sundermeyer1983}, which considered metric-dependent diffeomorphisms $\bs\psi(g)\in \Diff(M)$ with $M$ the ``spacetime" manifold.\footnote{The quotes here are intended to remind that it is incorrect to identify spacetime with $M$: Not only because spacetime is modeled by a Lorentzian manifold $(M , g)$, but more crucially because $M$ is not even the bare manifold of true \emph{physical} ``events", or of ``spacetime points". For a general discussion of this point in the context of gRGFT see  \cite{JTF-Ravera-Foundation2025, JTF-Ravera2024gRGFT}, also \cite{JTF-Ravera_NoBdryPb_2025, Berghofer-et-al2025}.
We  discuss its equivalent in MFT next, in section~\ref{The relationality of MFT}. }

\subsection{The relationality of MFT: the Hole and Point-Coincidence arguments}
\label{The relationality of MFT}

 The local symmetries of a gRGFT encode the relational character of the physics it describes, as reminded in the introduction \ref{Introduction}.
 MFT is no exception: its $\Diff(I)$-covariance signals that physical d.o.f. are  best represented as relations among the fields $\phi=\{x, t\}$. 
This insight arises in exactly the same way as in General-Relativistic physics (GR in particular), via the conjunction of a ``Hole argument", raising a problem, and a ``Point-Coincidence argument", that solves it. 
See \cite{JTF-Ravera2024gRGFT} for a complete technical treatment, and \cite{JTF-Ravera-Foundation2025} for an in-depth conceptual analysis.  
For the sake of pedagogy and clarity, let us spell out the logic in detail for MFT.
\medskip

An a priori look at the kinematics of MFT may let one think that $I$ is the physical timeline,\footnote{With $I$ understood to be endowed with its ``standard" metric $\textsf{g}: TI \times TI \rarrow \RR^+$, $(X, Y) \mapsto \textsf{g}(X, Y) =\b X \b Y$ 
for $X=\b X \tfrac{\d}{\d\tau}$ and $Y=\b Y \tfrac{\d}{\d\tau}$. 
This metric may be loosely said ``canonical" since $TI\simeq \RR$ as vector spaces.}
call it Time, parametrizing the evolution of both the fields $x(\tau)=x^a(\tau)$, with (internal) ``spatial" d.o.f., and of the clock field $t(\tau)$. The~situation appears congruent with  Newton's famous definition of time in the \emph{Scholium} of the 
\emph{Principia} (1687): 
\begin{quote}
``\emph{Absolute, true, and mathematical time, of itself, and from its own nature, flows equably without relation to anything external, and by another name is called duration:} [...]"
\end{quote}
quoted from \cite{sep-newton-stm}.
We may interpret this passage as referring to $I$. It is immediately followed by 
\begin{quote}
``\emph{[...] relative, apparent, and common time, is some sensible and external (whether accurate or unequable) measure of duration by the means of motion, which is commonly used instead of true time; such as an hour, a day, a month, a year.}"
\end{quote}
That would be (abstractly) the clock field $t$ understood as a physical object.  
Yet, a challenge to this simple view, based on a naive reading of the mathematical framework, arises from the $\Diff(I)$-covariance of MFT, i.e. from the fact that the space of solutions is a $\Diff(I)$-principal bundle $\S \rarrow \M_\S$. 

This challenge is set by the ``Hole argument": 
The $\Diff(I)$-covariance of the field equations $\bs E_{|\phi}=E(\bs d\phi; \phi)=0$ implies that if $\phi=(x, t)$ is a solution, so is $\psi^* \phi$ for any $\psi \in\Diff(I)$. 
Now,  considering two solutions $\phi, \phi' \in \O_\phi\subset \S$, i.e. $\phi'=\psi^*\phi$,  where $\psi$ is a compactly supported diffeomorphism whose support $D_\psi \subset I$ is the ``hole", 
we have that $\phi=\phi'$ on $I/D_\psi$, but  $\phi \neq \phi'$ on the time interval $D_\psi$. 
Therefore, $\bs E=0$ have an ill-defined Cauchy problem; i.e. the field equations cannot uniquely determine time evolution, so that MFT seems to be badly undeterministic. 
To~avoid~this, one has to admit that all solutions within the same $\Diff(I)$-orbit $\O_\phi$ represent the \emph{same} physical state. 
This implies that MFT, unable to physically distinguish between $\Diff(I)$-related solutions of $\bs E=0$, consequently cannot physically distinguish $\Diff(I)$-related points of $I$ either. 
In other words, $I$ is not Time, the physical timeline. 
But then, how are Time and deterministically evolving mechanical d.o.f. encoded in MFT?

\medskip
The solution comes from the ``Point-Coincidence argument", a term coined by John Stachel to refer to Einstein's 1915 conceptual breakthrough by which he understood the meaning of diffeomorphisms covariance in GR.
In~MFT, it is the assertion that its physical content (thus its observables) is exhausted by the set of coincidences between the ``spatial" fields $x$ and the clock field $t$, and that the description of these coincidences is invariant under $\Diff(I)$. 
The~statement of the $\Diff(I)$-invariance of pointwise mutual relations $\R$ between the mechanical fields $\phi=\{x, t\}$,
 may be formally written as,
 \vspace{-1mm}
 \begin{equation}
 \label{PC-arg}
 \begin{aligned}
 \R :  \S \times I\  &\ \rarrow \ \   \S \times I /\sim \quad  \ \xrightarrow{\simeq} \  \text{Relational mechanical physical d.o.f.,} \\
 \big(\phi, \tau \big) \  &\  \mapsto  \ \big(\phi, \tau \big) \sim \big(\psi^*\phi, \psi\-(\tau)\big)   \ \mapsto\   \R \big(\phi; \tau \big) =\R \big(\psi^*\phi; \psi\-(\tau)\big),
 \end{aligned} 
  \end{equation}
  where $\S \times I /\sim$ is the quotient of the direct product space $ \S \times I$ by the equivalence relation $\big(\phi, \tau \big) \sim \big(\psi^*\phi, \psi\-(\tau)\big)$, i.e. an associated bundle to the bundle of solution $\S$ as defined in section \ref{Associated bundle of region and integration} -- for which see also 
 \cite{JTF-Ravera2024gRGFT} section 3.5.
 Remark~that the above is actually also true on $\Phi$, i.e. kinematically (off-shell). 
 The true physical mechanical d.o.f., subject to deterministic Time evolution, are the coincidental relations between $x$ and $t$ -- standardly denoted $x(t)$. 
 This reciprocally implies that, in MFT, a point of physical time, a moment of Time,  is \emph{defined} by the coincidence of the fields $x$ and $t$. 
 Time is thus relationally defined, encoded in the totality of relations, actual (on-shell) and potential (off-shell), between the fields $x$ and $t$ as described by \eqref{PC-arg}. 
 The manifold $I$ drops out of the physical picture, precisely as  does $M$ in GR. 

 This has us revisit the above second part of the Newton quote: contra the idea that ``common time, is some sensible and external [...] measure of duration \emph{by the means of motion}, which is commonly used instead of true~time" (our emphasis), it appears that in Mechanics, true Time is indeed defined ``by the means of motion", i.e. by physical Mechanical d.o.f., and that there is no need for an abstract notion of ``mathematical time [...] without relation to anything external". 
 All that is needed are clocks,
 which is indeed consistent with  actual experimental practice.\footnote{The only caveat is that in non-relativistic  MFT one is able to define a distinguished class  of ``good" clocks. More on this later.}
The~$\Diff(I)$-covariance of MFT thus encodes  the relational co-definition of physical mechanical d.o.f. and of Time.


\vspace{-2mm}

\paragraph{The Problem of Time and the Boundary Problem}
This relational core of MFT is  fundamental,
and provides an intuitive grasp on how the relational physics of the broader class of gRGFT may alleviate, or outright dissolve,  ``problems" accepted as valid issues in the literature.
 Let us highlight the two more prominently discussed. 
 \medskip
 
The first is the ``Problem of time" in general-relativistic physics (and its ramifications for {QG}):
It~is the concern that
if time evolution is understood as time coordinate change on $M$, itself interpreted as a special diffeomorphism, 
then observables cannot evolve in time since they must be $\Diff(M)$-invariant by definition. 
This is known as the ``frozen formalism" or ``frozen time" problem in GR. See e.g.  \cite{Isham1993, Kuchar2011, Thebault2022}. 
Its analogue in the context of MFT is that, interpreting $I$ as the physical Time and understanding Time evolution as change in the parameter $\tau$, observables in Mechanics have trivial dynamics, they are ``timeless", since they are $\Diff(I)$-invariant by definition.

We know this reasoning to be obviously incorrect in Mechanics, where there is no such ``problem of time". 
The correct understanding in MFT is as stated above, following the Point-Coincidence argument: 
The invariant mechanical d.o.f. \eqref{PC-arg} are the relations between the fields $x$ and $t$, and physical Time evolution is not defined by change of both w.r.t. to $\tau \in I$, but by change of one w.r.t. to the other; in that case, variation of the ``spatial" field compared to variation of  the clock field $t$, as is expressed by the very familiar functional form $x(t)$. 
Said otherwise, physical Time evolution is the change of the relational structure of the field $\phi=\{x, t\}$,
a $\Diff(I)$-invariant quantity that has a well-defined Cauchy problem for the field equations $\bs E=0$, i.e. a deterministic  Time evolution. 

The same resolution in principle applies in general-relativistic physics and gRGFT  -- granted the caveats by Kucha\v{r} in \cite{Kuchar2011} section 3 -- whereby relational invariant physical d.o.f. have well-defined physical spatiotemporal 
evolution 
understood, not as variation of individual (mathematical) fields across the manifold $M$ (which disappears from the physical picture), 
but as the variation of their relational structure. 
The  relational understanding of gRGFT physics should defuse the ``Problem of time" (at least its classical, i.e. non-quantum, variant). 
\medskip

  The second problem, conceptually closely related, is the ``Boundary problem":
 In the form often encountered in the literature, it is the statement that
 ``spacetime boundaries break local symmetries (diffeomorphisms and/or gauge transformations)" \cite{DonnellyFreidel2016, Speranza2018, Geiller2018, Freidel-et-al2020-1, Kabel-Wieland2022, Ciambelli2023}. 
 It is sometimes followed, e.g. in the covariant phase space literature, by the claim that  ``Goldstone modes", a.k.a. ``edge modes", must arise at spacetime boundaries as a result of such symmetry breaking when studying gRGFT over bounded  regions. 
 Further still, some claim that these boundary modes are key to a deeper understanding of quantum gRGFT. 
 The above statement, though, is flatly wrong. 
 
 What is actually meant by it is the similar yet  non-equivalent statement that
 ``boundaries $\d U$ of bounded regions $U\! \subset\! M$ are not preserved by local symmetries", which is trivially true but physically inconsequential, since it is logically equivalent to the Hole argument. The Point-Coincidence argument has us realise that the manifold of physical events (spacetime points) is not $M$, but is represented by the $\Diff(M)$-invariant relational structure among fields d.o.f.
 Boundaries of spacetime regions are thus relationally defined, in a $\Diff(M)$-invariant way. 
 See \cite{JTF-Ravera_NoBdryPb_2025}.
 
 In MFT, the implausibility of such ``boundary problem" is transparent. 
 It would imply to say that $\Diff(I)$ is ``broken" at the boundary $\d i=\{\tau_0, \tau_1\}$ of an interval $i \subset I$ over which we study MFT, i.e.  that 
 boundary clock-time readings $t_0=t(\tau_0)$ and  $t_1=t(\tau_1)$ -- and  boundary field configurations $x_0=x(\tau_0)$ and  $x_1=x(\tau_1)$  -- 
 are not $\Diff(I)$-invariant. 
 One would then claim that ``edge modes" need to be introduced, or arise as ``Goldstone modes", at $\d i$ as a result of this $\Diff(I)$-breaking, and perhaps that they are key to understand MFT on bounded regions $i \subset I$ (i.e. Mechanics in finite time interval) and its quantization (i.e. Quantum Mechanics). 
 
It is certainly clear enough that this logic is quite foreign to what we know to be the correct understanding of both Classical and Quantum Mechanics, where ``Time boundaries" are of no particular physical consequence -- aside from setting arbitrary boundary conditions -- 
and there is no need for extra d.o.f. at the initial and final instant of the Time interval considered. 
Again, this ``Boundary problem" is logically equivalent to the Hole argument, thus solved by the Point-Coincidence argument that has us realise that Time, thus boundaries of Time intervals (which we may write $\{x_0[t_0], x_1[t_1]\}$, suggestively), are relationally defined in a $\Diff(I)$-invariant way.



\subsection{Model of MFT: Non-relativistic Mechanics}
\label{Models of MFT}

Let us consider, for concreteness, 
the simplest  model of the MFT framework: the non-relativistic case in (flat) 3-space, where a single clock field is used.
{This example plays the role of the fundamental concrete realization of the general framework developed in the present work. In particular, it allows one to exhibit explicitly the parametrized theory, the clock dressing procedure,  and, consequently, the resulting dressed classical action and PI (see sections \ref{Dressed non-relativistic mechanics} and \ref{QM as the relational quantization of MFT}, respectively), thereby providing a direct comparison with the standard deparametrization and gauge-fixing treatments of parametrized mechanics -- moreover, the link with the standard approach in \emph{configuration space-time} is provided. The purpose is not merely to recover ordinary QM, but to show geometrically that its standard PI formulation already possesses an intrinsically relational and dressed structure.}

The construction of the Lagrangian requires a proper use of Cartan calculus on $I$ and the structure on the target space $\sf T$ of the fields (geometry),  as well as  simple dimensional analysis (physics). 
Let us then start by further generalities about the former.
  
On $I$, the de Rham derivative  of the fields $\phi = \big(x, t \big)$ are, 
\begin{equation}
\label{d-phi-on-I}
\begin{aligned}
d\phi = \big(dx, dt\big)&= d \tau\, \frac{\d \phi}{\d\tau} \rdefeq \dot{\phi\,}\, d \tau 
=\big(\dot x, \dot t\, \big) \, d\tau \quad \in \Omega^1\big(I\big).
\end{aligned}
\end{equation}
Let us suppose we are in the case $\phi:I \rarrow {\sf T}= \EE^{3(N)} \times \T$, then this is:
\begin{equation}
\label{d-phi-on-I-bis}
\begin{aligned}
dx = d \tau\, \frac{\d x}{\d\tau} \rdefeq \dot x\, d \tau  \ \ \in \Omega^1\big(I, \RR^{3(N)}\big),
\qquad \text{and}\qquad 
dt =d \tau \, \frac{ \d t}{\d\tau} \rdefeq \dot t\, d \tau  \ \ \in \Omega^1\big(I, \RR\big),
\end{aligned}
\end{equation}
where $dt$ is the ``clock 1-form field".
It has a natural dual, the ``clock vector field", 
\begin{align}
\label{clock-vect-field}
\d_t \in \Gamma(TI)\quad s.t. \quad \iota_{\d_t} dt =1, 
\qquad \text{so} \qquad
\d_t= \frac{1}{\dot t} \frac{\d}{\d\tau} =  \frac{1}{\dot t} \, {\d_\tau}.
\end{align}
Remark that it is of the form $\bs X=\bs X(\phi) \in \Gamma(I)$, i.e. it is a field-dependent vector field of $I$. 
The action of $\Diff(I)$ on 
\eqref{d-phi-on-I} is, by the naturality of $d$, easily found to be
\begin{align}
 \label{psi-trsf-dphi}   
 \psi^*d\phi 
 = d\psi^* \phi 
 = (\dot{\phi\,} \circ \psi) \,\dot \psi\, d\tau.
\end{align}
We shall denote $\phi'\defeq\psi^*\phi$, so $d\phi'=\psi^*d\phi$.
The action of $\Diff(I)$ on  the dual clock vector field \eqref{clock-vect-field} is thus 
\begin{equation}
\label{psi-trsf-t-field}   
\begin{aligned}
\d_{t'}=\d_{\psi^*t} \defeq&\,  
  [ \tfrac{\d}{\d\tau}( t \circ \psi) ]\- \d_\tau
= [(\dot t \circ \psi )\,\dot \psi ]\- {\d_\tau}
\\
=&\, (\dot t \circ \psi )\-\, \dot\psi\-  \d_\tau
= (\dot t \circ \psi )\-\, [\d_\tau(\psi\-) \circ \psi ] \d_\tau
= (\dot t \circ \psi )\-\, [(\psi\-)_*\, \d_\tau \circ \psi] \\
=&\, (\psi\-)_* \,\d_t \circ \psi.
\end{aligned}
\end{equation}
 The first line being consistent with  $\iota_{t'}dt'=\iota_{\d_{\psi^*t}} \psi^* dt = 1$, the last showing the $\phi$-dependent vector field $\bs X(\phi)=\d_t$ to have equivariance $R^\star_\psi \bs X = (\psi\-)_* \,\bs X \circ \psi$, thus to be an element of the gauge Lie algebra of $\Phi$, $\bs X(\phi)=\d_t \in \bs\diff(I)$. 
One defines the velocity and acceleration fields of $x$ as
\begin{equation}
\label{v-and-a}
\begin{aligned}
v\defeq \mathfrak L_{\d_t} x &= \d_t x = \tfrac{\dot x}{\dot t} \ \in \Omega^0(I, \RR^{3(N)}), 
\quad \text{ so that } 
\quad dx = v dt, \\
a\defeq \mathfrak L_{\d_t} v &= \d_t v = \tfrac{\dot v}{\dot t}\ \in \Omega^0(I, \RR^{3(N)}), 
\quad \text{ so that } 
\quad dv = a dt, 
\end{aligned}
\end{equation}
where indeed, in terms of $x$, we have $a=(\ddot x \dot t - \dot x \ddot t\,) \,{\dot t}^{-3}$. 
Their transformation under $\Diff(I)$ is, by a well-known property of Lie derivatives and by \eqref{psi-trsf-t-field},
\begin{equation}
\label{psi-trsf-v-a}
\begin{aligned}
v'\defeq \psi^* v 
&= \psi^*( \mathfrak L_{\d_t} x) 
= \mathfrak L_{( {\psi\-}_* \d_t \circ \psi) } (\psi^*x)
=\mathfrak L_{\d_{t'}} x' 
= \d_{t'} x', \\
&= \frac{\dot x \circ \psi}{\dot t \circ \psi} = v \circ \psi. \\
a'\defeq \psi^* a 
&= \psi^*( \mathfrak L_{\d_t} v) 
= \mathfrak L_{( {\psi\-}_* \d_t \circ \psi) } (\psi^*v)
=\mathfrak L_{\d_{t'}} v' 
= \d_{t'} v', \\
&= \frac{\dot v \circ \psi}{\dot t \circ \psi} = a \circ \psi. 
\end{aligned}
\end{equation}
From this we have $dx'=v'dt'$ and $dv'=a'dt'$, as one would heuristically expect.



\medskip

The Lagrangian must be an $\RR$-valued top form $L(\phi) \in \Omega^1(I, \RR)$. 
And since $I$ is considered dimensionless, $L$ must have the dimension of an action, Mass $\times$ Length$\!{\phantom{|}}^2$ $\times$ Time$\!{\phantom{|}}^{-1}$,  ML$\!{\phantom{|}}^2$T$\!{\phantom{|}}^{-1}$.
The dimensions of the spatial and clock fields are $|x|=$ L and $|t|=$ T, so that since again $I$ is dimensionless, $|dx|=$ L and $|dt|=$ T. It follows that $|\d_t|=$ T$\!{\phantom{|}}^{-1}$. 
The only intrinsic physical parameters associated to $N$ point particles are their masses $m=\{m_1, \ldots, m_N\}$. 
Thus, using the standard scalar product $\langle\_\,, \_\rangle$ on $\RR^3$, we may write  kinetic terms for the spatial fields: $\tfrac{m}{2} \langle \iota_{\d_t} dx, dx\rangle = \tfrac{m}{2}\langle \mathnormal{v} , dx\rangle \defeq \Sigma_{i=1}^N \tfrac{m}{2}\langle \mathnormal{v}_i , dx_i\rangle $.
Given a potential functional $V=V(\phi)$ with the dimension of an energy, ML$\!{\phantom{|}}^2$T$\!{\phantom{|}}^{-2}$, we may build a potential term:  $Vdt$. 
The simplest Lagrangian we may write is then
\begin{equation}
 \begin{aligned}
  \label{NR-Lagrangian} 
  L(\phi)=L(x, t)
  &=
  \tfrac{m}{2} \langle \iota_{\d_t} dx, dx\rangle - V(x, t)\,  dt
  \quad \in \Omega_\text{tens}^0\big(\Phi, \Omega^1(I) \big)\\[1mm]
  &=
  \Big(
  \tfrac{m}{2} \langle \dot x, \dot x\, \rangle \,  \dot t\- - V(x, t)\, \dot t\,    \Big)\, d\tau \\
  &=
  \Big(
  \tfrac{m}{2} \langle v, v\, \rangle    - V(x, t)\Big)\, dt.
 \end{aligned}   
\end{equation}
{
As we shall see shortly, this describes
precisely the standard non-relativistic particles with potential, which shall next be understood within the dressed MFT framework. From this perspective, the usual quantum-mechanical PI will arise as the dressed expression associated with the relational clock sector, rather than from an \emph{ad hoc} gauge-fixing prescription. The dressing of the classical theory is provided in section \ref{Dressed non-relativistic mechanics} -- where we also remark the difference between dressing via the DFM and gauge fixing -- and its relational PI quantization presented and discussed in section \ref{QM as the relational quantization of MFT}.}

By what {precedes}, the $\Diff(I)$-covariance, or equivariance, of $L$ is easily shown:  $\psi^* L(\phi)=\Big(
  \tfrac{m}{2} \langle v', v'\, \rangle    - V(x', t')\Big)\, dt'=L(\phi')=L(\psi^*\phi)$, otherwise written as $R^\star_\psi L = \psi^* L$. 
Notice that we cannot write a kinetic term for the clock field $t$: only with the help of a physical constant with the dimension of a velocity could we possibly write one. This is how, from the MFT perspective, one arrives at relativistic mechanics; it hinges upon the introduction of what we should call the \emph{Einstein constant} $\bs c$, s.t. $|\bs c|=$~LT$\!{\phantom{|}}^{-1}$  (a.k.a. the ``speed of light").
This will  be further elaborated elsewhere.

To write the variational principle \eqref{dS-dL} as geometrically as possible, i.e. by starting from the first line of \eqref{NR-Lagrangian}, let us write a useful lemma.
Define the variation of the clock vector field by $\d_{\bs d t} \defeq \bs d \d_t = (\bs d \dot t\-)\, \d_\tau = - (\dot t^{-2} \bs d t) \, \d_\tau$. 
We~have also that $\dot t\- \bs d \dot t = \dot t\- \d_\tau (\bs d t) = \d_t(\bs d t)= \iota_{\d_t} d(\bs d t) = \mathfrak L_{\d_t} (\bs dt)$.
Then, we find that 
\begin{align}
 \iota_{\bs d t} dx
 = - \dot x \dot t^{-2} \bs d \dot t   
 = - \dot x \dot t^{-1} \, \dot t\- \bs d \dot t
 = v \mathfrak L_{\d_t} (\bs dt).
\end{align}
From this, and by the observing that 
$\langle \iota_{\d_t}d (\bs d x), dx \rangle = \langle \iota_{\d_t}d x, d(\bs d x) \rangle$  -- both terms giving $\langle v, \tfrac{\d}{\d\tau} (\bs d x) \rangle\, d\tau $ --
we find
\begin{equation}
\label{bare-var-princ-NR}
\begin{aligned}
\bs d L
&=
\tfrac{m}{2} \langle \iota_{\d_{\bs dt}} dx, dx \rangle 
+\tfrac{m}{2} \langle \iota_{\d_t}d (\bs d x), dx \rangle 
+\tfrac{m}{2} \langle \iota_{\d_t}d x, d(\bs d x) \rangle
- \bs d V(x, t )\, dt
- V(x, t )\, d(\bs d t) \\
&=
-\tfrac{m}{2} \langle v , dx \rangle  \mathfrak L_{\d_t} (\bs d t) 
+ m \langle \mathfrak L_{\d_t} \bs dx, dx \rangle
- \left( \tfrac{\delta V}{\delta t}  \bs d t + \langle\tfrac{\delta V}{\delta x} , \bs d x \rangle \right)\, dt
- d\left(V(x, t)\, \bs d t \right) + dV(x, t) \bs d t  \\
&= 
-\tfrac{m}{2} \mathfrak L_{\d_t} \Big( \langle v , dx\rangle \bs d t \Big)
+ \tfrac{m}{2} \mathfrak L_{\d_t} \Big(\langle v, dx \rangle \Big) \bs d t
\ \,  + \ \, 
m\mathfrak L_{\d_t} \Big( \langle \bs d x, dx \rangle \Big)
- 
m \langle \mathfrak L_{\d_t} d x,\bs dx \rangle
\ \,  - \ \,  \ldots \\
&= 
-\tfrac{m}{2} d \Big( \langle v , \iota_{\d_t} dx\rangle \bs d t \Big)
+ \tfrac{m}{2}d\Big(\langle v, \iota_{\d_t}dx \rangle \Big) \bs d t
\ \, + \ \, 
m \, d \Big( \langle \bs d x, \iota_{\d_t}dx \rangle \Big)
-
m \, \langle d  (\iota_{\d_t}d x),\bs dx \rangle \ \,  - \ \,  \ldots \\
&=
-\langle d(mv) + \tfrac{\delta V}{\delta x} dt, \bs d x   \rangle 
+ \left[
d\Big( \tfrac{m}{2} \langle v, v\rangle + V(x, t) \Big)  - \tfrac{\delta V}{\delta t}dt
\right]\bs d t 
\ \ +\ \ 
d \left[ \langle mv, \bs d x\rangle
- \Big(  \tfrac{m}{2} \langle v, v\rangle + V(x, t)   \Big)\,\bs dt
\right] \\
&\rdefeq E(\bs dx; \phi) +  E(\bs dt; \phi) 
\ \ + \ \ 
d \left[\theta(\bs dx; \phi) + \theta(\bs d t; \phi) \right]
= E(\bs d\phi; \phi) + d\theta(\bs d\phi; \phi) = \bs E + d\bs \theta,
\end{aligned}
\end{equation}
where we used the fact that  $\mathfrak L_X \beta = d\iota_X \beta$ for $\beta$ a 1-form on $I$. 
The presymplectic potential can be rewritten as
\begin{align}
  \bs\theta = \langle p_x, \bs d x \rangle +  p_t \, \bs d t 
  = \langle\langle p_\phi \, ,\bs d\phi \rangle\rangle,
\end{align}
where $p_x \defeq mv$ is the canonical momentum associated to the spatial fields $x$, i.e. the linear momentum,  while  $p_t \defeq -\big( \tfrac{m}{2} \langle v, v\rangle + V(x, t) \big)$ is the canonical momentum associated to the clock field $t$, i.e. the mechanical energy, and  $\langle\langle\_\, 
 ,\_ \rangle\rangle$ denotes the scalar product on $\RR^{3(N)} \times \RR$.  
Notice that an identity holds between momenta: 
\begin{align}
\label{constraint-momenta}
p_t + \tfrac{1}{2m}\langle p_x, p_x \rangle + V=0,
\end{align}
which is often called an Hamiltonian ``constraint" in the canonical analysis. 
This relates to the fact that, in the usual approach, mechanical energy is given by the standard Hamiltonian $H(p_x, x) = \tfrac{1}{2m}\langle p_x, p_x \rangle + V(x, t)= -p_t$.

It is then interesting to notice that the field equations $\bs E=0$ may be seen as continuity equations controlling the non-conservation of the canonical momenta -- the gradient of the potential acting as a source of momenta: 
\begin{align}
\label{Field-Eq-NR-Mechanics}
\bs E = E(\bs d\phi; \phi) =- \langle\langle d p_\phi  + \tfrac{\delta V}{\delta \phi} dt ,\bs d \phi\rangle\rangle=0, \ \  \forall \, \bs d\phi 
\quad \Rightarrow \quad 
d p_\phi  =-  \tfrac{\delta V}{\delta \phi} dt.
\end{align}
Defining  the \emph{force} 1-form as $Fdt \defeq - \tfrac{\delta V}{\delta x} dt$, and remembering that the 3-acceleration is $a \defeq \d_t v$, so that $dv = d\tau \, \d_\tau v = d\tau \, \dot t\, {\dot t}\- \d_\tau v =dt \, \d_t v = dt\, a$, the field equation for the spatial field $x$ is
\begin{equation}
\label{E-NR-Mechanics-bis}
\begin{aligned}
E(\bs dx; \phi) =0, \ \  \forall \, \bs dx \quad   \Rightarrow \quad  
dp_x = F dt, 
\quad \text{i.e. }\ 
&m\,\d_\tau v \ d\tau= F \dot t\,  d\tau \, 
\\  
&\hookrightarrow  
F = m\,  \dot t\- \d_\tau v = m\, \d_t v = m\,a.
\end{aligned}
\end{equation}
The standard form of Newton's equation is recognised.
The field equation $E(\bs dt; \phi) =0$ for the clock field $t$ states that mechanical energy is not conserved for $t$-dependent potentials (i.e. non-conservative forces).
It may be noticed that the field equations for $x$ and $t$ are not independent here, $E(\bs dx; \phi) =0$ implies $E(\bs dt; \phi) =0$.\footnote{
{One only needs to show} that $\langle d(mv)+\tfrac{\delta V}{\delta x}dt, v \rangle = d\big( \tfrac{m}{2}\langle v, v \rangle + V \big) - \tfrac{\delta V}{\delta t}dt$. 
The same computation needed to prove the horizontality of $\bs E$, next.}
\medskip

We can verify explicitly the properties \eqref{vertic-E-theta} in the case at hand: 
First we find that 
\begin{align}
\label{theta-L-relation}
\iota_{X^v} \bs\theta 
=
 \langle p_x, \mathfrak L_X x \rangle 
+
p_t\, \mathfrak L_X t
=
\big(
\langle p_x, v  \rangle 
+
p_t \big)\, \mathfrak L_X t
=
\big(
\tfrac{1}{m}\langle p_x, p_x  \rangle 
+
p_t \big)\, \iota_X dt
=
\iota_X L,
\end{align}
where we use the fact that $\mathfrak L_X x = v \mathfrak L_X t $ following from $dx = v dt$, and  \eqref{constraint-momenta}. 
As we have seen in section \ref{The variational principle for MFT, and its covariance group}, the property $\iota_{X^v} \bs\theta = \iota_X L$ is  key to 
the tensoriality of the field equations by the identity \eqref{id-dL-E-theta}. We may also prove it directly, using the form \eqref{Field-Eq-NR-Mechanics} as well as \eqref{constraint-momenta}:
\begin{equation}
\begin{aligned}
\iota_{X^v}\bs E
&=
-\langle dp_x + \tfrac{\delta V}{\delta x} dt, \mathfrak L_X x  \rangle 
+ \left[
dp_t  - \tfrac{\delta V}{\delta t}dt
\right]\, \mathfrak L_X t \\
&=
\left[ -\langle dp_x + \tfrac{\delta V}{\delta x} dt ,  v    \rangle 
+ 
d\Big(\tfrac{1}{2m} \langle p_x, p_x\rangle + V(x, t) \Big)  - \tfrac{\delta V}{\delta t}dt
\right]\, \mathfrak L_X t \\
&= \Big[-\tfrac{1}{m}\langle dp_x, p_x \rangle 
   + 
   \tfrac{1}{2m}d \langle p_x, p_x\rangle
   \ \ 
 - \langle\tfrac{\delta V}{\delta x}, v dt     \rangle - \tfrac{\delta V}{\delta t}dt + dV(x, t) \Big]\,\mathfrak L_X t
 =0.
\end{aligned}
\end{equation}
Either way, the tensoriality of the field equation 1-form,  $\bs E \in \Omega^1_\text{tens}(\Phi)$ so  $\bs E^{\bs \psi} =\bs\psi^* \bs E$, 
 implies not only that the space of solutions  of Newtonian Mechanics is a $\Diff(I)$-bundle, $\S\rarrow \M_S$, but that its group of covariance is $\bs\Diff(I)$; 
 i.e. Newtonian Mechanics is covariant under \emph{field-dependent} reparametrization.

\paragraph{Link with standard approaches}  
The values of the fields $\phi(\tau)=\big( x(\tau), t(\tau)\big)$ are coordinatisations of a point $p$ in
the target space ${\sf T}=\EE^{3N} \times \T$, which is called  in \cite{JTF-Ravera2025NRrelQM}
 the (non-relativistic) \emph{configuration  space-time} and noted $\mC$ -- {also called ``extended configuration space" or ``space of partial observables" in \cite{Rovelli2002, Rovelli2004}.}
 In \cite{JTF-Ravera2025NRrelQM}, it is stressed as an essential fact that $\mC$ is
furthermore a principal bundle over $\T$ -- with structure group $H=\RR^{3N}$ when one considers, as we do here, structureless particles whose only d.o.f. are their position.
The fibers of $\mC$ are isomorphic to 
the  configuration space $Q$ of the standard approach to mechanics, which may then be called the ``typical fiber" of $\mC$. In a bundle coordinate system for $\mC$, we may write $p=(q, t)$ with $q \in Q$ and $t\in \T$,  i.e. indeed $(q, t)=\big(x(\tau), t(\tau)\big)$ or $p=\phi(\tau)$. 
The standard ``phase space"  $T^*Q$ is thus seen to be the isomorphic, at each time $t$,  to the vertical cotangent to the fibers $V^*_p\mC$ for $p=(q, t)$, so that the cotangent $T^*\mC$ is the ``extended phase space" as mentioned e.g. in \cite{Rovelli2004, Rovelli-Vidotto2014}.
The latter is often closely associated to the so-called ``parametric approach" \cite{Lanczos2012}, which MFT developed here generalizes. 

The mechanical fields $\phi=(x, t) \in \Phi$, which are a \emph{point} in the MFS $\Phi$, are a parametric coordinatization of a curve $\gamma \in \mC$.  
But since such a curve is an intrinsic geometrical object in $\mC$, indifferent to parametrization, 
it corresponds to the full  $\Diff(I)$-orbit $[\phi]=[x, t]$ of $\phi$: so that  $\gamma = [\phi]$. 
Which means that the base space $\M$ of the MFS bundle $\Phi \rarrow \M$ is exactly the space of curves, or paths, $\P(\mC)$ in the configuration space-time $\mC$: $\M=\P(\mC)$.
We may say that the latter is a good coordinatization of the former. 
Curves $\gamma \in \P(\mC)$  then ``projects" as a curves in the standard fiber $Q$, $\gamma_q \in \P(Q)$:
hence the standard approach on $Q$ is often said to be ``de-parametrized" or ``unparametrized".
See Fig. \ref{configuration_space-time-fig} below.
\vspace{-2mm}
\begin{figure}[H]
\begin{center}
\includegraphics[width=0.63\textwidth]{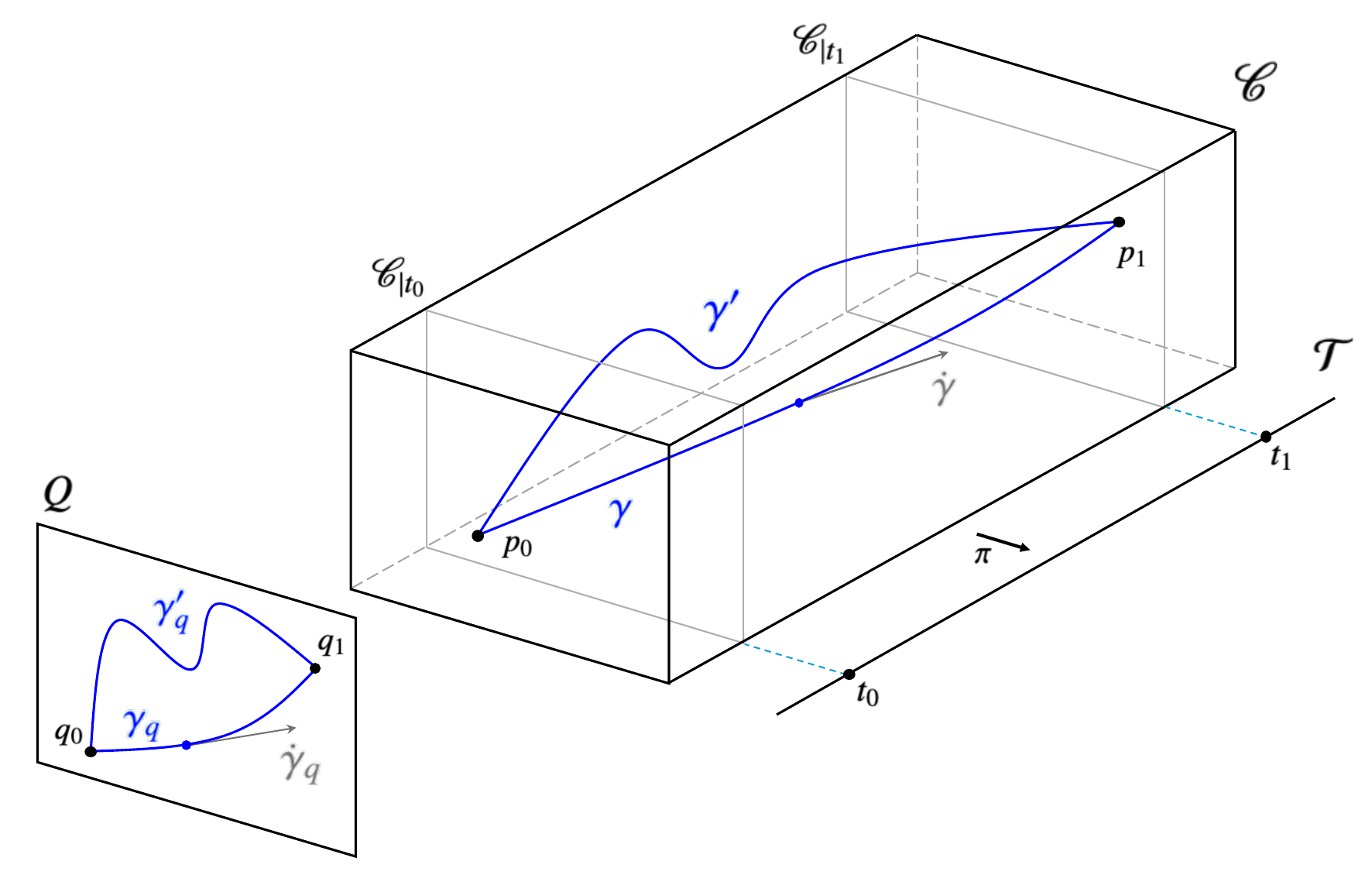}
\caption{The configuration space-time bundle $\mC$  \emph{vs} the standard configuration space $Q$.}
\label{configuration_space-time-fig}
\end{center}
\end{figure}
\vspace{-3mm}
It is worth noting that the variational principle \eqref{dS-dL} of MFT, performed on $\Phi$, is thus conceptually distinct from those one may perform on either $\mC$ or  $Q$: In the latter cases, one is  working on the spaces $\P(\mC)$ or $\P(Q)$ containing the $\Diff(I)$-invariant kinematical \emph{physical} d.o.f. only. 
In other words, in the standard approaches one already works on the moduli space $\M=\P(\mC)$, representing the \emph{relational} d.o.f. among the mechanical fields $\phi=(x, t)$.
Therefore, 
the set of critical curves $\gamma^c \in \P_{\!\!c}(\mC)$, or $\gamma_{q}^c \in \P_{\!\!c}(Q)$, i.e. the physical histories, is the moduli space  $\M_\S$ of the bundle of solutions $\S \rarrow \M_\S$ of MFT. 
This observation will motivate the notion of \emph{relational quantization}, and its implementation via the DFM.

\subsection{Quantization of MFT} 
\label{Quantum Mechanics on Phi}  

As a 1D classical gRGFT, MFT can be quantized via the standard toolbox of QFT, yielding a 1D QFT of the bosonic fields $\phi=\{x, t\}$ on $I$.
The latter's relation to standard QM can then be examined. 
It may be remarked that the idea of (non-relativistic) Quantum Mechanics being a 1D QFT is not new, it is e.g. explicitly stated by Zee \cite{Zee2010} to argue that the formalism of QFT naturally extends standard QM: 
but it is the standard unparametrized version, without $\Diff(I)$-covariance, that is talked about.
The parametrized version of the relativistic particle, the relativistic model of MFT, is mentioned multiple times by Polchinski \cite{Polchinski1998}, but more as a motivational example for string theory than as a full treatment. 
Typically, Lagrangian QFT uses path integral (PI) quantization. 
The PI for a MFT with action $S(\phi)=\int L(\phi)$ is 
  $Z(\phi) = \int \!\mathcal D  \phi\  e^{\sfrac{i}{\hbar}\, S(\phi)}$,
where $\mathcal D \phi=\mathcal Dx \mathcal D t$ is a (formal) measure on $\Phi$ allowing to define integration.

Considering the PI as a functional $Z: \Phi \rarrow \CC$, $\phi \mapsto Z(\phi)$,
 the most general transformation  compatible with the right action of $\Diff(I)$ on $\Phi$ is 
\begin{align}
\label{PI-anomalous}
 R^\star_\psi Z = C(\ ; \psi)\- Z, \quad
 \text{i.e.} \quad
  Z(\psi^*\phi) = C(\phi ; \psi)\- Z(\phi),
\end{align}
where $C(\phi; \psi)=e^{\sfrac{i}{\hbar}\int c(\phi; \psi)}$  is a 1-cocycle for $\Diff(I)$ as described in eqs. \eqref{cocycle}-\eqref{cocycle-bis}.
The latter has to come from the lack of invariance of the measure, as the action $S$ is $\Diff(I)$-invariant, and its linearization  $a(X; \phi)$ is a 1-cocycle for $\diff(I)$  which is the theories' $\diff(I)$-\emph{anomaly}:  its defining property \eqref{inf-cocycle} reduces, since $\mathfrak u(1)$ is Abelian, to 
$X^v\! \cdot a(Y; \phi)- Y^v\!\cdot a(X; \phi) =a([X,Y]_{\text{{\tiny $\diff(I)$}}}; \phi)$ and is otherwise known as the \emph{Wess-Zumino consistency condition}.
Meaning that, in general, a PI is a cocyclic tensorial 0-form, $Z \in \Omega^0_\text{tens}(\Phi, C)$.

% ...

 But, typically, for MFT the measure is also $\Diff(I)$-invariant, the PI has trivial equivariance $R^\star_\psi Z=Z$, i.e. it is invariant, and thus basic $Z \in \Omega^0_\text{basic}(\Phi, \CC)$. 
It therefore uniquely represents a PI $\t Z \in \Omega^0(\M, \CC)$ on the physical moduli space $\M$,
formally written as $\t Z([\phi])=\int \mathcal D[\phi] e^{\sfrac{i}{\hbar}\, S([\phi])}$. 
Yet, approached the standard way, the  PI $Z(\phi)$ on $\Phi$ would need to be \emph{gauge-fixed}.
As stated in section \ref{The Mechanical Field Space and its natural transformation groups}, a gauge-fixing is a choice of local section $\s: \U \rarrow \Phi_{|\U}$.
Concretely, it is  specified by a gauge condition taking the form of an algebraic and/or differential equation on the field variables, $\C(\phi)=0$, whose solutions constitute by definition the image of the local section $\s$: 
the~gauge-fixing slice is 
$\N \defeq \{ \phi \in \Phi\, |\, \C(\phi)=0 \}=\s(\U) \subset \Phi$
is the image of the local section $\s$.\footnote{In general, there exists $\N =\s(\M)$  iff the field space bundle $\Phi$ is trivial. The fact that no global gauge-fixing exists for $S\!U(n)$-gauge theories over compact regions of spacetime is known as the Gribov-Singer obstruction (or Gribov ambiguity), \cite{Singer1978, Singer1981, Fuchs1995}.}
The gauge-fixed path integral is thus equivalent to integration of the pullback by $\s$ of the integrand over the pre-image of $\N$ by $\s$, the region $\U \defeq\s\-(\N) \subset \M$ containing only physical d.o.f.:
\begin{equation}
\label{GF-Z}
\begin{aligned}
Z(\phi)_{\,|\,\text{{\tiny GF}}}\defeq \int_{\N=\s(\U)} \mathcal D \phi   \, e^{\sfrac{i}{\hbar}S( \phi)} 
= \int_\U  \s^* \big( \mathcal D \phi   \, e^{\sfrac{i}{\hbar}S(\phi)} \big) 
= \int_\U   \mathcal D [\phi]_\s   \, e^{\sfrac{i}{\hbar}S([\phi])_\s} \rdefeq \t Z( [\phi])_{\s\,|\,\text{{\tiny{$\U$}}}}.       
\end{aligned}
\end{equation}
Note that $Z(\phi)_{\,|\,\text{{\tiny GF}}}$ depends on $\sigma$, and it is not invariant.
That is the conceptual picture. 
Practically, the constraint $\C(\phi)$ may be enforced via BRST methods, implying the introduction of a ghost field $\xi$, Grassmann odd parameter place holder for the gauge parameter $X \in \diff(I)$, and of a BRST differential $s$, s.t. $sd+ds=0$ and  generating the bigraded algebra:
$s\phi =\mathfrak L_\xi \phi = \iota_\xi d\phi $ 
and $s\xi = -\tfrac{1}{2} \mathfrak L_\xi \xi= -\tfrac{1}{2}[\xi, \xi]$. 
Clearly, it encodes the $\diff(I)$-transformation of $\phi$, and it is easily checked that $s^2=0$.
The result is that $Z(\phi)_{|\text{GF}}$ is expressed in terms of an effective action $S_\text{eff}(\phi, \xi)$  in the same $s$-cohomology class as the original $S(\phi)$, and has thus the same observables. 
See e.g. \cite{Polchinski1998} section 4.2. 

It is interesting and conceptually important to realise that the PI $\t Z \in \Omega^0\big(\M, \CC \big)$ avoids these  technicalities needed for gauge-fixing by directly defining quantization on the space $\M$ of relational physical d.o.f. 
As observed above, the latter is precisely the space of paths  in the configuration space-time $\mC$,  $\M=\P(\mC)$, which indeed represents the invariant relational d.o.f. among the mechanical fields $\phi$.
Therefore, $\t Z \in \Omega^0\big(\P(\mC), \CC \big)$ is the \emph{standard} PI of QM, and represents the quantization of the \emph{relational} d.o.f.
of MFT. 

From this follows the first key message we want to convey: The PI typically used in GFT \emph{is not} the analogue, neither conceptual nor technical, of the PI of standard QM; the former is defined on the field space $\Phi$ and is non-relational, the latter is defined on the moduli space $\M$ and is fundamentally relational. 
This, in our view, strongly suggests that the most natural quantization scheme for (gR)GFT is what we shall call \emph{Relational Quantization}. 

Of course, the main issue is to find as good a representation (``coordinatization") of $\M$ as the one existing in MFT, $\P(\mC)$, whose points are relational variables (relational histories, actually). 
Indeed, one argument often provided as to why to resort to gauge-fixing in (gR)GFT is that one cannot work  with the moduli space $\M$. 
This is where basic (or equivariant) cohomology of field space  $\Phi$ may play a key role: as observed in section \ref{Differential structures}, it is isomorphic to the de Rham cohomology of $\M$, $\Omega^\bullet_\text{basic}(\Phi)\simeq \Omega^\bullet(\M)$, and yet allows manipulation of fairly concrete expressions on $\Phi$. 
Our main concern then is this: given objects (forms) on $\Phi$, how can one extract its ``basic component"? Or rather, how to produce its ``basic counterpart"? 
This is the question answered to by the Dressing Field Method.

\section{Dressing Field Method and Relational Quantization} 
\label{Dressing Field Method and Relational Quantization}  

The Dressing Field Method (DFM) is a formal yet systematic algorithm to construct basic forms on the field space  $\Phi$. 
In favorable situations, the elementary ``dressed" variables it produces may be understood to encode the invariant relations among the d.o.f. of the ``bare" fields $\phi \in \Phi$, hence its natural \emph{relational} interpretation. 
{It encompasses and generalizes various constructions of the literature, e.g. edge modes \cite{DonnellyFreidel2016, Speranza2018, Chandrasekaran_Speranza2021, Speranza2022, Chandrasekaran-et-al2022} (see appendix \ref{Covariant phase space formalism for MFT} for a brief explanation of their introduction in covariant phase space), ``gravitational dressings" \cite{Giddings-Donnelly2016, Giddings-Donnelly2016-bis, Giddings-Kinsella2018, Giddings2019, Giddings-Weinberg2020, Witten-et-al2023, Speranza-et-al2023, Kabel-Wieland2022}, ``dynamical reference frames"  \cite{carrozza-et-al2022, Hoehn-et-al2022}, or  ``embedding fields"   \cite{Speranza2019, Ciambelli-et-al2022,  Speranza2022, Ciambelli2023}.}
The most complete technical and conceptual presentation of the DFM to date is \cite{JTF-Ravera2024gRGFT}, and we draw from it the following simplified account giving 
the minimum necessary to formulate next the dressed MFT and its relational quantization.



\subsection{Building basic forms via dressing}  
\label{Building basic forms via dressing}  

The DFM involves a conditional proposition: if one can identify, or build, a \emph{dressing field} from the kinematics of a theory, i.e. its field space $\Phi$, then it shows how basic forms on $\Phi$ can be built.  
Let us then define the pivotal notion in the context under consideration.
Given $J$ a reference $1D$ manifold,
one defines the space of $\Diff(I)$-\emph{dressing fields} 
\begin{align}
\label{Dressing-fields-space}
\D r[J, I] \defeq \left\{\ \upupsilon: J \rarrow I\  |\ \upupsilon^\psi \defeq \psi\- \circ \upupsilon, \ \text{ with }\  \psi \in \Diff(I) \right\} \ \subset\  \text{Hom}(J, I).
\end{align} 
The linearisation of their defining property is then $\delta_X \upupsilon \defeq  -X \circ \upupsilon$ for $X \in \diff(I)$.
The dressing map is defined as: 
\begin{equation}
\begin{aligned}
\label{Dressed-field}
|^\ups : \Phi \ &\rarrow\ \Phi^\upupsilon,  \\
\phi \ &\mapsto \ \phi^\upupsilon\defeq  \ \upupsilon^*\phi,
\end{aligned}
\end{equation}
where $\phi^\upupsilon=(x^\upupsilon, t^\upupsilon)$ are the \emph{dressing} of  the bare fields $\phi=(x, t)$.
By definition, dressed fields in  $\Phi^\upupsilon$ ``live" on Im$(\upupsilon^*) \subset J$ and are thus  expected to be
$\Diff(I)$-invariant: explicitly,   $(\upupsilon^*\phi)^\psi \defeq (\upupsilon^\psi)^*(\phi^\psi) = (\psi\- \circ \upupsilon )^*(\psi^*\phi)= \upupsilon^*\phi$.
We next define a \emph{field-dependent $\Diff(I)$-dressing field} as 
\begin{equation}
\begin{aligned}
\label{Field-dep-dressing}
\bs \ups : \Phi \ &\rarrow\ \D r[J, I] ,  \\
\phi \ &\mapsto \  \bs \ups(\phi), \qquad \text{ s.t. }\quad  R^\star_\psi \bs \ups = \psi\- \!\circ \bs \ups \quad\text{ i.e. } \quad \bs \ups(\phi^\psi)=\psi\-\! \circ \bs \ups(\phi). 
\end{aligned}
\end{equation} 
The infinitesimal equivariance is thus  $\bs L_{X^v} \bs \ups = -X \circ \bs \ups$.
Such $\phi$-dependent dressing fields are thus tensorial (equivariant) $0$-form on $\Phi$, 
therefore their $\bs\Aut_v(\Phi)\simeq \bs\Diff(I)$ and  $\bs\aut_v(\Phi)\simeq \bs\diff(I)$ transformations are,  respectively,
\begin{align}
\label{Vert-trsf-dressing}
\bs {\ups^\psi}\defeq \Xi^\star \bs \ups = \bs\psi\- \circ \bs \ups, \qquad \text{ and }\qquad \bs L_{\bs X^v} \bs \ups= -\bs X \circ \bs \ups.
\end{align}
Now, a $\phi$-dependent dressing field induces the map  
\begin{equation}
\begin{aligned}
\label{F-map}
F_{\bs \ups} : \Phi \ &\rarrow\ \M \simeq \P(\mC) , \\
\phi \ &\mapsto \  F_{\bs \ups}(\phi) \defeq\phi^{\bs\ups} \defeq \bs \ups(\phi)^* \phi \sim [\phi]\simeq \gamma,  \qquad \text{ s.t. }\quad  F_{\bs \ups} \circ R_\psi = F_{\bs \ups}. 
\end{aligned}
\end{equation}
The constancy along $\Diff(I)$-orbits is easy to show: $F_{\bs \ups}(\phi^\psi) = \bs \ups(\phi^\psi)^*(\phi^\psi)=\big( \psi\-\! \circ \bs \ups(\phi) \big)^*\psi^*\phi= \bs \ups(\phi)^* \phi \rdefeq F_{\bs \ups}(\phi)$,
meaning that  dressed fields $\phi^{\bs\ups}=\bs \ups(\phi)^* \phi$ indeed represents a whole $\Diff(I)$-orbits $[\phi] \in \M$ of the bare fields $\phi \in \Phi$. 
The image of $F_{\bs \ups}$ can thus be seen as a ``coordinatisation" of $\M$, i.e. of the space of kinematical histories $\P(\mC)$.
 Furthermore, $\phi^{\bs \ups}=\phi^{\bs \ups(\phi)}$ is explicitly a $\phi$-dependent coordinatisation of the physical d.o.f. embedded in the bare fields $\phi$. 
 In other words, (kinematical) physical d.o.f. $[\phi]\simeq \gamma$ are represented in a manifestly \emph{relational} way.\footnote{Dressed fields $\phi^{\bs \ups}$ are then ``relational Dirac variables",  or  ``complete observables" in the terminology of \cite{Rovelli2002, Tamborino2012}. }

The map \eqref{F-map} is manifestly a $\phi$-dependent realisation of $\Phi$'s projection, $F_{\bs \ups} \sim \pi$, and thus in principle
 it allows to build basic forms on $\Phi$ by 
$\Omega^\bullet_{\text{basic}}(\Phi) =$ Im$\,\pi^\star \simeq$ Im$\,F_{\bs \ups}^\star$.  
The  basic counterpart of a form $\bs\alpha=\alpha\big(\!\w^\bullet\! \bs d\phi; \phi \big) \in \Omega^\bullet(\Phi)$ is built as follows: Consider first its formal functional analogue on the base space $\b{\bs \alpha}=\alpha\big( \!\w^\bullet\! \bs d[\phi]; [\phi] \big) \in  \Omega^\bullet(\M)$, and then define the \emph{dressing} of $\bs\alpha$ by
\begin{align}
\label{Dressing-form}
\bs \alpha^{\bs \ups}\defeq F_{\bs \ups}^\star  \b{\bs \alpha} = \alpha \big(  \!\w^\bullet\! F_{\bs \ups}^\star \bs d[\phi]; F_{\bs \ups}(\phi)  \big) \quad \in \ \Omega^\bullet_{\text{basic}}(\Phi).
\end{align}
 It is basic by construction, so has trivial gauge transformation:  $(\bs{\alpha^\ups})^{\bs \psi} = \bs{\alpha^\ups}$ for $\bs\psi \in \bs\Diff(I)\simeq\bs\Aut_v(\Phi)$.
Its concrete form 
depends on the expression of $F_{\bs \ups}^\star \bs d[\phi]$, the dressing of $\bs d\phi$ and a basis for basic forms, in terms of $\bs d \phi$ and $\bs \ups$. 
It may be obtained  via the pullback/pushforward duality computing ${F_{\bs \ups}}_\star : T_{\phi}\Phi \rarrow T_{F_{\bs \ups}(\phi)}\M$, $\mathfrak X_{|\phi} \mapsto {F_{\bs \ups}}_\star \mathfrak X_{|\phi}$, for which we direct the reader to \cite{JTF-Ravera2024gRGFT}. The result is 
\begin{align}
\label{Dressing-basis-1-form}
\bs d\phi^{\bs \ups}\defeq F_{\bs \ups}^\star \bs d [\phi] =  \bs \ups^* \big( \bs d\phi  +  \mathfrak L_{\bs{d\ups} \circ \bs \ups\-} \phi \big) \quad \in \ \Omega^1_{\text{basic}}(\Phi).
\end{align}
The dressing of $\bs\alpha$ may then be written as
\begin{align}
\label{Dressing-form-bis}
\bs \alpha^{\bs \ups} =  \alpha \big(  \!\w^\bullet\! \bs d\phi^{\bs \ups};\, \phi^{\bs \ups}  \big) \ \  \in \ \Omega^\bullet_{\text{basic}}(\Phi).
\end{align}
One may notice from the formal similarity between the action of vertical automorphism $\Xi(\phi)=\bs\psi(\phi)^*\phi$ and of the dressing operation $F_{\bs \ups}(\phi)=\bs \ups(\phi)^*\phi$ follows the analogue expressions for $\bs d\phi^{\bs \psi}$  \eqref{GT-basis-1-form} and $\bs d\phi^{\bs \ups}$ \eqref{Dressing-basis-1-form}, and thus for  $\bs\alpha^{\bs \psi}$  \eqref{GT-general} and $\bs\alpha^{\bs \ups}$ \eqref{Dressing-form}-\eqref{Dressing-form-bis}. 
This grounds the DFM ``rule of thumb" to obtain the dressing of a form $\bs\alpha$ on $\Phi$: 
First, compute its  gauge transformation $\bs\alpha^{\bs \psi}$, then substitute $\bs\psi \rarrow \bs \ups$ in the resulting expression to obtain $\bs\alpha^{\bs \ups}$. 
We shall make systematic use of this rule. 
It should be stressed however that despite this formal resemblance, dressings are not gauge transformations (and thus not gauge fixings), since a dressing field $\bs\ups$ is manifestly not an element of the gauge group $\bs\Diff(I)$, and $F_{\bs \ups}$ is clearly not a vertical automorphism of $\Phi$. 

Remark that a dressed (basic) form $\bs\alpha^{\bs\ups}$ on $\Phi$ can be seen as a form on the space of dressed fields $\Phi^{\bs\ups}$. 
So when acting on them, the exterior derivative can, by \eqref{Dressing-basis-1-form}, we written as 
$\bs d 
=\bs d \phi^{\bs\ups} \tfrac{\delta}{\delta \phi^{\bs\ups}} 
= \bs d x^{\bs\ups} \tfrac{\delta}{\delta x^{\bs\ups}} 
+
\bs d t^{\bs\ups} \tfrac{\delta}{\delta t^{\bs\ups}}$.
This is just a manifestation of its naturality $[\bs d, F^\star_{\bs\ups}]=0$, and the fact that it is a covariant derivative for basic forms.

A dressed form \eqref{Dressing-form-bis} is  a relational representation on $\Phi$ of a form 
$\t{\bs\alpha}=\alpha \big(  \!\w^\bullet\! \bs d\gamma;\, \gamma \big) \   \in \ \Omega^\bullet\big(\P(\mC)\big)$ on the space of physical kinematical histories $\P(\mC)\simeq \M$: in particular, the variational principle of Mechanics involves exact forms in $\Omega^1\big(\P(\mC)\big)$, and we have $\bs d \phi^{\bs\ups} \simeq \bs d \gamma$, the variational operator being the ``dressed" version of $\bs d$ just mentioned. 
As~noted below Fig. \ref{configuration_space-time-fig}, the latter is distinct from a variational principle on the space of bare fields $\Phi$, which involves exact forms in $\Omega^1(\Phi)$, and whose variational operator is the ``bare" $\bs d=\bs d\phi\tfrac{\delta}{\delta \phi}$. 
This stresses again a theme of this paper, i.e. that dressed, relational quantities and operations are the ones that are ``standard" in textbook Mechanics, so that the latter are dis-analogous to their  counterparts on fields space $\Phi$ which are the ones typically used in gRGFT (thus in MFT):
 closer attention should be paid to these conceptual and technical dis-analogies.

 \subsubsection{Dressed regions and integrals} 
 \label{Dressed regions and integrals} 

Dressed forms are defined above via Im$F_{\bs\ups}^\star \simeq$ Im$\,\pi^\star$, i.e. as basic on $\Phi$.  
One similarly  defines dressed integrals as being basic on $\Phi \times \bs i(I)$, i.e. via Im$\,\b\pi^\star$ with $\b\pi: \Phi \times \bs i(I) \rarrow \mathcal R(I)$ and $\mathcal R(I)$ the associated bundle of regions of $I$ defined by  \eqref{Assoc-bundle-regions} in section \ref{Associated bundle of region and integration}.
The projection  $\b\pi$ is realised as:
\begin{equation}   
\begin{split} 
    \b F_{\bs u}: \Phi \times \bs i(I) &\rarrow \mathcal R(I) \simeq \Phi^{\bs \ups} \times \text{Im}\bs\ups\-, \\
    (\phi, i) &\mapsto    \b F_{\bs \ups}(\phi, i)\defeq \big(F_{\bs \ups}(\phi), \bs \ups\-(i) \big) =  \big( \phi^{\bs \ups}, \bs i^{\bs\ups} \big).
\end{split}    
\end{equation}    
Remark that a \emph{dressed interval} $i^{\bs\ups} \defeq  \bs \ups\-(i)$ is a map $i^{\bs\ups}: \Phi \times \bs i(I) \rarrow $ Im$\ups\-$, $(\phi, i) \mapsto i^{\bs\ups(\phi)}$. 
s.t. for $\psi \in \Diff(I)$:
\begin{align}
\label{inv-dressed-region}
(i^{\bs \ups})^\psi \defeq \b R^\star_\psi \, i^{\bs  \ups} 
= i^{\bs  \ups}  \circ  \b R_\psi
= (R^\star_\psi \bs  \ups)\- \circ \psi\-(i) = (\psi\- \circ \bs  \ups)\-\circ \psi\-(i) = \bs  \ups\-(i) \rdefeq i^{\bs\ups},
\end{align}
where the action $\b R_\psi$ on $\Phi \times \bs i(I)$ is defined by \eqref{right-action-diff} -- similarly   for $\bs \psi \in \bs\Diff(I)\simeq  \bs\Aut_v(\Phi)$, so that \mbox{$(i^{\bs \ups})^{\bs\psi} = i^{\bs\ups}$}.
In~particular,  a \emph{dressed point} is also written via the invariant evaluation operation \eqref{ev-op}: $\tau^{\bs\ups}\defeq \bs\ups\-(\tau)={\sf ev}(\bs\ups\-, \tau)$.  

In section \ref{The relationality of MFT} it was established that by the $\Diff(I)$-covariance of MFT, and 
 by conjunction of the hole and the  point-coincidence arguments,  physical time is   defined via the mechanical fields $\phi$ themselves in a $\Diff(I)$-invariant \emph{relational} way.
 This relational notion of physical time, which we denoted Time, is only tacit in MFT but made explicit via the DFM:
 $i^{\bs \ups}$ is a formal implementation of such a $\phi$-dependent $\Diff(I)$-invariant interval of Time, on which relationally defined and $\Diff(I)$-invariant fields $\phi^{\bs \ups}\defeq \bs \ups^*\phi$ ``live", 
 and may then be integrated over.\footnote{In this 1D context, one may equivalently say that $\phi^{\bs \ups}$ are parametrized by, or evolve against, $i^{\bs \ups}$.}
In particular, it is clear that the (0-dimensional) boundary of a Time interval $\d (i^{\bs \ups})$  is $\Diff(I)$-invariant, so that as advertised in section \ref{The relationality of MFT} and made manifest via the DFM \cite{JTF-Ravera_NoBdryPb_2025}, there is no grip for a ``boundary problem". 

Next, one defines the dressing operation on the space $\Omega^\bullet \big(\Phi, \Omega^1(i) \big) \times \bs i(I)$ by
\begin{equation}   
\begin{split} 
    \t F_{\bs \ups}: \Omega^\bullet \big(\Phi, \Omega^1(i) \big) \times \bs i(I) &\rarrow \Omega^\bullet_\text{\text{basic}}\big(\Phi, \Omega^1(\text{Im}\bs\ups \-) \big) \times  \text{Im}\bs\ups \- \\
    (\bs \alpha, i) &\mapsto 
    \t F_{\bs \ups} (\bs \alpha, i)\defeq 
    \big(F_{\bs \ups}^\star \bs \alpha, \bs \ups\-(i) \big) 
    = \big(\bs \alpha^{\bs \ups}, i^{\bs\ups} \big). 
\end{split}    
\end{equation}  
Then, by the DFM rule of thumb, using \eqref{Equiv-GT-Int},  the dressing of an integral $\bs\alpha_i \defeq \langle \bs\alpha, i \rangle=\int_i \bs\alpha$ is
\begin{equation}   
\label{Dressed-integral}
\begin{split} 
({\bs\alpha_i})^{\bs \ups} \defeq \langle\ ,\ \rangle \circ  \t F_{\bs \ups}(\bs \alpha, i) = \langle \bs\alpha^{\bs\ups} , i^{\bs\ups}  \rangle 
= \int_{i^{\bs\ups}} \bs\alpha^{\bs \ups}. 
\end{split}    
\end{equation}  
We may then also write it as $(\bs\alpha^{\bs\ups})_{i^{\bs \ups}}$
It must be taken into account that the action of $\bs d$ on a dressed integral will now affect the dressed, $\phi$-dependent, region $i^{\bs\ups}$: we have the analogue of \eqref{commut-bXi-d-int},
\begin{align}
\label{commut-tF-d-int}
\bs d\big({\bs\alpha_i}^{\bs\ups} \big)
&=
\left( \bs d\bs\alpha_i\right)^{\bs\ups}
- \langle \mathfrak L_{ \bs\ups
\-_* \bs d\bs\ups}\, \bs\alpha^{\bs\ups}, 
i^{\bs\ups}  \rangle.
\end{align}
This result, $[\t F_{\bs\ups}^\star, \bs d]\neq 0$, stems from the fact that $\bs d$ is not a natural operator on $\Phi \times \bs i(I)$, and implies that the $\Diff(I)$-invariance of a dressed integral is not in general preserved by the action of $\bs d$. 
However, the commutator yields a boundary term, $[\t F_{\bs\ups}^\star, \bs d]=\langle \iota_{ \bs\ups
\-_* \bs d\bs\ups}\, \bs\alpha^{\bs\ups} , \d i^{\bs\ups}  \rangle$, a key fact to get a well-behaved \emph{relational variational principle}. 

Now, for $\bs \alpha \in \Omega^\bullet_\text{tens} \big(\Phi, \Omega^1(i) \big)$ one has 
\begin{equation}   
\label{dressed-bare-int-eq}
\begin{split} 
({\bs\alpha_i})^{\bs \ups} = \langle \bs\alpha^{\bs \ups} , i^{\bs\ups}   \rangle = \langle \bs \ups^*\bs\alpha, i^{\bs\ups}   \rangle = \langle \bs\alpha, i   \rangle = \bs\alpha_i,
\end{split}    
\end{equation}  
by the invariance property \eqref{invariance-action} of the integration pairing. 
Which means that, as one might expect, invariant bare ``$I$-integrals" (of tensorial integrand) are numerically equal to their dressed (physical) counterparts ``Time integrals". 
For such integrals we have thus, ${\bs d \bs\alpha}_i = \langle \bs d \bs\alpha, i \rangle =  \bs d ({\bs\alpha_i})^{\bs \ups}=\bs d  \langle \bs \ups^*\bs\alpha, i^{\bs\ups}   \rangle $.
This, combined with \eqref{commut-tF-d-int} (specialized for a tensorial $\bs\alpha$, so that $\bs\alpha^{\bs\ups}=\bs\ups^*\bs\alpha$)
yields  the dressed analogue of  \eqref{Vert-trsf-int-dalpha}: 
\begin{equation}
\label{Dressing-int-dalpha}
\begin{aligned}
 \langle  \bs{d\alpha}, i \rangle ^{\bs \ups} = \langle\bs{d\alpha}, i \rangle + \langle \mathfrak L_{\bs{d\ups}\circ \bs \ups\-} \bs\alpha,  i \rangle.
\end{aligned}
\end{equation}
This too is key to the relational variational principle we discuss next in section \ref{Basic Classical Mechanics}. 

\subsubsection{Residual transformations of the second kind}  
\label{Residual transformations of the second kind}  

If a dressing field is defined for only a normal subgroup $\Diff(I)_{\text{\tiny$0$}} \subset \Diff(I)$, then one should expect dressed fields $\phi^{\bs\ups}$ and dressed forms $\alpha^{\bs\ups}$ to display transformations under the residual group $\Diff(I)_{\text{\tiny$r$}}\defeq \Diff(I)/\Diff(I)_{\text{\tiny$0$}}$ -- idem for dressed integrals. 
These are usually called  \emph{residual transformations of the 1st kind} in the DFM \cite{JTF-Ravera2024gRGFT}.
For example, one may take $\Diff(I)_{\text{\tiny$0$}}=\Diff(I)_{\text{\tiny c}}$ the subgroup of diffeomorphisms with compact supports,   involved in the Hole argument. 
But we shall not  pursue this further here, but rather focus on another type of residual transformations.   

Given the definition \eqref{Dressing-fields-space}, two dressing fields may be related by the action of a smooth invertible map on their source space, that is, there is a priori a right action of $\Diff(J)$ 
\begin{equation}
\begin{aligned}
\D r[J, I] \times \Diff(J) &\rarrow \D r[J, I] \\
(\bs \ups, \vphi) &\mapsto  \bs \ups \circ \vphi = \vphi^*\bs\ups \rdefeq \bs \ups^\vphi.  \label{dressing-ambiguity}
\end{aligned}
\end{equation}
The group $\Diff(J)$ then parametrizes the ``ambiguity" in the choice of the dressing field: this we call the group of \emph{residual transformations of the 2nd kind} \cite{JTF-Ravera2024gRGFT}.

\emph{If} it does \emph{not} act on $\Phi$, which we write $\phi^\vphi=\phi$,  
then its action on dressed fields is 
\begin{equation}
\label{2nd-trsf-dressed-field}
\begin{aligned}
\Phi^{\bs\ups} \times \Diff(J) &\rarrow \Phi^{\bs\ups}, \\
(\phi^{\bs \ups}, \vphi) &\mapsto   (\phi^{\bs \ups})^\vphi \defeq (\bs \ups^\vphi)^*\phi = ( \bs \ups \circ \vphi)^*\phi =  \vphi^* ({\bs \ups}^*\phi) =  \vphi^* \phi^{\bs \ups}. 
\end{aligned}
\end{equation}
The space  of dressed fields  would then be a priori a $\Diff(J)$-bundle,  $\Phi^{\bs \ups} \xrightarrow{\pi}   \Phi^{\bs \ups}/\Diff(J) \rdefeq \M^{\bs u}$,
with  SES  
 \begin{align}
 \label{SESgroup2}
\bs\Diff(J) \simeq \bs{\Aut}_v(\Phi^{\bs \ups})  \rarrow   \bs{\Aut}(\Phi^{\bs \ups}) \rarrow \bs{\Diff}(\M^{\bs \ups}),
 \end{align}
where $\bs\Diff(J)$ 
is the gauge group of $\Phi^{\bs \ups}$, 
acting on
 a dressed form $\bs\alpha^{\bs \ups}=\alpha\big( \! \w^\bullet \! \bs d\phi^{\bs \ups}; \phi^{\bs \ups}\big) \in \Omega^\bullet(\Phi^{\bs \ups})$ as:
\begin{align}
\label{2nd-kind-trsf-general-bis} 
(\bs\alpha^{\bs \ups})^{\bs\vphi}  = \alpha\big(\! \w^\bullet \! (\bs d\phi^{\bs \ups})^{\bs \vphi};\,  (\phi^{\bs \ups})^{\bs\vphi} \big),  
 \end{align}
 in exact analogy with the action  \eqref{GT-general} and \eqref{GT-basis-1-form} of the gauge group $\bs\Diff(I)$ on a bare form $\bs\alpha$. 

Since $(\phi^{\bs \ups})^{\vphi}$ is $\Diff(I)$-invariant for all $\vphi\in \Diff(J)$,  all representatives in the $\Diff(J)$-orbit $\O_{\Diff(J)}[\phi^{\bs \ups}]$ of $\phi^{\bs \ups}$  are valid coordinatisations of $[\phi]\in \M$. 
So, a priori $\O_{\Diff(J)}[\phi^{\bs \ups}] \simeq  \O_{\Diff(I)}[\phi]$, and $\M^{\bs \ups} \simeq \M$.  
Such a situation, where little seems to be gained by dressing since $\bs\Diff(J)$ replaces the eliminated $\bs\Diff(I)$, obtains notably (but not only) for $\phi$-\emph{independent} dressing fields $\upupsilon$, or \emph{ad hoc} dressing fields,  introduced by \emph{fiat} in a theory as new d.o.f. (thereby creating a \emph{distinct} theory).
We direct the reader to the discussion in \cite{JTF-Ravera2024gRGFT} section 5.2.5 for more on this~point.
In~concrete and favorable situations, the constructive process yielding a $\phi$-\emph{dependent} dressing field $\bs \ups(\phi)$ from the fields $\phi$ of the theory may be such that the arbitrariness reflected by \eqref{dressing-ambiguity} is parametrized by only a  discrete subgroup of $\Diff(J)$ -- e.g. just counting the restricted options for candidates dressing fields presented by the list of fields $\phi$ --
or even by the trivial group (implying the uniqueness of the dressing field).
\medskip

But there is another important case, overlooked in earlier presentations of the DFM and yet, as it turns out,  essential here -- and important in gRGFT more generally (e.g. when the DFM reproduces ``scalar coordinatizations" of GR):
The case when $\Diff(J)$ acts on $\Phi$, typically because one of the fields in the collection $\phi=\{\phi_0,\ldots ,\phi_N\}$ is itself the dressing field. 
Then   $\phi^\vphi \neq \phi$, and there is no guarantee in general that $(\phi^{\bs\ups})^\vphi = \vphi^* \phi^{\bs\ups}$. 
Transformations \eqref{2nd-kind-trsf-general-bis} still hold, just with $(\phi^{\bs\ups})^\vphi$ computed from the specifics of the situation at hands rather that defined by \eqref{2nd-trsf-dressed-field}.
For example, if the dressing $\bs\ups(\phi)=\bs\ups(\phi_i)$ is built from a field $\phi_i: I \rarrow J \in \phi$ so that 
\begin{equation}
\label{Res-2nd-kind-new}
\begin{aligned}
  \phi_j^\vphi &= \phi_j \quad \text{for } j\neq i, 
  \quad \text{ and } \quad
  \phi_i^\vphi =  \vphi\- \circ \phi_i, 
 \\
   \text{then }\quad 
 (\phi_j^{\bs\ups})^\vphi &= {\bs\ups}^*\phi_j^{\bs\ups}, \quad \text{for } j\neq i, 
  \quad \text{ and } \quad
  (\phi_i^{\bs\ups})^\vphi =  \vphi\- \circ \phi_i^{\bs\ups} \circ \vphi.
\end{aligned}
\end{equation}

In either cases, $(\phi^{\bs\ups})^\vphi = \vphi^* \phi^{\bs\ups}$ or $(\phi^{\bs\ups})^\vphi \neq \vphi^* \phi^{\bs\ups}$, $\Diff(J)$ acts   on dressed integrals $\langle \bs\alpha^{\bs\ups}, i^{\bs\ups}\rangle$  as 
\begin{align}
 \label{trsf-2nd-kind-dress-int}
\langle \bs\alpha^{\bs\ups}, i^{\bs\ups} \rangle^{\,\bs\vphi}
\defeq
\int_{ \bs\vphi\-(i^{\bs\ups})} (\bs\alpha^{\bs\ups})^{\bs\vphi},  
\end{align}
 in  analogy with the action \eqref{Equiv-GT-Int} of the gauge group $\Diff(I)$ on bare integrals $\bs\alpha_i=\langle \bs\alpha, i\rangle$.
But in the 
case 
$(\phi^{\bs\ups})^\vphi \neq \vphi^* \phi^{\bs\ups}$, the dressed objects $\bs\alpha^{\bs\ups}$ do not a priori transform under $\Diff(J)$ in a way analogous to bare quantities $\bs\alpha$ transforming under $\Diff(I)$.
This means that the residual transformations of the second kind can carry a physical meaning that the original $\Diff(I)$ transformations did not have. 
Indeed, covariance under $\Diff(J)$ is not automatic, but may hold for some subgroup(s) -- perhaps trivial -- revealing non-trivial physical features of the theory: as they typically encode \emph{frame transformations}, requiring invariance/covariance distinguishes a special class of frames.

\subsection{Basic Classical Mechanics} 
\label{Basic Classical Mechanics}  

We now apply the DFM to obtain the basic, i.e. manifestly invariant and relational, version of a MFT. 
First, naturally we write the dressed action integral 
\begin{equation}
\label{dressed-Action-functional}
\begin{aligned}
 S^{\bs\ups}\defeq \langle L^{\bs\ups}, i^{\bs\ups} \rangle 
 = \int_{i^{\bs\ups}} L^{\bs\ups}
 =\int_{{\bs\ups}\-(i)} {\bs\ups}^*L, 
\end{aligned}
\end{equation}
involving the dressed Lagrangian $L^{\bs\ups}(\phi)= {\bs\ups}^*L(\phi)=L(\phi^{\bs\ups})$, i.e. the Lagrangian expressed in terms of the dressed fields $\phi^{\bs\ups}$ which, as  observed earlier, ``live on" (are parametrized by) the physical Time interval $i^{\bs\ups}$. 

Since $L \in \Omega_\text{tens}\big(\Phi,  \Omega^1(I)\big)$, we have that  $S^{\bs\ups}=S$ as a special case of \eqref{dressed-bare-int-eq}, meaning that the bare ($I$-integral) action is numerically equal to the dressed physical (Time integral) action.  
It is then tempting to think the variational principle remains unchanged, as  $\bs d S = \bs d(S^{\bs\ups})$. 
But we should remind that $[\t F_{\bs\ups}^\star, \bs d]\neq 0$ by \eqref{commut-tF-d-int}, and stress that $\bs d(S^{\bs\ups})$ would not be $\Diff(I)$-invariant.
To get an invariant, relational variational principle, we must correct the bare one, as indicated by \eqref{Dressing-int-dalpha}, as 
\begin{align}
\label{relat-var-princ}
(\bs d S)^{\bs\ups} 
= \bs d S + \int_i \mathfrak L_{\bs{d\ups}\circ \bs\ups \-} L
= \bs d S + \int_{\d i} \iota_{\bs{d\ups}\circ \bs\ups \-} L.
\end{align}
This indicates that bare and dressed variational principles lead to  equations of motion with the same functional form, which we can  flesh out right away.
By \eqref{Dressed-integral}, and by definition of the bare variational principle \eqref{dS-dL}, we have
\begin{align}
\label{relat-var-princ-bis}
(\bs d S)^{\bs\ups} 
\defeq 
\int_{i^{\bs\ups}} (\bs dL)^{\bs\ups} 
=
\int_{i^{\bs\ups}} \bs E^{\bs\ups} + d (\bs \theta^{\bs\ups}),
\end{align}
where  $\bs E^{\bs\ups}=E(\bs d\phi^{\bs\ups}; \phi^{\bs\ups}) \in \Omega^1_\text{basic}(\Phi)$ are the  dressed equation of motion, and $\bs \theta^{\bs\ups}=\theta(\bs d\phi^{\bs\ups}; \phi^{\bs\ups}) \in \Omega^1_\text{basic}(\Phi)$ is the dressed symplectic potential.
By the DFM rule of thumb, using their gauge transformations \eqref{GT-E-theta} we find their expressions in terms of their bare counterparts,
\begin{align}
\label{dressed-E-theta} 
 \bs E^{\bs\ups} = \bs\ups^* \bs E, 
\quad \text{and} \quad 
 \bs \theta^{\bs\ups} = \bs\ups^* \big(  \bs \theta + \iota_{\bs{d\ups}\circ \bs\ups\-} L \big).
\end{align}
The relational equations of motion $\bs E^{\bs\ups}$ being $\bs\Diff(I)$-invariant, and manifestly implementing the point-coincidence argument, they are immune to the hole argument (both discussed in section \ref{The relationality of MFT}) and have thus a well-posed Cauchy problem. 
The space of dressed/relational solutions $\S^{\bs\ups}\defeq \big\{\,\phi^{\bs\ups} \in \Phi^{\bs\ups} \ |\ \bs E^{\bs\ups}=0 \, \big\}$ is thus a \emph{coordinatization of the base space} $\M_\S$ of the bundle of bare solutions $\S \rarrow \M_\S$, which -- as discussed below Fig. \ref{configuration_space-time-fig} 
-- is  itself isomorphic to the space $\P_{\!\!c}(\mC)$ of physical dynamical histories (critical curves) in configuration space-time $\mC$ -- which, in turn, projects to the space $\P_{\!\!c}(Q)$ of critical curves  in the configuration space $Q$: 
we have $\S^{\bs\ups} \simeq \M_\S \simeq \P_{\!\!c}(\mC) \rarrow \P_{\!\!c}(Q)$, 
and 
$\phi^{\bs\ups} \simeq [\phi]\simeq \gamma^c \rarrow \gamma_{q}^c$.

Assessment of the presence and meaning of residual transformations of the second kind cannot be made in general, beyond what was already said in section \ref{Residual transformations of the second kind}. But next, we shall flesh out the case of NR MFT, where indeed they will show to relate to a notion of clock change and allows to single out a class of ``good" clocks.  

\subsubsection{Dressed non-relativistic mechanics}
\label{Dressed non-relativistic mechanics}

Section \ref{Models of MFT} started with considerations on the bare kinematics, we shall then start here with the  dressed kinematics. 
First, let us denote by $d\b t$ the basis 1-form of Im$(\ups\-)$ 
($\subset J$ a priori)
and $\d_{\b t}$ the dual vector basis, so that $\iota_{\d_{\b t}} d\b t =1$ and the exterior derivative on dressed fields is $d=d\b t\, \d_{\b t}$. 
We have first 
\begin{equation}
\begin{aligned}
\ups^*d \phi = d\ups^*\phi &=d\phi^\ups 
= \left(dx^\ups ,\, dt^\ups \right)\\
&= d\b t\, \d_{\b t} \,\phi^\ups
= d\b t \, \left( \d_{\b t} \,x^\ups,\, \d_{\b t} \,t^\ups\right).
\end{aligned}
\end{equation}
So, analogous to \eqref{psi-trsf-t-field}, we have
\begin{equation}
\label{dressed-t-field}   
\begin{aligned}
\d_{t^\ups}=\d_{\ups^*t} \defeq&\,  
  [ \d_{\b t}\, t^\ups ]\- \d_{\b t}
\\
=&\, (\dot t \circ \ups )\-\, (\d_{\b t}\ups)\-  \d_\tau
= (\dot t \circ \ups )\-\, [\d_\tau(\ups\-) \circ \ups ] \d_\tau
= (\dot t \circ \ups )\-\, [(\ups\-)_*\, \d_\tau \circ \ups] 
= (\ups\-)_* \,\d_t \circ \ups.
\end{aligned}
\end{equation}
The result of the first line ensures that $\iota_{\d_{t^\ups}} d\, t^\ups=1$, that of the second allows to crosscheck using \eqref{psi-trsf-t-field} the $\Diff(I)$-invariance of $\d_{t^\ups}$ -- which is first manifest from that of $t^\ups$. 
The dressed velocity and acceleration are then 
\begin{equation}
\label{dressed-v-a}
\begin{aligned}
v^\ups\defeq \ups^* v 
&= \ups^*( \mathfrak L_{\d_t} x) 
= \mathfrak L_{( {\ups\-}_* \d_t \circ \ups) } (\ups^*x)
=\mathfrak L_{\d_{t^\ups}} x^\ups 
= \d_{t^\ups} x^\ups \\
&=\frac{\d_{\b t}\, x^\ups}{\d_{\b t}\, t^\ups}
= \frac{\dot x \circ \ups}{\dot t \circ \ups} = v \circ \ups. \\
a^\ups\defeq \ups^* a 
&= \ups^*( \mathfrak L_{\d_t} v) 
= \mathfrak L_{( {\ups\-}_* \d_t \circ \ups) } (\ups^*v)
=\mathfrak L_{\d_{t^\ups}} v^\ups 
= \d_{t^\ups} v^\ups \\
&=\frac{\d_{\b t}\, v^\ups}{\d_{\b t}\, t^\ups}
= \frac{\dot v \circ \ups}{\dot t \circ \ups} = a \circ \ups. 
\end{aligned}
\end{equation}
From which follows that $dx^\ups=v^\ups\,dt^\ups$ and $dv^\ups=a^\ups\, dt^\ups$. 
Let us now consider next the dressed dynamics.

Given the bare Lagrangian \eqref{NR-Lagrangian}, the dressed Lagrangian is 
\begin{equation}
\label{dressed-L-NR}
\begin{aligned}
 L^{\ups}(\phi)
  =\ups^* L(\phi)
  = L(\phi^{\ups}) 
  = L( x^\ups, t^\ups)
  &=
  \Big(
  \tfrac{m}{2} \langle  v^\ups,  v^\ups\, \rangle    - V( x^\ups, t^\ups)\Big)\, d t^\ups \\
  &=\Big(
  \tfrac{m}{2} \langle  \d_{\b t} x^\ups, \d_{\b t} x^\ups\, \rangle\, (\d_{\b t} t^\ups)\-    - V( x^\ups, t^\ups)\ \d_{\b t} t^\ups\,\Big)\, d \b t.
\end{aligned}
\end{equation}
The relational variational principle \eqref{relat-var-princ-bis} gives the dressed field equations 1-form: 
\begin{equation}
\label{dressed-E-NR}
\begin{aligned}
\bs E^{\bs\ups}=E(\bs d\phi^{\bs\ups}; \phi^{\bs\ups}) &=
E(\bs d x^\ups; x^\ups, t^\ups ) + E(\bs d t^\ups; x^\ups, t^\ups ) \\
&= -\langle d(mv^\ups) + \tfrac{\delta V^\ups}{\delta x^\ups}\, dt^\ups, \bs d x^\ups \rangle 
 + 
\left[
d\Big( \tfrac{m}{2} \langle v^\ups, v^\ups\rangle + V(x^\ups, t^\ups) \Big)  - \tfrac{\delta V^\ups}{\delta t^\ups}dt^\ups
\right]\bs d t^\ups \\
&\rdefeq - \langle d p_{x^\ups}  - F^\ups dt^\ups ,\bs d x^\ups \rangle - 
\left[d p_{t^\ups} +   \tfrac{\delta V^\ups}{\delta t^\ups}dt^\ups
\right]\bs d t^\ups,
\end{aligned}
\end{equation}
where the dressed conjugate momenta  $p_{\phi^\ups}= \ups^* p_\phi$ can also be denoted $(p_\phi)^\ups$.
The dressed field equations have a functionally identical form as the bare ones \eqref{bare-var-princ-NR}-\eqref{Field-Eq-NR-Mechanics};
thus, as in the bare case, $E(\bs d x^\ups; x^\ups, t^\ups )=0$ implies $E(\bs d t^\ups; x^\ups, t^\ups )=0$.
By the above dressed kinematics, and \eqref{Dressing-basis-1-form}, 
one checks that   \eqref{dressed-E-theta} holds:
$\bs E^{\bs\ups}=E(\bs d\phi^{\bs\ups}; \phi^{\bs\ups})=\ups^*\bs E$. 
In particular, $E(\bs d x^\ups; x^\ups, t^\ups )=0$ yields $0=(ma^\ups - F^\ups)\, dt^\ups =  \ups^*(ma - F)\, \d_{\b t} t^\ups\,  d\b t$. 
From that same variational principle \eqref{relat-var-princ-bis}
one finds the dressed presymplectic potential 
\begin{equation}
\label{dressed-theta-NR}
\begin{aligned}
\bs\theta^\ups 
&=
\theta(\bs d x^\ups; x^\ups, t^\ups ) 
+
\theta(\bs d t^\ups; x^\ups, t^\ups )\\
&=
\left[\langle mv^\ups, \bs d x^\ups\rangle
- \Big(  \tfrac{m}{2} \langle v^\ups, v^\ups\rangle + V(x^\ups, t^\ups)   \Big)\,\bs dt^\ups
\right]\\
&\rdefeq
\langle p_{x^\ups}, \bs dx^\ups \rangle + p_{t^\ups} \bs d t^\ups.
\end{aligned}
\end{equation}
Again, by the dressed kinematics above and \eqref{Dressing-basis-1-form}, it is easy to verify that \eqref{dressed-E-theta} holds, i.e. $\bs \theta^{\bs\ups} = \bs\ups^* \big(  \bs \theta + \iota_{\bs{d\ups}\circ \bs\ups\-} L \big)$. 

All this shows that  that the DFM rule of thumb applies, but still only \emph{formally}. 
Indeed, as things stand the dressing field $\ups$ is introduced by \emph{fiat}: it is not constructively built from the existing mechanical fields $\phi=(x, t)$, but added as a separate d.o.f. thereby extending the MFS $\Phi$ to $\Phi' =\Phi +\ups = \{x, t , \ups\}$. 
Such are \emph{ad hoc} dressing fields, as mentioned in section \ref{Residual transformations of the second kind} -- see section 5.2.5 of \cite{JTF-Ravera2024gRGFT} for a discussion. 
As a matter of fact, the $\bs\Diff(I)$-invariant dressed Lagrangian  \eqref{dressed-L-NR},  resembles the textbook one, but is not quite it yet. 
To achieve fleshing out the physics of the dressed formalism,  the dressing field must be extracted from $\Phi$.

\subsubsection{Clock field as dressing field}
\label{Clock field as dressing field}

The first thing to do to either identify or built a $\phi$-dependent dressing field \eqref{Field-dep-dressing} is to inspect the (bare) kinematics so as to check if a candidate stands out among the existing fields.  
In MFT, such a natural candidate is none other than the clock field: 
Since $t^\psi = R_\psi t = t\circ \psi$, we have
\begin{equation}
\label{clock-fied-dressing}
\begin{aligned}
\bs\ups(\phi)=\bs\ups(x, t)\defeq t\-\ : \ \T &\rarrow I, \\ 
\b t &\mapsto [\bs\ups(\phi)](\b t) \rdefeq \bs\ups(\b t), \\[1mm]
(R^\star_\psi \bs\ups)(\phi) =\bs\ups(\phi^\psi) \defeq (t^\psi)\-=&\ \psi\- \circ t\- =\psi\- \circ \bs\ups(\phi).
\end{aligned}
\end{equation}
In that case we have thus $J=\T$ and $\b t$ 
is a reading of the clock field $t$.
The corresponding dressed fields are
\begin{align}
  \label{dressed-fields-t}
  \phi^{\bs\ups}
  = \big( x^{\bs\ups}, t^{\bs\ups} \big) 
  = \big( x \circ t\-, t\circ t\- \big)
  = \big( \b x, \id_\T \big),
\end{align}
where $\b x \defeq x \circ t\-: \T \rarrow \RR^{3N}$, $\b t\mapsto \b x(\b t)$, so that 
$ \phi^{\bs\ups}(\b t) = \big(\b x(\b t), \b t  \,\big)$ represents a point  $p$ in the configuration space-time (bundle) $\mC\simeq (\RR^{3N} \times \T)$.
It is then clear that the $\Diff(I)$-invariant dressed fields $\phi^{\bs\ups}$ represent  a curve $\gamma \in \mC$, and  also the graph of the function $\b x$ which is a curve $\gamma_{q} \in Q\simeq \RR^{3N}$. 

The dressed kinematics of the previous section correspondingly simplifies somewhat. 
We get first
\begin{equation}
\begin{aligned}
d\phi^\ups 
= \left(dx^\ups ,\, dt^\ups \right)
= d\b t \, \left( \d_{\b t} \,x^\ups,\, \d_{\b t} \,t^\ups\right)
= d\b t \, \left( \d_{\b t} \,\b x,\, 1 \right) 
= \left(d\b x ,\, d \b t \right),
\end{aligned}
\end{equation}
and the clock vector field \eqref{dressed-t-field} reduces to $\d_{t^\ups}= [ \d_{\b t}\, t^\ups ]\- \d_{\b t}=\d_{\b t}$. 
The dressed velocity and accelerations simplify as
\begin{equation}
\label{dressed-v-a-clock}
 \begin{aligned}
v^\ups = \d_{t^\ups}\, x^\ups = \d_{\b t}\, \b x &\rdefeq \b v 
\qquad \qquad \text{ and} \qquad
a^\ups = \d_{t^\ups}\, v^\ups = \d_{\b t}\, \b v \!\!\!\!\! &&\rdefeq \b a \\[-.5mm]
&= v\circ t\-,  &&=  a\circ t\-,
 \end{aligned}   
\end{equation}
so $d\b x = \b v\, d\b t$ and $d\b v = \b a\, d\b t$. 
Therefore, as expected, the $\bs\Diff(I)$-invariant dressed velocity $v^{\bs\ups} = \b v$ is the variation of the $\bs\Diff(I)$-invariant dressed spatial field $x^{\bs\ups} = \b x$  measured against the (value set of the) dressed clock field $t^{\bs\ups} =\id_\T$.

\medskip

The basic basis 1-form \eqref{Dressing-basis-1-form} is then $\bs d \phi^{\bs\ups}=  \big( \bs d\b x, 0\big)$, seing that $\bs d (\id_\T) =0$, and basic forms \eqref{Dressing-form-bis} are 
\begin{equation}
\label{Dressing-form-bis-t}
\begin{aligned}
\bs \alpha^{\bs \ups} &=  \alpha \big(  \!\w^\bullet\! \bs d\b x\,;\, \b x, \id_\T \big) 
\quad  \in \Omega^0_\text{basic} \big(\Phi, \Omega^1(\T) \big)=\Omega^0\big(\Phi^{\bs\ups}, \Omega^1(\T) \big), \\[.5mm]
 \text{or }\quad 
\bs \alpha^{\bs \ups} &=  \alpha \big(  \!\w^\bullet\! \bs d\b x\,;\, \b x(\b t), \b t\, \big).
\end{aligned}
\end{equation}
The exterior derivative thus acts on them as 
$\bs d=\bs d \b x \tfrac{\delta}{\delta \b x}$,  
reflecting indeed the variational operation $\bs d=\bs d \gamma \tfrac{\delta}{\delta  \gamma}$ on $\P(\mC)$.
The  expression  \eqref{Dressing-form-bis-t} highlights that we are now dealing with $\T$-dependent, i.e. clock-time dependent quantities. 
Dressed integrals \eqref{Dressed-integral} are thus $\T$-integrals: $({\bs\alpha_i})^{\bs \ups} = \langle \bs\alpha^{\bs\ups},  \b i \,\rangle $
with $\b i \defeq i^{\bs\ups} = \bs\ups\-(i) = t(i) \subset \T$.\footnote{Remember that in that definition of $\b i =t(i)$ as a $\T$-valued functional on $\Phi \times \bs i(I)$,  $\Diff(I)$ acts via $\b R_\psi^\star \b i$ with $\b R_\psi$ defined by \eqref{right-action-diff}, i.e. not only on $t$ -- i.e. by $R_\psi$ as one would posit intuitively -- but also on the domain $i \in \bs i(I)$, according to \eqref{inv-dressed-region}. }

The dressed action \eqref{dressed-Action-functional} is thus the standard clock-time integral  
\begin{equation}
\label{dressed-Action-functional-clock}
\begin{aligned}
 S^{\bs\ups}\defeq \langle L^{\bs\ups}, \b i \,\rangle 
 =\int_{\b i} L(\phi^{\bs\ups}) 
 =\int_{\b i} L(\b x, \id_\T),
\end{aligned}
\end{equation}
with the dressed Lagrangian 
\eqref{dressed-L-NR} 
\begin{equation}
\begin{aligned}
\label{dressed-NR-Lagrangian-clock} 
  L(\phi^{\bs\ups}) 
  = L(\b x, \id_\T)
  &=
  \Big(
  \tfrac{m}{2} \langle \b v, \b v\, \rangle    - V(\b x, \id_\T)\Big)\, d\b t \quad \in \Omega^0_\text{basic} \big(\Phi, \Omega^1(\T) \big),
  \\[1mm]
  \text{or } \quad 
  L\big(\b x(\b t\,), \b t\,\big)
  &=
  \left(
  \tfrac{m}{2} \langle \b v, \b v\, \rangle    - V\big (\b x(\b t\,), \b t\, \big) \right) d\b t,
 \end{aligned}
 \end{equation}
 being  precisely the standard textbook Lagrangian formulation of NR Mechanics. 
The dressed variational principle \eqref{relat-var-princ}-\eqref{relat-var-princ-bis}, then involving varying $\b x$ via $\bs d=\bs d \b x \tfrac{\delta}{\delta \b x}$, 
yields the dressed field equation 
\begin{align}
 \label{dressed-NR-E-clock}
 \bs E^{\bs \ups} 
 =
 E(\bs d \b x; \b x, \id_\T)
 &=
 -\langle d(m \b v) + \tfrac{\delta \b V}{\delta \b x} d \b t, \bs d \b x   \rangle \quad \in \Omega^0_\text{basic} \big(\Phi, \Omega^1(\T)\big)\\
 &\rdefeq
 - \langle d p_{\b x}  - \b F d\b t ,\bs d \b x \rangle,  \notag
\end{align}
as a reduction of \eqref{dressed-E-NR}, which we may also denote 
$\bs E^{\bs\ups}=E(\bs d\b x\,; \b x(\b t\,), \b t\, )$. 
Clearly, $\bs E^\ups=0$ is the just the textbook form of Newton's second law. 
The space of relational solutions is then $\S^{\bs\ups}\defeq \big\{\, \phi^{\bs\ups}= (\b x, \id_\T)  \in \Phi^{\bs\ups} \ |\ \bs E^{\bs\ups}=0 \, \big\}$, and 
the isomorphisms and projection  
 $\S^{\bs\ups} \simeq \M_\S \simeq \P_{\!\!c}(\mC) \rarrow \P_{\!\!c}(Q)$ discussed earlier, below \eqref{dressed-E-theta}, reads then  explicitly 
$\phi^{\bs\ups}= (\b x, \id_\T)  \simeq [\phi]\simeq \gamma^c \rarrow \gamma_{q}^c \simeq \b x$. 
It~appears a part of  the field equations is lost as  $\bs dt^\ups = \bs d \, \id_\T=0$, 
but as noticed earlier,
it is fortunately already a consequence of  $E(\bs d x^\ups; x^\ups, t^\ups )=E(\bs d \b x; \b x, \id_\T )=0$ and is just a continuity equation for the non-conservation of $\bs\Diff(I)$-invariant mechanical energy:
\begin{align}
\label{dressed-cont-eq-energy}
dp_{\b t} - \tfrac{\delta V}{\delta \b t}d\b t 
=
d\Big( \tfrac{m}{2} \langle \b v, \b v\rangle +  V(\b x, \id_\T) \Big)  - \tfrac{\delta \b V}{\delta \b t}d\b t = 0 .
\end{align}
So, no information is lost by switching to the dressed/invariant formulation.
Likewise the dressed presymplectic potential is the standard one: 
\begin{align}
 \label{dressed-NR-theta}
\bs \theta^{\bs \ups} 
 =
 \theta(\bs d \b x; \b x, \id_\T)
 =
 \langle p_{\b x}, \bs d\b x \rangle
 =
 \langle m \b v, \bs d \b x\rangle 
 \quad \in \Omega^0_\text{basic} \big(\Phi, \Omega^0(\T)\big),
\end{align}
as a reduction of \eqref{dressed-theta-NR}, and  may be denoted 
$\bs \theta^{\bs\ups}=\theta(\bs d\b x\,; \b x(\b t\,), \b t\, )$.
It ``lives", or is evaluated at, the invariant, physical 0-dimensional boundary $\d i^{\bs\ups}=\d \b i= \{\b t_0, \b t_1 \} \in \T$ with $\b t_i \defeq \bs\ups\-(\tau_i)= t(\tau_i) ={\sf ev}(t, \tau_i)$ -- where the bare (unphysical) boundary is $\d i=\{\tau_0, \tau_1\}$. 
Which is just the initial and final clock field readings (setting the boundary conditions $\{\b x(\b t_0), \b x(\b t_1)\}$ for the physical spatial field $\b x$).\footnote{Notice that, as stated in section \ref{The relationality of MFT} and reiterated in section \ref{Dressed regions and integrals}, there is no ``boundary problem" -- and no ``edge modes" on the bare boundary $\d i$  are needed to make sense of the physics at play.}

\paragraph{Remark on dressing \emph{vs} gauge-fixing}
As alluded to  in section \ref{Models of MFT}, 
the bare formulation of NR Mechanics in terms of $\phi(\tau)=\big(x(\tau), t(\tau)\big)$ is often referred to as the ``parametrized" formulation, by opposition to the ``de-parametrized" (or ``unparametrised") standard formulation via $x(t)$, which we obtain above as the dressed formulation in terms of  $\phi^{\bs\upsilon}(\b t)=\big(\b x(\b t\,), \b t\, \big)$. 
It is almost as often said that one may go from the former to the latter formulation via  ``gauge-fixing"; by which it is meant that the clock field may be required to satisfy the condition $\dot t=1$, heuristically interpreted as identifying $t$ with the parameter $\tau$, so that $x(\tau)=x(t)$. 

While such a choice is of course allowed,  one should notice that, as stressed in section \ref{The Mechanical Field Space and its natural transformation groups}, like any gauge-fixing it is \emph{not} $\Diff(I)$-invariant: 
as if $t$ satisfies $\dot t=1$,  $t^\psi = t\circ \psi$  would not.
One may then consider only the subgroup of $\Diff(I)$ preserving the gauge-fixing condition; but since 
$\d_\tau{(t^\psi)}=(\dot t \circ \psi)\, \dot \psi$, this subgroup is trivial, $\psi=\id_I$. 
Meaning that by imposing this condition, we have ``broken by hand" the group $\Diff(I)$. 
As a matter of general principles, it is thus conceptually incorrect to state that the $\Diff(I)$-covariant  parametrized (bare) formulation can be ``gauge-fixed" to yield the $\Diff(I)$-invariant de-parametrized (dressed) formulation. 

We remark also that the condition lacks precise characterization, both mathematical and physical, if we are to keep a clear conceptual distinction between the source and target spaces of the clock field $t$ -- even acknowledging in practice the (non-canonical) isomorphisms $I\simeq \RR \simeq \T$.
Regarding the physical aspect, we shall revisit {it in the next section after} discussing the notion of ``good" clocks.
Regarding the mathematical aspect, 
the kernel of truth contained in this  condition is that, properly understood, it is allows to identify the clock field as dressing field.

Consider indeed the \emph{composite field} $\phi^{\bs\ups} \defeq \phi\circ \bs\ups$, where $\bs\ups$ is a smooth map with target space $I$ (it is not yet said that it is a dressing field). Now,
imposing the condition $\tfrac{\d}{\d\b t}(t^{\bs\ups})=1$, for $\tfrac{\d}{\d\b t}$ the basis of vector fields on the target space $\T$ of $t$, implies that $\T$ is also the source space of $\bs\ups$. 
So, $\bs\ups : \T \rarrow I$ is not an element of $\Diff(I)$, and $t^{\bs\ups}$ is not a gauge-transform of $t$ but a smooth map on~$\T$.
The condition -- clearly not a gauge-fixing -- 
may then be written as
\begin{align}
 \label{cond-dressing=clock}   
 \C(\phi^{\bs\ups})&\defeq \tfrac{\d }{\d \b t}( t^{\bs\ups} )  - 1 =0, 
 \quad \Rightarrow \quad
 \tfrac{\d }{\d \b t}\,  \bs\ups - \frac{1}{\dot t \circ \bs\ups}=0.
\end{align}
Solving for $\bs\ups$ in terms of $\phi$, one gets $\bs\ups(\phi)=t\-$; 
  consequently, one finds that $\bs\ups(\phi^\psi)=(t^\psi)\- = \psi\- \circ t\-=\psi\- \circ\bs\ups(\phi)$, i.e. $\bs\ups(\phi)\defeq t\-$ is a $\phi$-dependent dressing field (and manifestly not an element of $\bs\Diff(I)$), 
our starting point of section \ref{Clock field as dressing field}.
One has therefore finally that $\phi^{\bs\ups}=(x^{\bs\ups}, t^{\bs\ups})=(\b x, \id_\T)$ are  $\Diff(I)$-invariant dressed fields,   satisfying \eqref{cond-dressing=clock} by construction, and
 furthermore \emph{invariantly so}.

We slightly belabor the point so as to highlight the important technical and conceptual distinction between  dressing, yielding an invariant formulation, and gauge-fixing, which does not. 
This is a key point of the DFM, and the above circumstance -- whereby a ``gauge-fixing condition" when solved for the parameter (analogue of $\bs\ups$) yields a field-dependent dressing field (instead of a field-dependent gauge group element, analogue of $\bs\psi$), so that one ends-up building invariant fields instead of gauge-fixed ones -- has several echoes in the (gR)GFT literature: 
It~e.g. is the case of the Coulomb or Lorenz conditions as well as for the ``unitary gauge" in the electroweak model \cite{Berghofer-Francois2024, Berghofer-et-al2023, Francois2018}, and instances are found in supersymmetric field theory \cite{JTF-Ravera2024-SUSY, JTF-Ravera-UnconventionalSUSY}.

\subsubsection{Residual transformations of the second kind: clock change covariance, and ``good" clocks}
\label{Clock change covariance and good clocks}

In section \ref{Residual transformations of the second kind} we have discussed the possible ambiguity in the choice of dressing field, and established that it is parametrized by what we called the group $\Diff(J)$ of ``residual transformations of the second kind": $\ups^\vphi = \ups \circ \vphi$ for $\vphi \in \Diff(J)$. 
Two cases have been distinguished. 
\medskip

The first 
assumes that $\phi^\vphi=\phi$, so that the group of second kind transformations {acts on the dressed field} simply by pullback \eqref{2nd-trsf-dressed-field}:
$(\phi^\ups)^\vphi=\vphi^* \phi^\ups$. 
If it is so, the whole dressed formulation transforms under $\Diff(J)$ exactly as the bare formulation transforms under $\Diff(I)$.

This typically happens when 
the dressing field is \emph{ad hoc}, i.e. introduced by hand as separate d.o.f., so that indeed $\phi^\vphi=\phi$ {follows}: 
But in this case it is rather unclear what is gained by dressing, seeing that one appears to just trade $\Diff(I)$ for $\Diff(J)$.
The gain, if any, would seem to  rest entirely on some interpretive effort: 
For example,  on the understanding that the $\Diff(I)$-invariance of $\phi^\ups$ {means that the dressed fields} are physical d.o.f., one may think that $\Diff(J)$ is a transformation group somehow more directly ``physical" than $\Diff(I)$ (which was gauge),\footnote{This is exactly the angle adopted by the ``edge modes" literature (following \cite{DonnellyFreidel2016}): there \emph{ad hoc} dressing {fields are called} edge modes, and residual transformations of the second kind ($\Diff(J)$) are coined ``surface/corner symmetries" and claimed to be ``physical symmetries".} and whose  direct significance is then tied to its interpretation as a physical operation. 

Considering thus the dressing of NR Mechanics of section \ref{Dressed non-relativistic mechanics}, 
one would have that the dressed Lagrangian \eqref{dressed-L-NR} and field equations \eqref{dressed-E-NR} are $\Diff(J)$-covariant. 
{Now comes} an attempt at interpreting:
Say one considers $J$ to be the physical timeline, the action of $\Diff(J)$ on the dressed clock $t^\ups$, $(t^\ups)^\vphi=t^\ups \circ \vphi$,  seems to plausibly instantiate a change of clock, which is then implemented on the dressed spatial field by $(x^\ups)^\vphi=x^\ups \circ \vphi$. 
 But then that would mean that NR Mechanics \eqref{dressed-L-NR}-\eqref{dressed-E-NR} is covariant under arbitrary change of clocks. 
 This conclusion is incongruent with our usual understanding of classical mechanics. 

Alternatively, one may counter-argue  that in such a case (\emph{ad hoc} dressing field), at best, one has actually simply traded a gauge symmetry ($\Diff(I)$) for another, $\Diff(J)$.\footnote{As a matter of fact, in the gauge theoretic setup this has been shown to happen  even when the dressing field is not \emph{ad hoc} and still $\phi^\vphi=\phi$ {follows} (this is what is meant by the ``at best" caveat): See the treatment via DFM of (pure) gauge gravity in section 4.3.1.b of \cite{Francois2021}, where the (co-)tetrad field is a (field-dependent) Lorentz dressing field and the Lorentz gauge group (analogue of $\Diff(I)$) is eliminated but the group of  transformations of the second kind is  exactly the gauge group of general coordinate transformations (analogue of $\Diff(J)$).}
And as a matter of fact, the SES  \eqref{SESgroup2}, being isomorphic to the SES \eqref{SESgroup}, could be pressed in support of this alternative view -- as hinted at in the paragraph following \eqref{SESgroup2}. 
But then we are back to wondering what is gained by dressing, in an \emph{ad hoc} manner. 
This discussion highlights the danger of relying on sheer formality and how it can lead to puzzling and/or wrongheaded physical conclusions.\footnote{In that respect, we again draw attention to the edge mode literature, and related, which  by now has attracted non-negligible attention, but also to the literature on Metric-Affine and/or Poincaré gravity  \cite{JTF-RaveraMAG2025}}

\medskip
As previewed at the end of section \ref{Residual transformations of the second kind}, the second case whereby $\phi^\vphi\neq \phi$ implies that covariance of the dressed formalism under  transformations of the second kind is not automatic, meaning that they may have a distinctive interpretation and potentially encode relevant physical information. 
This is just the case for the dressed NR Mechanics of the previous section \ref{Clock field as dressing field}.
There indeed, the clock field being the dressing field, the group of residual transformations of the second kind  is $\Diff(\T)$, and acts by definition as $\bs\ups^\vphi = \vphi^* \bs\ups = \bs\ups \circ \vphi$ for  $\vphi \in \Diff(\T)$, i.e. $t^\vphi = \vphi\- \circ t $. 
In other words, residual transformations of the second kind are quite clearly \emph{clock changes}: $t \mapsto t'=t^\vphi$. 
We~are indeed in the  case described in  \eqref{Res-2nd-kind-new}: 
the action of $\Diff(\T)$ on the bare fields being $\phi^\vphi = (x^\vphi, t^\vphi)= (x, \vphi\- \circ t) \neq \phi$, 
its action on the dressed fields $\phi^{\bs\ups}=(x^\ups, t^\ups)=(\b x, \id_\T)$ 
is thus
\begin{align}
\label{dressed-fields-2nd-trsf-NR}
(\phi^\ups)^\vphi  
= \big( \,\b x^\vphi, \ \vphi\- \circ \id_\T \circ \vphi \big)
= \left( \,\vphi^* \b x,\ \id_\T  \right)
\neq \vphi^* \phi^\ups.
\end{align}
Defining ${\b t\,}'\! \defeq  \vphi(\b t\,)$, a reading (from the value set) of the new clock $t'=t^\vphi$, and ${\b x\,}'\! \defeq \vphi^* \b x =\b x^\vphi$, 
we then have that $[(\phi^{\bs\ups})^\vphi](\b t\,)
=
\big( {\b x\,}'(\b t\,), \b t\, \big)
=
\big( \b x ({\b t\,}'),  {\b t\,}\big)$.
This shows that we still keep track of the old clock after the change $t \mapsto t'$, allowing to compare the two.
The relation \eqref{dressed-fields-2nd-trsf-NR} implies that the dressed Lagrangian and field equations are not  $\Diff(\T)$-covariant:  
\begin{align}
 \label{2nd-kind-trsf-S}
 (S^{\bs\ups})^{\vphi} 
 = \int_{\bs\vphi\-(\b i)} (L^{\bs\ups})^{\vphi} 
 \neq \int_{\vphi\-(\b i)} {\vphi}^*(L^{\bs\ups})
 =\int_{\b i} L^{\bs\ups} = S^{\bs\ups},
  \qquad \text{ so that } \qquad
  (\bs E^{\bs\ups})^{\bs\vphi} 
 \neq {\bs\vphi}^*(\bs E^{\bs\ups}).
\end{align}
Meaning that 
\emph{NR Mechanics is not covariant under arbitrary clock change}, as we expect.
Its possible covariance under a subgroup $\G_c$ of $\Diff(\T)$ would mean that $\G_c$ defines a distinguished class of clocks: the ``$\G$ood clocks".

To ascertain the facts of the matter one needs only to find the $\Diff(\T)$-transformations of the dressed kinematics. 
But first,  for reference, let us take note that by \eqref{dressed-NR-Lagrangian-clock} we have
\begin{align}
\label{ref-cov-dressed-L-2nd-trsf}
\vphi^*(L^\ups)
=
\vphi^* 
\Big(
  \tfrac{m}{2} \langle \b v, \b v\, \rangle    - V(\b x, \id_\T)\Big)\, \vphi^*d\b t 
=
\Big(
  \tfrac{m}{2} \langle {\b v\,}', {\b v\,}'\, \rangle    - V({\b x\,}', \vphi)\Big)\, d{\b t\,}', 
\end{align}
where  ${\b v\,}'= \b v \circ \vphi$ can  be defined via the dual clock vector field $\d_{{\b t\,}'} \defeq (\d_{\b t}\vphi)\- \d_{\b t}$ associated to $d {\b t\,}' = (\d_{\b t}\vphi)\ d\b t$, by
${\b v\,}' \defeq \mathfrak L_{\d_{{\b t\,}'}}   {\b x\,}' = \d_{{\b t\,}'}  {\b x\,}' = (\d_{\b t}\vphi)\- \d_{\b t}\, (\b x \circ \vphi)
=(\d_{\b t}\vphi)\- \, \left[(\d_{\b t} \b x \circ \vphi) \, \d_{\b t} \vphi \right]= \b v \circ \vphi\,$ -- using \eqref{dressed-v-a-clock} in the last equality --
so that $d {\b x\,}'= {\b v\,}'d{\b t\,}'$. 
Correspondingly, we define 
${\b a\,}' \defeq \mathfrak L_{\d_{{\b t\,}'}}   {\b v\,}' = \d_{{\b t\,}'}  {\b v\,}' = (\d_{\b t}\vphi)\- \d_{\b t}\, (\b v \circ \vphi)
=(\d_{\b t}\vphi)\- \, \left[(\d_{\b t} \b v \circ \vphi) \, \d_{\b t} \vphi \right]= \b a \circ \vphi\,$ -- using \eqref{dressed-v-a-clock}  again --
so that $d {\b v\,}'= {\b a\,}'d{\b t\,}'$. 
Now, 
 by \eqref{dressed-fields-2nd-trsf-NR} we have,
\begin{align}
(d\phi^\ups)^\vphi
\defeq d (\phi^\ups)^\vphi
= d\left( \, \vphi^* \b x,\  \id_\T\right)
= 
d\b t\, 
\big( \, (\d_{\b t}\, \b x \circ \vphi)\, \d_{\b t}\, \vphi, \ 1 \, 
\big)
=\left( d{\b x\,}', d\b t\,\right),
\end{align}
so that indeed $(d \phi^\ups)^\vphi \neq  \vphi^* d\phi^\ups = \vphi^*   \left( d\b x, \, d\b  t\, \right)
= d\b t\, 
\big( \, (\d_{\b t}\, \b x \circ \vphi)\, \d_{\b t}\, \vphi, \, \d_{\b t}\, \vphi \, 
\big) = \big(d{\b x\,}',\, d{\b t\,}' \big)$.
Correspondingly, the $\Diff(\T)$-transformation of the dressed clock field is indeed found to be
\begin{align}
(\d_{t^\ups})^\vphi 
\defeq
\d_{(t^\ups)^\vphi} 
=
[ \d_{\b t}\ (t^\ups)^\vphi ]\- \d_{\b t}
=\d_{\b t}. 
\end{align}
The $\Diff(\T)$-transformations of the dressed velocity $\b v$ and dressed acceleration $\b a$ 
\enlargethispage{.55\baselineskip}
\eqref{dressed-v-a-clock}
are then:
\begin{equation}
\label{2nd-trsf-v-a}
\begin{aligned}
 \b v^\vphi 
 \defeq&\
 \mathfrak L_{\d_{(t^\ups)^\vphi} } \ \b x^\vphi   
 =
\d_{\b t} \, \b x^\vphi 
=
(\d_{\b t}\, \b x \circ \vphi)\, \d_{\b t}\, \vphi
= 
(\b v \circ \vphi )\ \d_{\b t}\, \vphi
=
{\b v\,}'\ \d_{\b t}\, \vphi,  \\[1mm]
\b a^\vphi 
 \defeq&\
 \mathfrak L_{\d_{(t^\ups)^\vphi} } \ \b v^\vphi   
 =
\d_{\b t} \, \b v^\vphi 
=
(\b a \circ \vphi)\, (\d_{\b t}\vphi)^2+ (\b v \circ \vphi)\, \d^2_{\b t} \vphi
=
{\b a\,}'\, (\d_{\b t}\, \vphi)^2 + 
 {\b v\,}'\, \d^2_{\b t} \vphi.
\end{aligned}
\end{equation}
So, we have $d\b x^\vphi=\b v^\vphi\, d\b t$ (which is also equal to $ {\b v\,}' d{\b t\,}'=d{\b x\,}'$, as expected from ${\b x\,}'\defeq \b x^\vphi$), 
and
$d\b v^\vphi=\b a^\vphi\, d\b t$.
From this we have the $\Diff(\T)$-transformation of the dressed Lagrangian \eqref{dressed-NR-Lagrangian-clock}
is
\begin{align}
\label{dressed-NR-L-clock-2nd-trsf} 
  (L^\ups)^\vphi
  \defeq
  L\big((\phi^{\bs\ups})^\vphi\big) 
  = 
  \Big(
  \tfrac{m}{2} \langle {\b v\,}', {\b v\,}'\, \rangle\, (\d_{\b t}\, \vphi)^2    - V({\b x\,}', \id_\T)\Big)\, d{\b t} 
  =
  \Big(
  \tfrac{m}{2} \langle {\b v\,}', {\b v\,}'\, \rangle\, (\d_{\b t}\, \vphi)    - V({\b x\,}', \id_\T) \, (\d_{\b t}\, \vphi)\-\Big)\, d{\b t\,}' .
  \end{align}
Comparing \eqref{dressed-NR-L-clock-2nd-trsf} and \eqref{ref-cov-dressed-L-2nd-trsf} one may  identify the covariance subgroup $\G_c \subset \Diff(\T)$ of the dressed formalism. 

The first thing we may notice is that if the potential 
is explicitly  (clock-)time dependent, then the covariance group reduces to the identity $\G_c=\{\id_\T\}$:   covariance thus singles out the clock field $t$ as a \emph{unique} dressing field. 
This echoes the remarks made in section \ref{Residual transformations of the second kind}, pointing out that the procedure of identification/construction of $\phi$-dependent dressing fields may turn out to be s.t. their arbitrariness is parametrized by a  subgroup of transformations of the second kind that is discrete or  trivial.

When the potential is (clock-)time independent $\b V=V(\b x)$, we remark that mechanical energy $p_{\b t}=\tfrac{m}{2} \langle \b v, \b v\rangle +  V(\b x)$ is conserved by \eqref{dressed-cont-eq-energy}, and 
\eqref{ref-cov-dressed-L-2nd-trsf}-\eqref{dressed-NR-L-clock-2nd-trsf}
reduce to 
\begin{align}
\label{...}
\vphi^*(L^\ups)
&=
\Big(
  \tfrac{m}{2} \langle {\b v\,}', {\b v\,}'\, \rangle    - V({\b x\,}')\Big)\, d{\b t\,}', 
  \quad \text{and} \quad
(L^\ups)^\vphi
= 
\Big(
  \tfrac{m}{2} \langle {\b v\,}', {\b v\,}'\, \rangle\, (\d_{\b t}\, \vphi)    - V({\b x\,}') \, (\d_{\b t}\, \vphi)\-\Big)\, d{\b t\,}' .
  \end{align}
Notice the (dis)similarity with the  expression of the bare Lagrangian \eqref{NR-Lagrangian}. 
The covariance group is then 
\begin{align}
\label{cov-grp-2nd-trsf}
\G_c=\big\{\, \vphi \in \Diff(\T)\ |\ \d_{\b t}\vphi=1 \, \big\}.
\end{align}
Now, from the defining relation 
$t'=\vphi\- \circ t$, 
or $\ups'= \ups \circ \vphi$, or equivalently $t=\vphi(t')$,
we have that  
\begin{align}
\label{ratio-clock-rate}
\d_{\b t} \vphi
=
{\bs\ups'}^* \left( \frac{\dot{t}}{\dot{t'}} \right),
%
\qquad \text{ so that } \qquad
\d^2_{\b t} \vphi
=
{\bs\ups'}^* \left( \big(\, \ddot t\, \dot{t'} - \dot t \, \ddot{t'}\, \big)\, (\dot{t'})^{-3} \right).
\end{align}
So, $\d_{\b t}\vphi$  encodes in a $\Diff(I)$-invariant manner the \emph{ratio of distinct clocks' rates}, 
and $\d^2_{\b t} \vphi$ is the 
variation of this~ratio. 
Both quantities 
are  observables.
It follows that the class of clocks defined by the covariance group $\G_c$ \eqref{cov-grp-2nd-trsf},  \emph{good~clocks}, are those whose relative rates are identical: 
Solving $\d_{\b t}\vphi=1$ yields $\vphi(\b t)= \b t + s$ with $s$   a constant (shift), which translates as the relation among clock fields: $t= \vphi(t')= t'+ s$. 
NR Mechanics (for time independent potentials) is covariant under exchanges of clocks thus related, i.e. it is invariant under ``time translations": 
we thus recover the transformations of the time coordinate given by the {Galilean} (and Bargmann) group(s).\footnote{Such groups  would be, from the perspective of the bare MFT on $I$,   "internal" (gauge) groups acting on the fields $\phi=(x, t)$. 
The Gauge Principle would have  required covariance under those,  constraining the form of the bare Lagrangian $L(\phi)$, leading to \eqref{NR-Lagrangian}. }

The $\Diff(\T)$-transformation of the field equation 1-form is $(\bs E^\ups)^\vphi = E\big(\bs d{\b x}^\vphi; {\b x}^\vphi, \id_\T \big) = \langle (\, m {\b a}^\vphi - {\b F}^\vphi)\, d\b t, \, \bs d{\b x}^\vphi  \rangle$. 
Then, while in the free case ($\b V=0=\b F$) $\bs E^\ups=0$ yields $m\b a= 0$, by \eqref{2nd-trsf-v-a}  $(E^\ups)^\vphi=0$ yields
$m{\b a\,}' = m{\b v\,}' \d_{\b t}\left( (\d_{\b t} \vphi )\-\right)$.
Clocks defined by the covariance group $\G_c$ \eqref{cov-grp-2nd-trsf} are thus those with which there are no ``fictitious/kinematic forces". 
This is  analogue to the (secondary) characterization of \emph{inertial frames} as those in which no ``fictitious/inertial forces" arise: 
The \emph{good clocks} defined by $\G_c$ may then also be called suggestively \emph{inertial clocks}.\footnote{Remark that the free case allows to relax slightly the covariance requirement: for $m {\b a\,}'=0$ to hold it is enough that $\d_{\b t}\vphi=r$ with $r\neq 0$ (the ratio of clock rates) constant. This leads to an affine relation among clocks: $t=\vphi(t')= r\, t' +s$. 
Such  rescalings  feature in generalisations of the {Galilean group}, e.g. the Schrödinger group.}

\paragraph{On the definition of good clocks}
In the bare formulation of MFT, we have  called $t$ a clock field. But the \emph{rate} $\dot t$ of a clock is as important as its readings (its value set), and so does the change $\ddot t$ in its rate (and so on). 
One could say that a clock is (at least) the $2$-jet $(t, \dot t, \ddot t\,)$. 

On the intuition expressed in Newton's \emph{Scholium} cited in section \ref{The relationality of MFT}, 
according to which  ``Absolute, true, and mathematical time, of itself, and from its own nature, \emph{flows equably} [...]" (our emphasis), 
and given an a priori understanding of $I$ as representing this ``Absolute, true, and mathematical time", one may define a \emph{good clock} as one ``keeping Time accurately", 
i.e. as one whose rate is \emph{stable} (in the metrological sense of the term):
meaning a good clock $t$ is s.t. $(t, \dot t=\rho, \ddot t=0)$, with $\rho\neq 0$  constant. 
The ``gauge-fixing" condition mentioned at the end of section \ref{Clock field as dressing field} could then be interpreted as picking up a good clock:  setting $\rho=1$ could be understood as specifying a choice of time unit, which identifies $I$ (Newton's time) with the value set $\T$ of a good clock $t$.
%
Then, any other good clock $t'$ would be s.t. $(t', \dot{t'}=\rho, \ddot{t'}=0)$, so that the dimensionless ratio of their {rates is 
$r={\dot t}/{\dot t'}=1$}.

The problem with this definition of good clocks within the bare formalism, echoing the relational picture of section \ref{The relationality of MFT}, is that
$I$ is unobservable, so how would one operationally verify the condition $\dot t=\rho$ in the first place?\footnote{The argument that it could be ascertained via the fact that when it holds the equation of motion $m a = F$ of $x$ reduces to the familiar $m \ddot x = F$ -- resembling the above discussion of ``inertial clocks" -- does not hold since $\ddot x$ is not an observable either. }
Relatedly, this definition is in tension with the requirement of $\Diff(I)$-covariance of the bare MFT formalism (or with the notion that it is an important feature) as it trivializes it: as discussed in
\ref{Clock field as dressing field}, such a definition is obviously not $\Diff(I)$-invariant, it is preserved only by $\psi=\id_I$. 
\smallskip

By contrast, the definition of ``good clocks" arising from the dressed formalism via the covariance group $\G_c$ \eqref{cov-grp-2nd-trsf} is $\Diff(I)$-invariant, and
essentially \emph{relational} -- in keeping with, but not identical with, the \emph{relational} structure of MFT discussed in section \ref{The relationality of MFT}. 
Indeed, it does not rely on unobservable quantities:
by \eqref{ratio-clock-rate}, the (unobservable) ``rates" of any two good clock fields $\dot t$ and $\dot{t'}$ can be arbitrary as long as their ratio is one, $\d_{\b t}\vphi =1$.  
This is in line with the actual (metrological) practice, whereby one can never define a good, stable, physical clock in itself, in isolation, but 
only always as part of a comparative network of clocks whose rates are defined w.r.t. each other's. 
In other words, one does not have to (actually cannot) start by defining one good clock to then propagate the definition to others.
Rather, one must define the whole relative network of (both actual and potential) good clocks all at once, via the covariance group $\G_c$ itself. And a good clock is not detachable from the network in which it partakes: good clocks and their relations co-define each other, relata and relations are coextensive.\footnote{
For the philosophically minded reader, we remark that this has unmistakable ontic structural realist undertones, and is especially strongly reminiscent of Eddington's group theoretic `moderate' structural realism  \cite{French2003, French-Ladyman2003} -- see also \cite{French} section 4.5 and 4.6. --
according to which physical objects and their network of relations come as a (metaphysical) package, analogously to the elements a group and their relations: 
A~group element is only defined by its ``position" relative to others w.r.t. the abstract group structure (i.e. the composition law) relating~them. 
Analogously, 
the precise position of a physical object within the (lawful)  relational network is constitutive of its  ``identity" and properties. Objects are then defined as nodes within the network, and undetachable from it.}

\subsection{Relational Quantization} 
\label{Basic Quantum Mechanics: Relational Quantization}  

We are, at last, in position to flesh out the notion of \emph{relational quantization} alluded to at the end of section \ref{Quantum Mechanics on Phi}. 
It~shall be first defined 
in the context of MFT, specifically NR MFT as discussed above. 
This will give the template to follow for the relational quantization for gRGFTs, which we outline to conclude the section. 

\subsubsection{QM as the relational quantization of MFT} 
\label{QM as the relational quantization of MFT}  

Let us consider first the case of
non-anomalous quantum MFTs:
those defined by PI that are basic $Z \in \Omega^0_\text{basic}(\Phi, \CC)$, i.e. $R^\star_\psi Z=Z$ or $Z(\phi^\psi)= Z(\phi)$, for 
 $Z(\phi) = \int \!\mathcal D  \phi\  e^{\sfrac{i}{\hbar}\, S(\phi)}$
with $\mathcal D \phi=\mathcal Dx \mathcal D t$ is a formal integration measure on $\Phi$.
By definition of a basic form, it must induce (or arises from the pullback of) a 
corresponding 0-form $\t Z \in \Omega^0\big(\M, \CC \big)$ on the moduli base space $\M$ of physical d.o.f., i.e. the space of paths (kinematical histories) in $\mC$ the configuration space-time, $\M = \P(\mC)$. 
It is a priori better to work with $\t Z$ which involves only physical  relational  d.o.f., instead of $Z$ which involves ``gauge" d.o.f. 
Yet, as we observed in section \ref{Quantum Mechanics on Phi}, typically in field theory one does not  work on $\M$ directly, but rather with a gauge-fixed version 
$Z(\phi)_{\,|\,\text{{\tiny GF}}(\s)}$ defined by \eqref{GF-Z} via a section $\s: \M \rarrow \Phi$. 
The drawback is that this manifestly breaks $\Diff(I)$-covariance. 
The choice $\dot t=1$ discussed in section \ref{Clock field as dressing field} could be interpreted as such a section, via Im$(\s) = \big\{ \, \phi=(x, t) \in \Phi\, |\, \dot t=1   \big\}$. 

By contrast the DFM yields a manifestly $\Diff(I)$-invariant quantization scheme via the \emph{dressed PI}, 
\begin{align}
 \label{dressed-PI}
 Z^\ups(\phi )\defeq Z(\phi^\ups) 
 = \int \!\mathcal D  \phi^\ups\  e^{\sfrac{i}{\hbar}\, S(\phi^\ups)},  \quad \in \Omega^0_\text{basic}(\Phi, \CC)= \Omega^0 (\Phi^\ups, \CC), 
\end{align}
with dressed action integral $S(\phi^\ups) = S^\ups(\phi)$ given by \eqref{dressed-Action-functional}, 
and $\D\phi^\ups=\D x^\ups \D t^\ups$. 
Contrary to $Z$, $Z^\ups$ is invariant not by virtue of its functional form but {because it is built} from invariant fields. 
As the above indeed expresses already, the dressed PI can simply be seen as a functional on the space of dressed fields,  $Z^\ups: \Phi^\ups \rarrow \CC$ with  $\D\phi^\ups$ the integration measure on  $\Phi^\ups$. 
And, as the latter is a good coordinatization of the moduli space,
$\Phi^\ups \sim \M\simeq \P(\mC)$, we indeed have that $Z^\ups\simeq \t Z$. 
Furthermore, since for a $\phi$-dependent dressing field $\ups$, $\phi^\ups$ are explicitly relational variables, we may say that the dressed PI \eqref{dressed-PI} defines a \emph{relational quantization} scheme. 

{Issues related to gauge reduction, boundary conditions, measure factors, and possible global obstructions are naturally tied to the geometric structure underlying the MFT configuration space -- and to the covariant phase space formulation discussed in appendix \ref{Covariant phase space formalism for MFT}. 
In the present setting, once the dressed PI is shown to reproduce the standard quantum-mechanical PI, these aspects reduce correspondingly to their standard counterparts in ordinary QM. 
From this viewpoint, the dressing procedure does not introduce additional structures in the quantization, but rather provides a geometric organization of the underlying redundancy and its reduction, together with an automatic anomaly cancellation mechanism, discussed in the following.}

{This analysis also clarifies why the DFM becomes particularly powerful when extended beyond the MFT setting to field-space constructions for gRGFTs, where the geometric organization of gauge, relational, and anomaly structures become essential for the formulation of an invariant relational PI quantization. 
In such contexts, physically meaningful information from anomalies is not lost, but rather reappears through transformations of the dressing fields themselves, encoding changes of physical reference frames -- cf. section \ref{Relational Quantization for general-relativistic gauge field theories}.}

\medskip
Considering NR MFT with Lagrangian
 $ L(\phi)=L(x, t)
 =
\Big(
\tfrac{m}{2} \langle v, v\, \rangle    - V(x)\Big)\, dt
=
\Big(
\tfrac{m}{2} \langle \dot x, \dot x\, \rangle \,  \dot t\- - V(x)\, \dot t\,    \Big)\, d\tau$, 
 then NR quantum MFT is given by the bare PI $Z(\phi)=Z(x, t)$ as above. 
It is easy to see that the condition $\dot t=1$, viewed as a gauge-fixing, leads to $Z(\phi)_{\,|\,\text{{\tiny GF}}(\s)}$ which much resembles the standard Feynman-Dirac PI of NR QM. 
Yet it cannot truly be the same, as it is not defined on the space of $\Diff(I)$-invariant d.o.f. (paths) $\M=\P(\mC)$. 
We claim that the latter is actually reproduced via DFM.
Indeed, the dressed PI for NR MFT is given by \eqref{dressed-PI} with the dressed Lagrangian \eqref{dressed-L-NR} 
$L(\phi^{\ups}) 
= L( x^\ups, t^\ups)
=
\Big(
\tfrac{m}{2} \langle  v^\ups,  v^\ups\, \rangle    - V( x^\ups)\Big)\, d t^\ups 
=\Big(
\tfrac{m}{2} \langle  \d_{\b t} x^\ups, \d_{\b t} x^\ups\, \rangle\, (\d_{\b t} t^\ups)\-    - V( x^\ups)\ \d_{\b t} t^\ups\,\Big)\, d \b t$. 
But,~as we have seen, this formal result 
must be fleshed out by producing an explicit $\phi$-dependent dressing field.\footnote{As is, it holds in particular for \emph{ad hoc} dressing fields -- e.g. \emph{edge modes} -- but as  discussed in section \ref{Clock change covariance and good clocks}, those typically lead to a wrongheaded physical picture.}
This~was done in section \ref{Clock field as dressing field}, by identifying the clock field as the dressing field \eqref{clock-fied-dressing} through the condition  \eqref{cond-dressing=clock}, $\d_{\b t} \ups =1$. 

\noindent
One then gets the final result that the $\Diff(I)$-invariant relational quantization of NR MFT is 
\begin{align}
 \label{dressed-PI-clock}
Z(\phi^\ups) 
=
Z(\b x, \id_\T)
= \int \!\mathcal D  \b x \  e^{\sfrac{i}{\hbar}\, S(\b x, \id_\T)}  
\quad \in  
\Omega^0 (\Phi^\ups, \CC)\simeq 
\Omega^0 \big(\P(\mC), \CC\big), 
\end{align}
with the dressed Lagrangian \eqref{dressed-NR-Lagrangian-clock}
$L(\phi^{\bs\ups}) 
= L(\b x, \id_\T)
=
\Big(
\tfrac{m}{2} \langle \b v, \b v\, \rangle    - V(\b x)\Big)\, d\b t$:
this is 
the textbook PI for NR QM. 
As~one might have expected, the relational structure of classical NR mechanics carries over to the quantum theory.

Of course,  the dressed PI  \eqref{dressed-PI-clock}
is not invariant under transformations of the second kind $\Diff(\T)$, i.e. arbitrary clock changes, since as we have seen in section \ref{Clock change covariance and good clocks} the dressed action $S^\ups=S(\b x, \id_\T)$ is not. 
Yet, the latter's covariance group $\G_c  \subset \Diff(\T)$ defines the relational network of good clocks.
It is interesting to inquire if the $\Diff(\T)$-transformation of the formal measure $\D \b x$ concurs, or give additional information: one finds
\begin{align}
\label{2nd-trsf-anomaly-NR-clock}
\D {\b x\,}'= \D {\b x}^\vphi = \D \b x \ \exp \Big\{\,   -\tfrac{1}{2} \int \ln (\d_{\b t} \vphi ) \, d\b t \, \Big\}.
\end{align}
So, $Z^\ups$ and $S^\ups$  have the same covariance group $\G_c$. 
One may call the exponential term in 
 \eqref{2nd-trsf-anomaly-NR-clock} a transformation of the second kind anomaly, not to be conflated with 
\emph{gauge} anomalies, which are automatically canceled via the DFM.


\subsubsection{Automatic $\Diff(I)$-anomaly cancellation via dressing} 
\label{Automatic anomaly cancellation}  

As observed in section \ref{Quantum Mechanics on Phi}, the PI on the bare field space $\Phi$ is in general a cocyclic tensorial 0-form $Z \in \Omega^0_\text{tens}(\Phi, C)$ whose $\Diff(I)$-equivariance is given by \eqref{PI-anomalous}:  $R^\star_\psi Z = C(\ ; \psi)\- Z$. 
If so the PI is anomalous and the 1-cocycle~$C(\phi; \psi)$, satisfying \eqref{cocycle}, 
is the integrated $\Diff(I)$-anomaly,  arising possibly from the non-invariance of the integration measure on $\Phi$:
the $\Diff(I)$-anomaly being, 
$a(X; \phi) = \tfrac{d}{ds}  C(\phi; \psi_s) \big|_{s=0}$ 
with $\psi_{s=0}=\id_I$ and $X \defeq \tfrac{d}{ds}  \psi_s \big|_{s=0} \in \diff(I)$, 
and satisfies the Abelian version of the $\diff(I)$ 1-cocycle property \eqref{inf-cocycle}.\footnote{
For example, some 1D MFTs may behave like  conformal field theories (CFT) and exhibit an anomaly proportional to the Schwarzian derivative of $\psi \in \Diff(I)$.}

Suppose that the 1-cocycle formally extends as 
$C: \Phi \times \D r[J, I] \rarrow U(1)$, $(\phi, \ups)\mapsto C(\phi; \upupsilon)$, so that for a $\phi$-dependent dressing field $\ups$ we introduce the notation $C(\ups)\defeq C\big(\phi; \ups(\phi) \big)$, so $C(\ups):\Phi \rarrow U(1)$. 
By the defining 1-cocycle property \eqref{cocycle} and the defining property of the dressing field $\ups$, we have that
\begin{align}
\label{cocyclic-dressing}
&C\big(\psi^* \phi; \ups(\psi^* \phi) \big)
= C\big(\psi^* \phi; \psi\- \circ \ups(\phi) \big)
=
C\big(\psi^* \phi; \psi\-\big) \, 
C\big({\psi\-}^{*}\psi^* \phi;  \ups(\phi) \big)
=
C\big( \phi; \psi\big)\-  
C\big(\phi;  \ups(\phi) \big) \notag \\[1mm]
& \Leftrightarrow \quad 
R^\star_\psi C(\ups) = C(\_\,;\psi)\-
C(\ups).
\end{align}
That is, $C(\ups)$ is a \emph{cocyclic dressing field}. 
The linear version of its equivariance property is  
$
\bs L_{X^v} C(\ups) = - a(X; \phi)\, C(\ups)
$.
The dressed PI \eqref{dressed-PI} for an anomalous theory is thus
\begin{align}
\label{dressed-PI-anomalous}
Z^\ups(\phi)=Z(\phi^\ups)=
C(\ups)\- Z(\phi),
\end{align}
illustrating again the DFM {rule of thumb}.
It is by construction $\Diff(I)$-invariant, $R^\star_\psi Z^\ups = Z^\ups$, or $\bs L_{X^v} Z^\ups=0$, as can be easily checked. 
Meaning that in the dressed PI,  anomalies are automatically canceled via the cocyclic dressing field~$C(\ups)$. 
This encompasses as a special case the construction of so-called Wess-Zumino (WZ) counter-terms, usually introduced by hand to {cancel anomalies}. 
See next section for more on this. 

We may observe that since $Z^\ups \in \Omega^0_\text{basic}(\Phi)$, the variation $\bs d Z^\ups \in \Omega^1_\text{basic}(\Phi)$ is geometrically well-defined, while $\bs d Z$ is not; 
as mentioned in section \ref{Differential structures},  $\bs d$ is a covariant derivative for basic, not tensorial, forms. 
Relatedly, remark that \eqref{cocyclic-dressing}, together with its linearisation, imply that $\bs\varpi_0 \defeq - \bs dC(\ups)\, C(\ups)\-$ is a flat cocyclic connection on $\Phi$ as defined by \eqref{Variational-twisted-connection} in section \ref{Connections on Mechanical Field Space}. 
Then,  the cocyclic covariant derivative of the anomalous PI is the twisted tensorial 1-form $\bs D_0 Z \defeq \bs d Z + \bs \varpi_0 Z \in \Omega^1_\text{tens}(\Phi, C)$. 
It is easy to see that $(\bs D_0 Z)^\ups=C(\ups)\- \bs D_0 Z = \bs d Z^\ups$. 
This indicates a link between $\bs D_0 Z$ and WZ counter-terms. 

Say one defines the anomalous effective quantum functional 
$W \defeq -i \ln Z$ 
as well as 
$c(\_\, ;\psi) \defeq -i \ln C(\_\, ;\psi)$  
so that $R^\star_\psi W = W -\, c(\_\, ; \psi)$, or $\bs L_{X^v}  W= a(X; \phi)$. 
Suppose  we may similarly define $c(\ups) \defeq -i \ln C(\ups)$, which  by \eqref{cocyclic-dressing} is an Abelian cocyclic dressing field: i.e. $R^\star_\psi c(\ups) = c(\ups) - c(\_\, ; \psi)$, or $\bs L_{X^v}  c(\ups)= a(X; \phi)$. 
Then $\bs D_0 Z = iZ \, \bs d\big(W-c(\ups)\big)$, and
since $Z$ and $\bs D_0 Z$ have identical (twisted) equivariance, we have $W^\ups \defeq W  - c(\ups) \in \Omega^0_\text{basic}(\Phi)$. 
This is (the DFM generalization of) a version of the WZ construction: $c(\ups)$  cancels the anomaly of $W$. 
Clearly the horizontality condition $\iota_{X^v} \bs D_0 Z =0$ is equivalent to the $\diff(I)$-invariance of the  dressed (or WZ ``improved") effective quantum functional, $\bs L_{X^v} W^\ups =0$. 
Again, $\bs d W^\ups \in \Omega^1_\text{basic}(\Phi)$ is geometrically well-defined, contrary to $\bs d W$.

\subsubsection{Relational Quantization of general-relativistic gauge field theories} 
\label{Relational Quantization for general-relativistic gauge field theories}  

 Our premise has been to show that (NR) MFT is a special case of the broader framework of gRGFT. 
 In the latter, the field space $\Phi$ is generically made of fields $\phi=\{\omega, \varpi=(A, e), \uppsi, \varphi, \ldots \}$ comprising  Ehresmann connection(s) $\omega$ (Yang-Mills type gauge potentials), Cartan connection $\varpi$ (gravitational potential) made of $A$ a spin-type connection and $e$ the soldering (or vielbein) 1-form, spinor field(s) $\uppsi$ (matter), scalar field(s) $\vphi$, etc. 
 The structure group of $\Phi$ as a principal bundle, the covariance group of gRGFTs, is $\Diff(M) \ltimes \H$ with $\H$ an internal gauge group (e.g. $\SU(N)$ or $\SO(1,3)$) and $\Diff(M)$ the diffeomorphism group of a manifold $M$ \emph{representing} spacetime.\footnote{{The manifold $M$, which
we remind is actually unphysical} and drops out of the physical picture, as per Einstein's point-coincidence argument, {in the same way $I $ is unphysical and
drops out in MFT} as argued in section \ref{The relationality of MFT}. See e.g. \cite{JTF-Ravera_NoBdryPb_2025,JTF-Ravera-Foundation2025, JTF-Ravera2024gRGFT}. }
 The right action of the structure group on $\Phi$ is $R_{(\psi, \upgamma)}\phi \defeq \psi^*(\phi^\upgamma)\rdefeq \phi^{(\psi, \upgamma)}$, with $(\psi, \upgamma) \in \Diff(M) \ltimes \H$ and where $\phi^\upgamma$ generically denote the gauge $\H$-transformation of $\phi$. 
 This action induces a fibration of field space $\Phi \rarrow \M$ where the base $\M$ is the moduli space of fields, containing the invariant \emph{physical} d.o.f. 
The  bundle geometry of the field space $\Phi$ of gRGFTs was studied in detail in \cite{JTF-Ravera2024gRGFT}\footnote{See also \cite{Francois2023-a} for the case of general-relativistic theories, i.e. only $\Diff(M)$, and applications of the bundle geometry of their field space to the their  covariant phase space formalism.} -- yielding the specialisation to MFT described in sections \ref{Geometry of Mechanical Field Space: Kinematics} and \ref{The variational principle for MFT, and its covariance group} -- 
and general results were derived on the  transformations under the gauge group $\bs\Diff(M) \ltimes \bs\H$ (a.k.a. \emph{field-dependent} gauge transformations) of their Lagrangian, action, variational principle and field equations.\footnote{
These were essential steps toward the application of the DFM, as explained in section \ref{Building basic forms via dressing}, leading to an invariant relational formulation of gRGFTs, sketched below.}

 Now, the standard PI quantization of (gR)GFT is defined on $\Phi$:\footnote{We write "(gR)" since although standard PI for GR, or general relativistic theories, have been studied, they did not yet lead to satisfying (finite, renormalizable) quantum theories, contrary to the case of GFTs  encompassing the Standard Model.}
 one writes
$Z(\phi )\defeq  
 \int \!\mathcal D  \phi \ \,  e^{\sfrac{i}{\hbar}\, S(\phi)}$,  
with $\mathcal D \phi$ is a formal measure on $\Phi$. From this, one derives the
effective quantum functional 
$W(\phi) \defeq -i \ln Z(\phi)$. 
This is the textbook approach  we followed in section \ref{Quantum Mechanics on Phi} to discuss quantization of MFT.
As is well-known, it is supplemented by gauge-fixing, e.g. via the BRST formalism. 
While 
$Z(\phi)$ above is often though of as the direct analogue of the standard PI of (NR) QM, it should  be clear  that it actually is not: the latter, \eqref{dressed-PI-clock}, is the dressed PI defining a case of relational quantization.
The conceptual and technical landscape is summarized in Table \ref{Table1}.
\smallskip

\begin{center}
{\small{
\begin{tabular}{c||c|c|} 
          &\textbf{NR MFT}	&		\textbf{gRGFT}	\\[.2em] \hline\hline\\[-1em]
 \textbf{Quantization on Field space} $\Phi$	   &   \ding{52} ({\small parametrized PI}) 	 &         \ding{52} ({\small standard PI for (gR)GFT})	          \\[.2em] \hline\\[-1em]   \textbf{Relational quantization}, on $\M\simeq \P(\mC)$		 & 	\ding{52} ({\small standard PI for QM})			& 		?               \\[.1em] \hline
\end{tabular}
\captionof{table}{Bare \emph{vs} dressed PI quantization.}\label{Table1}}} 
\end{center}
\vspace{-1mm}

\noindent It is clear that the standard PI for GFT is the analogue of the ``bare" PI for MFT.
To complete the table, and find the true analogue of standard (NR) PI for gRGFT, one must define relational quantization for the latter.

To do so, one should 
start from the relational formulation of classical gRGFT via DFM as detailed in \cite{JTF-Ravera2024gRGFT}: 
{This presupposes
that we have found (built) a dressing field} $(\ups, \bs u)=\big(\ups(\phi), \bs u(\phi)\big)$ for the covariance group $\Diff(M)\ltimes \H$, with $\bs u$ a  dressing field for the internal gauge group $\H$, 
{and that we have defined the
invariant dressed fields}
$\phi^{(\ups, \bs u)} \defeq \ups^*(\phi^{\bs u})$ which, in view of the $\phi$-dependence of $(\ups, \bs u)$, manifestly are \emph{relational variables}. 
These then provide a relational coordinatization of the moduli space $\M$ (or possibly only a region thereof), since $\phi^{(\ups, \bs u)}$ maps to a $\Diff(M)\ltimes \H$-class $[\phi]\in \M$. 
The dressed classical action is then $S^{(\ups, \bs u)}=\int_{U^\ups} L^{(\ups, \bs u)}=\int_{U^\ups} \ups^*L(\phi^{\bs u})$, where $U^\ups$ is a $\Diff(M)$-invariant, $\phi$-dependent, region of \emph{physical spacetime}.\footnote{Meaning that   ``dressed regions" $U^\ups\defeq \ups^{-1}(U)$, with $U\subset M$,  are an 
implementation of Einstein's point-coincidence argument. See \cite{Berghofer-et-al2025}. 
}
One thus defines  relational quantization of gRGFTs via a dressed PI:
\begin{align}
\label{dressed-PI-gRGFT}
Z^{(\ups, \bs u)}(\phi)
\defeq 
Z\big(\phi^{(\ups, \bs u)}\big) 
 =
\int \!\mathcal D  \phi^{(\ups, \bs u)}\  e^{\sfrac{i}{\hbar}\, S^{(\ups, \bs u)}}  \quad \in \Omega^0_\text{basic}(\Phi, \CC)= \Omega^0 \big(\Phi^{(\ups, \bs u)}, \CC \big)\simeq \Omega^0(\M, \CC),  
\end{align}
with $\mathcal D  \phi^{(\ups, \bs u)}$ a formal measure on the space of invariant dressed fields $\Phi^{(\ups, \bs u)}$.  
Clearly, \eqref{dressed-PI} is now seen as a special case (in 1D MFT, and no gauge group $\H$) of \eqref{dressed-PI-gRGFT}. 
We may observe that this PI bypasses the notion of gauge-fixing, which it does not need. 
In particular, BRST gauge-fixing is not only superfluous but cannot be implemented since the BRST transformations of the dressed fields $\phi^{(\ups, \bs u)}$, mimicking $\Diff(M)\ltimes \H$ transformations, are trivial.\footnote{
This can be seen from the following fact: Given the BRST transformation $s\,\phi = f\big((\xi, c); \phi\big)$ for $s$ the BRST operator, $(\xi, c)$ the ghost of $\Diff(M)\ltimes \H$, and $f(\_\, ;\phi)$ a functional linear in its first argument, the DFM generically supplies the notion of \emph{dressed ghost} $(\b\xi, \b c)=(\xi, c)^{(\ups, \bs u)}$  -- with e.g. $c^{\bs u}\defeq \bs u^{-1} c \bs u + \bs u^{-1} s\bs u$ in the purely internal (e.g. Yang-Mills) case -- entering the dressed BRST algebra $s\,\phi^{(\ups, \bs u)} = f\big((\b\xi, \b c); \phi^{(\ups, \bs u)}\big)$. 
In case $(\ups, \bs u)$ are complete dressings for $\Diff(M)\ltimes \H$, leaving no \emph{residual transformations of the first kind}, the dressed ghost vanishes $(\b\xi, \b c)=0$, so that $s \, \phi^{(\ups, \bs u)}=0$ as expected.
In case $(\ups, \bs u)$ are dressings for a subgroup of  $\Diff(M)\ltimes \H$, the dressed ghost $(\b\xi, \b c)$ is associated with the subgroup $\K \subset \Diff(M)\ltimes \H$ that is not eliminated via dressing, and $s\,\phi^{(\ups, \bs u)} = f\big((\b\xi, \b c); \phi^{(\ups, \bs u)}\big)$ encodes the infinitesimal residual $\K$-transformations (of the first kind) of the dressed fields $\phi^{(\ups, \bs u)}$. 
For examples of applications e.g. to conformal Cartan geometry (including tractors and twistors) see \cite{FLM2016_I, Attard-Francois2016_I, Attard-Francois2016_II, Francois2019}, and to supersymmetric field theory see \cite{JTF-Ravera2025-AdP}. 
}

The definition \eqref{dressed-PI-gRGFT} applies to anomalous and non-anomalous theories alike. In the former case, by definition, the PI transforms under $\Diff(M)\ltimes \H$ as $R^\star_{(\psi, \upgamma)} Z = C\big(\_\, ; (\psi, \upgamma)\big)\- Z$, where  $C: \Phi \times \big(\Diff(M)\ltimes \H \big)\rarrow U(1)$, $\big(\phi, (\psi, \upgamma) \big) \mapsto C\big( \phi;(\psi, \upgamma)\big)$, is a 1-cocycle of the (structure) group $\Diff(M)\ltimes \H$, i.e. functionally satisfying 
\begin{align}
\label{C-gRGFT}
C\big( \phi;(\psi, \upgamma)\cdot (\psi', \, \upgamma')\big)
=
C\big(\phi; (\psi, \upgamma)\big)\
C\big( \phi^{(\psi, \upgamma)}; (\psi', \, \upgamma')\big),
\end{align}
where on the left-hand side ``$\cdot$" stands for the semi-direct product  of $\Diff(M)\ltimes \H$ given by: 
$(\psi,\,  \upgamma)\cdot (\psi',\,  \upgamma') \defeq \big(\psi  \circ \psi', \,   \upgamma\, (\psi^{-1*}\upgamma ')\big)$. 
The cocycle encodes the (integrated) combined diffeomorphism and gauge anomaly $a\big( (X, \lambda); \phi\big)$, with $(X, \lambda) \in \diff(M)\, \oplus$ Lie$\H$,
from which it is obtained by linearisation. 
As usual, the non-trivial equivariance, yielding the 1-cocycle $C$, comes from the non-invariance of the measure $\mathcal D \phi$.
This implies the gauge transformation
$Z^{(\bs\psi, \bs\upgamma)}= C(\bs\psi, \bs\upgamma)\- Z$, for  elements of the gauge group $(\bs\psi, \bs\upgamma) \in \bs\Diff(M) \ltimes \bs\H$, where we have defined the notation $C(\bs\psi, \bs\upgamma)\defeq C\left( \phi; \big( \bs\psi(\phi), \bs\upgamma(\phi) \big) \right)$. 
For such an anomalous theory, we have that \eqref{dressed-PI-gRGFT} is 
\begin{align}
  \label{Anomalous-dressed-PI-gRGFT}  Z^{(\ups, \bs u)} = C(\ups, \bs u)\- Z
\end{align}
by the DFM rule of thumb, where $C(\ups, \bs u)$ is shown to satisfy 
$R^\star_{(\psi, \upgamma)} C(\ups, \bs u) 
=
C\big(\_\, ; (\psi, \upgamma)\big)\-C(\ups, \bs u)  
$
using the defining property of the dressing field, $R^\star_{(\psi, \upgamma)} (\ups, \bs u)= (\psi, \upgamma)\-\!\cdot (\ups, \bs u)$, and \eqref{C-gRGFT} -- see eq.(349)-(351) in \cite{JTF-Ravera2024gRGFT} for a proof;
It is thus a \emph{cocyclic dressing field}, and
gauge transforms as 
  \begin{align}
\label{cocyclic-dressing-field-gRGFT}
 C(\ups, \bs u)^{(\bs\psi, \bs\upgamma)} 
=
C(\bs\psi, \bs\upgamma)\-C(\ups, \bs u).
 \end{align}
 The linearisation of its equivariance is $\bs L_{(X, \lambda)^v} C(\ups, \bs u) = - a\big((X, \lambda); \phi \big)\, C(\ups, \bs u)$.
 Manifestly, the cocyclic dressing compensates for the  transformation of the anomalous PI, so that \eqref{Anomalous-dressed-PI-gRGFT} remain fully invariant: 
 i.e. the dressed PI of relational quantization has an in-built anomaly cancellation mechanism.

A ``linear" version of this applies to the
effective quantum functional 
$W \defeq -i \ln Z$ which, defining similarly
$c\big(\_\, ;(\psi, \upgamma)\big) \defeq -i \ln C\big(\_\, ;(\psi, \upgamma)\big)$,  
is s.t. $R^\star_{(\psi,\upgamma)} W = W -\,c\big(\_\, ;(\psi, \upgamma)\big)$, 
or infinitesimally $\bs L_{(X, \lambda)^v}  W= -a\big((X, \lambda); \phi \big)$.\footnote{Otherwise known as the anomalous Ward Identity,  written in the BRST formalism as $sW(\phi) = a\big((\xi, c); \phi \big)$ 
and defining the anomaly.
}
Writing $c(\ups, \bs u) \defeq -i \ln C(\ups,  \bs u)$ 
, 
we have by \eqref{Anomalous-dressed-PI-gRGFT} (or by the DFM rule of thumb),
\begin{align}
\label{dressed-W}
W^{(\ups, \bs u)} \defeq W  - c(\ups, \bs u) \quad \in \Omega^0_\text{basic}(\Phi),
\end{align}
where $R^\star_{(\psi, \upgamma)} c(\ups,  \bs u) = c(\ups,  \bs u) - c\big(\_\, ; (\psi, \upgamma)\big)$, or $\bs L_{(X, \lambda)^v}  c(\ups, \bs u)= -a\big((X, \lambda); \phi\big)$;
i.e. the ``linear" cocyclic dressing field $c(\ups, \bs u)$  compensates for the anomalous transformation of $W$. 

The above encompasses as a special case the construction of Wess-Zumino (WZ) counter-terms, which are recovered as cocyclic dressing fields where $(\ups, \bs u)=(\upupsilon, u)$ are \emph{ad hoc},  introduced \emph{by hand} in a theory rather than built from its field content $\phi$. See e.g. section 12.3 in \cite{GockSchuck}, Chap.15 in \cite{Bonora2023}, or the end of Chap.4 in \cite{Bertlmann}.

Now, should transformations of the second kind (parametrizing the ambiguity in the choice of dressing field) -- such as described in section \ref{Residual transformations of the second kind} in the context of MFT -- be present, 
they might have anomalies of their own, giving rise to what we may call suggestively an ``anomaly seesaw mechanism".
 
\paragraph{Seesaw mechanism for anomalies}

As detailed in \cite{JTF-Ravera2024gRGFT} section 5.3.2., in the context of the DFM for gRGFTs,  transformations of the second kind may arise thus: 
Another dressing field has to be 
$(\ups', \,\bs u')= \big(\ups \circ \upvarphi,\, \bs u \upzeta \big)$, for $\upvarphi \in \Diff(N)$ with $N=$ Im$(\ups^*)$ the source space of the $\Diff(M)$-dressing field $\ups$, and $\upzeta \in \G\defeq\big\{ \upzeta: M \rarrow G\, |\, \upzeta^\gamma =\upzeta\big\}$ with $\gamma \in \H$ and $H$ (the target space of the dressing field $\bs u$) supporting a right action by $G$. 
This can be rewritten as 
\begin{align}
\label{gRGFT-ambig-dressing}
(\ups', \,\bs u')= \big(\ups \circ \upvarphi,\, \bs u\, (\ups^{-1*} \b\upzeta)\,  \big)
\rdefeq 
(\ups, \,\bs u) \cdot 
(\upvarphi, \,\b\upzeta\,),
\end{align}
where $\b\upzeta\defeq \ups^* \upzeta \in \b\G \defeq\big\{\b\upzeta: N \rarrow G\, |\, \b\upzeta^\gamma =\b\upzeta\big\}$, and the second equality defines the right action of $\Diff(N)\ltimes \b \G$ on the space of $(\Diff(M)\ltimes \H)$-dressing fields via a semi-direct product structure.\footnote{ 
The semi-direct product of $\Diff(N)\ltimes \b G$, formally analogue to that of $\Diff(M)\ltimes \H$, is given by: 
$(\upvarphi,\,  \b\upzeta)\cdot (\upvarphi',\,  \b\upzeta') \defeq \big(\upvarphi  \circ \upvarphi', \,  \b \upzeta\, (\upvarphi^{-1*}\b\upzeta ')\big)$.}
Now, two cases are to be distinguished.

The first, (A), is if transformations of the 2nd kind do not affect the bare fields,  $\phi^{(\upvarphi,\, \b\upzeta)}=\phi$. 
Then, the dressed fields $\phi^{(\ups,\, \bs u)}$ transform  as $(\phi^{(\ups, \,\bs u)})^{(\upvarphi, \,\b\upzeta)} 
\defeq 
\phi^{(\ups',\, \bs u  ')}
= \upvarphi^*\big[(\phi^{(\ups,\, \bs u)})^{\b\upzeta} \big]$ where $(\phi^{(\ups,\, \bs u)})^{\b\upzeta}$ has the same functional expression as the $\H$-transformation $\phi^\upgamma$ of the bare fields.
Analogously, the cocyclic dressing transforms as
\begin{align}
\label{C-2nd-kind}
C(\ups,\, \bs u)^{(\upvarphi,\, \b\upzeta\,)}
\defeq 
C\big(\phi; (\ups',\, \bs u  ') \big)
=
C\big(\phi; (\ups,\, \bs u) \cdot (\upvarphi, \b\upzeta\, )  \big)
=
C\big(\phi; (\ups,\, \bs u)\big) 
\ 
C\big(\phi^{(\ups,\, \bs u)};  (\upvarphi, \b\upzeta\, )  \big),
\end{align}
using again the defining property \eqref{C-gRGFT} of the cocycle $C$. 
Correspondingly, 
the linear cocycle $c(\ups, \bs u)$  transforms as
$c(\ups,\, \bs u)^{(\upvarphi,\, \b\upzeta\,)}=c(\ups,\, \bs u) +  c\big(\phi^{(\ups,\, \bs u)};  (\upvarphi,\, \b\upzeta\, )  \big)$.
Infinitesimally this yields $\delta_{(Y, \b\upxi)}\, c(\ups, \bs u)= - a\big((Y, \b\upxi\,); \phi^{(\ups,\, \bs u)}\big)$,
for $(Y, \b\upxi\,)\in \diff(N)\,\oplus$ Lie$\b\G$, 
which is the \emph{anomaly for transformations of the 2nd kind}.
The term $C\big(\phi^{(\ups,\, \bs u)};  (\upvarphi, \b\upzeta\, )  \big)$ is thus its ``integrated", or global, version. 
From all this follows  that the dressed PI \eqref{Anomalous-dressed-PI-gRGFT} and effective functional \eqref{dressed-W} transform as 
\begin{equation}
\begin{aligned}
\label{2nd-kind-dressed-PI}   
Z\big(\phi^{(\ups, \bs u)}\big)^{(\upvarphi, \b\upzeta)}
  &=
C\big(\phi^{(\ups,\, \bs u)};  (\upvarphi, \b\upzeta\, )  \big)\-\ 
  Z(\phi^{(\ups, \bs u)}), \\
W\big(\phi^{(\ups, \bs u)}\big)^{(\upvarphi, \b\upzeta)}
  &=
W(\phi^{(\ups, \bs u)})
-
c\big(\phi^{(\ups,\, \bs u)};  (\upvarphi, \b\upzeta\, )  \big).
\end{aligned}
\end{equation}
Infinitesimally, this gives in particular
$\delta_{(Y, \b\upxi)}\, W\big(\phi^{(\ups, \bs u)}\big) 
= 
-a\big((Y, \b\upxi\,); \phi^{(\ups,\, \bs u)}\big)$, 
i.e. the anomalous Ward Identity for the dressed effective functional. 
Remark that the anomaly for transformations of the 2nd kind has the same functional expression as the initial anomaly of $\Diff(M)\ltimes \H$, dressed fields replacing bare ones and the  parameters $(Y, \b\upxi)$ replacing the gauge parameters $(X, \lambda)$.

The second case, (B), is if transformations of the 2nd kind do indeed affect bare fields, 
$\phi^{(\upvarphi,\, \b\upzeta)}\neq\phi$.
Then, the transformations of dressed fields, $\big(\phi^{(\ups,\, \bs u)}\big)^{(\upvarphi, \b\upzeta)}$, have to be computed on a case by case basis,  no shortcut can a priori be given to find the $(\Diff(N)\ltimes \b\G)$-transformation of the dressed PI and/or effective functional. 
Generically, one should find 
\begin{equation}
\begin{aligned}
\label{2nd-kind-dressed-PI-bis}   
Z\big(\phi^{(\ups, \bs u)}\big)^{(\upvarphi, \b\upzeta)}
  &=
\b C\big(\phi^{(\ups,\, \bs u)};  (\upvarphi, \b\upzeta\, )  \big)\-\ 
  Z(\phi^{(\ups, \bs u)}), \\
W\big(\phi^{(\ups, \bs u)}\big)^{(\upvarphi, \b\upzeta)}
  &=
W(\phi^{(\ups, \bs u)})
-
\b c\big(\phi^{(\ups,\, \bs u)};  (\upvarphi, \b\upzeta\, )  \big), 
\end{aligned}
\end{equation}
where $\b C$, or $\b c$, is a 1-cocycle for the group $(\Diff(N)\ltimes \b\G)$, distinct from $C$, or $c$.
Infinitesimally, we get in particular
$\delta_{(Y, \b\upxi)}\, W\big(\phi^{(\ups, \bs u)}\big) 
= 
-\b a\big((Y, \b\upxi\,); \phi^{(\ups,\, \bs u)}\big)$; i.e. the anomaly for transformations of the 2nd kind has a functional expression a priori distinct from that of the initial $(\Diff(M)\ltimes\H)$-anomaly.
\medskip

The clock change transformations of section \ref{Clock change covariance and good clocks}, with anomaly given in relational quantization of MFT by \eqref{2nd-trsf-anomaly-NR-clock}, 
may be seen as an example of case (B). 
A famous example of case (A) is none other than the shift, achieved via the \emph{Bardeen-Zumino term}, between Lorentz and Einstein anomalies in theories $S(\phi)=S(\uppsi, A, e)$ of chiral fermions $\uppsi$ coupled to the gravitational field; the $(\so(1,3) \oplus \RR^4)$-valued Cartan connection $\b A=(A, e)$, with $A$ the spin connection and  $e={e^a}_\mu\,dx^{\,\mu} \rdefeq \bs e \cdot dx$ the soldering form.
The effective functional is $W=W(e)$, the fermions being integrated and $A=A(e)$ being taken torsion-free. 
See e.g. \cite{Bertlmann} chap.12 and \cite{Bonora2023} chap.15. 

The framework above can indeed specialize to $\H=\SO(1,3)$ the Lorentz gauge group and $\b \G=\GL(4)$ the group of local coordinate changes (or $\b G=\Diff(M)$). 
The $\phi$-dependent (i.e. non \emph{ad hoc}) dressing field is  the tetrad field, $\bs u(\phi)\defeq \bs e = {e^a}_\mu : M \rarrow GL(4)$, satisfying indeed the dressing field defining property $\bs e^\upgamma = \upgamma\- \bs e$ for $\upgamma \in \SO(1,3)$, and also subject to transformations of the 2nd kind    $\bs e^{\b\upzeta} \defeq \bs e \b\zeta$ where $\b\zeta \in \GL(4)$ is the Jacobian of a coordinate change (or of a diffeomorphism). 
The bare fields of the theory $\phi=\big\{ \uppsi, A, e \big\}$, being forms, are  all $\GL(4)$-invariant;  $\phi^{\b\upzeta}=\phi$, showing that we are indeed in case (A).\footnote{This case we mentioned in footnote 21, referring to  section 4.3.1.b  of \cite{Francois2021}. 
We remark that, there, a sort of classical analogue of the seesaw mechanism for anomalies is discussed; Noether charges exchange, whereby Lorentz charges are eliminated via dressing but replaced by charges for general coordinate transformations, or $\Diff(M)$. 
This bears in particular on the topic of the so-called ``dual" charges in Loop Quantum Gravity (LQG), see e.g. \cite{De-Paoli-Speziale2018, Oliveri-Speziale2020, Oliveri-Speziale2020-II}, which is similarly clarified by the DFM; see appendix G of \cite{Francois2021}.}
The bare gravitational field $\b A=(A, e)$ thus gets dressed as $\b A^{\bs u}=(A^{\bs u}, e^{\bs u}) \defeq \big(\bs e\- A \bs e + \bs e\-d\bs e,\, \bs e\- e \big) \rdefeq (\Gamma, dx)$, 
yielding $\SO(1,3)$-invariant fields, $(\b A^{\bs u})^\upgamma = \b A^{\bs u}$, 
but supporting transformations of the 2nd kind by $\GL(4)$; $(\b A^{\bs u})^{\b \upzeta} = \big(\Gamma^{\b \upzeta}, dx^{\b \upzeta}\big)=
\big(\b \upzeta\- \Gamma \b \upzeta + \b \upzeta\-d\b \upzeta,\, \b \upzeta\- dx\big)$. We may add that $e^t\defeq e^T \eta=dx \cdot \bs e^T \eta$, with $\eta$ the Minkowski metric and ``$(\_)^T$" the matrix transpose, $\SO(1,3)$-transforming as $(e^t)^\upgamma =e^t \upgamma$, is dressed as $(e^t)^{\bs u}\defeq e^t \bs e = dx\cdot \bs e^T \eta \bs e \defeq dx \cdot g$ where $g$ a metric on $M$. 
If $A=A(e)$ -- i.e. if the Cartan connection is \emph{normal} \cite{Sharpe, Cap-Slovak09, Attard-Francois2016_I, Attard-Francois2016_II, JTF-Ravera2024review} -- then $\Gamma=\Gamma(g)$.
Of course the fermions cannot be dressed via $\bs u=\bs e$, as ``$\uppsi^{\bs u} \defeq \rho(\bs e)\- \uppsi$", since $GL(4)$ has no (finite) spinor 
\enlargethispage{.55\baselineskip}representation,\footnote{Yet, the first published paper on the DFM \cite{GaugeInvCompFields}, section 4.3,  discusses  the possibility of extracting a $S\!O(r,s)$-valued dressing field out of the $GL(r+s)$-valued tetrad via  a (r,s)-signature extension of the Schweinler-Wigner orthogonalization procedure \cite{Schweinler-Wigner}. One may also consult Appendix B of \cite{Francois2014} for more details.
See also footnote 12 in \cite{Francois2018}, as well as section 4.3.2. of \cite{Francois-et-al2021}. 
This ``minimal dressing extraction" from the tetrad is the basis of a recent attempt to approach {QG} by Thiemann \cite{Thiemann2024} and called there the ``triangular gauge"; an unfortunate terminology given the clear   conceptual and mathematical distinction between dressing and gauge-fixing \cite{Berghofer-Francois2024}.
} but they are integrated out so that in the analysis of the transformation of the effective functional $W$
only the 
gravitational field 
contributes.
The $\SO(1,3)$-anomaly is  
$L_{\lambda^v} W(e, A)= -a\big(\lambda; A(e)\big)$
with $\lambda \in$ Lie$\SO(1,3)$,
 and explicit computation shows it to relate to an antisymmetric part for the effective energy-momentum tensor $\langle T_{ab} \rangle$. 
The integrated version is $R^\star_\upgamma W = W - c(\_; \upgamma)$. Specializing \eqref{dressed-W}, one has the dressed effective functional 
\begin{align}
\label{dressed-W-grav}
W^{\bs u}(e, A)=W(g, \Gamma) \defeq W(e, A) - c(\bs u) \quad \in \Omega^0_\text{basic}(\Phi),
\end{align}
where the linear cocyclic dressing $c(\bs u)=c\big(\phi; \bs u(\phi) \big) = c\big(\b A; \bs e \big)$
is the Bardeen-Zumino term, which by (the linear and infinitesimal versions of) \eqref{cocyclic-dressing-field-gRGFT} and \eqref{C-2nd-kind} satisfies, 
\begin{align}
\label{BZ-term-grav}
\bs L_{\lambda^v}  c(\bs u)= -a\big(\lambda; e, A\big)
\quad
\text{and}
\quad
\delta_{\b\upxi}\, c(\bs u)= - a\big(\b\upxi; g, \Gamma\big),
\end{align}
where $\b\upxi \in$ Lie$\GL(4)$ is the Jacobian of an infinitesimal coordinate change (or diffeomorphism); we can denote $\b \upxi =\d \upxi=\d_\mu \upxi^\alpha$ (itself noted $v_\upxi$ in \cite{Bertlmann} chap.12, or $\Xi$ in \cite{Bonora2023} chap.15).
The first relation above shows that $c(\bs u)$ cancels the Lorentz anomaly of the bare effective functional $W=W(e, A)$, while
the second relation implies, by the (linear, infinitesimal version of) \eqref{2nd-kind-dressed-PI-bis}, that the transformation of the 2nd kind of  dressed effective functional $W^{\bs u}=W(g, \Gamma)$ is 
$\delta_{\b\upxi} W(g, \Gamma) = - a\big(\b\upxi; \Gamma(g)\big)$, which is just the Einstein anomaly; explicit computation shows it to relate to the non-conservation the effective EM tensor, $\nabla_\mu \langle T^{\mu\nu}\rangle \neq 0$.
We see the anomaly seesaw mechanism at play.\footnote{
We may remind that anomalies can be determined through the Stora-Zumino descent equations, e.g. via BRST methods (see Chap.12 of \cite{Bertlmann}  or Chap.5 of \cite{Bonora2023}), and are 1-cocycles of the cohomologies of Lie algebras (e.g. of the gauge $\H$ or full covariance group $\Diff(M)\ltimes \H$). Their ``integrated", or non-linear, version given by  $C$, or $c$, are 1-cocycles of (Abelian) group cohomologies (e.g. of $\H$ or  $\Diff(M)\ltimes \H$), and yields what has been called ``global anomalies", see e.g. Chap.13 of \cite{Bonora2023}. See also  \cite{Francois2019_II} for the first presentation of \emph{cocyclic bundle geometry} and \emph{cocyclic connections}, showing how it underlies (global/local) consistent gauge anomalies.}
Compare \eqref{dressed-W-grav}-\eqref{BZ-term-grav} to eqs.(12.469)-(12.470) and eqs.(12.473)-(12.475) in \cite{Bertlmann}, and to eqs.(15.49)-(15.64) of \cite{Bonora2023}. 

\medskip
Let us observe that the DFM  extension to higher gauge theories features the same logic regarding 
anomaly cancellation and seesaw mechanism; we shall detail this in a forthcoming work.\footnote{There, when applied to string models, the automatic anomaly cancellation mechanism exposed above {may encompass} as a special case the Green-Schwarz mechanism; that is if the 2-form $B_2$ can be identified as an \emph{ad hoc} (higher) dressing field -- which as a matter of fact cannot be interpreted as a natural closed string mode, since its gauge transformation differs (and tuned specifically to cancel the anomaly).
} 
Let us further mention that we have recently presented the formal development of invariant path-integral quantization in the full gRGFT setting, together with the corresponding automatic anomaly cancellation mechanism and the seesaw mechanism for \mbox{anomalies}, in \cite{JTF-Ravera-InvariantPIQuant2026}, to which we refer the interested reader for further details. These results provide further support for the general framework based on the DFM advocated here.

We conclude the present work with perspectives on further developments and applications.

\section{Conclusion} 
\label{Conclusion}  

In line with the program developed in  \cite{JTF-Ravera2024gRGFT}, we have motivated the notion of relational quantization, achieved via the DFM, by analyzing the case of (NR) mechanics: Framing it as a 1D gRGFT, MFT, thereby generalizing the so-called ``parametrized" approach, we showed that its dressed, relational reformulation yields the well-known formulation when the clock field assumes the role of a dressing field. It follows that standard QM is a case of dressed, relational quantization of MFT.
Having ended with the above general outline of relational quantization for gRGFTs, in subsequent works we shall stress-test it via  applications to diverse models.  
As a simplest case, we shall perform first the relational quantization of scalar electrodynamics, then  of the Abelian Higgs model. 
It shall be a stepping stone to the relational quantization of the Electroweak (EW) model, which shall make contact with (and perhaps encompass) a closely related approach known as the Fröhlich-Morchio-Strocchi (FMS) mechanism \cite{Frohlich-Morchio-Strocchi81, Maas2019, Maas-et-al2021, Maas2023}: both the DFM and FMS approaches bypass the notion of spontaneous symmetry breaking \cite{Francois2018, Berghofer-et-al2023}. 
This  invites the analysis of the  issue of the interactions between relational quantization and renormalization (e.g. \cite{Kraus1998} for the EW model), which will be a separate contribution. 

\vspace{-3mm}
\paragraph{Relational Quantum Gravity} 
The latter is a key  input to our main goal of developing {RQG}, at first phenomenologically. 
Indeed, the PI \eqref{dressed-PI-gRGFT} could at least be applied to effective general-relativistic theories, notably models where gravity couples with matter modeled as a dust fluid. 
There, we will make contact notably with the literature on ``scalar coordinatization" of GR, e.g.
\cite{Kuchar1980, Hartle-Kuchar1984, Brown-Kuchar1995, Rovelli1991, Rovelli1991b, Rovelli2002b, Kuchar2011} --
which we have shown to be a special case of the DFM in
 \cite{JTF-Ravera2024gRGFT, JTF-Ravera_NoBdryPb_2025} (and where we  derive ``relational Einstein equations", applied to galaxy rotation curves in \cite{JTF-RaveraDarkMatter&DFM2025}) --
as well as with FMS approaches to {QG}
\cite{Maas2020}.
 Models with matter as dust notably  applies to cosmology, so we shall tackle relational quantum cosmology: it will be a case of relational MFT developed here, the
  relational minisuperspace being a case of dressed MFS $\Phi^\ups$. {RQG}, in a more fundamental guise, including fermionic matter -- thus approached via Cartan geometry \cite{JTF-Ravera2024review} -- will be investigated next.

\section*{Acknowledgment}  

We thank Philipp Berghofer for a careful reading of the manuscript and useful feedback.
J.F. is supported by the Austrian Science Fund (FWF), \mbox{[P 36542]} and by the Czech Science Foundation (GAČR), grant GA24-10887S.
L.R. is supported by the 
GrIFOS research project, funded by the Ministry of University and Research (MUR, Ministero dell'Università e della Ricerca, Italy), PNRR Young Researchers funding program, MSCA Seal of Excellence (SoE), 
CUP E13C24003600006, ID SOE2024$\_$0000103, of which this paper is part.



\appendix

\section{Covariant phase space formalism for MFT}
\label{Covariant phase space formalism for MFT}

{This appendix makes explicit the relation between the present formulation of MFT and the covariant phase space and edge-mode frameworks commonly used in gauge theories and generally covariant systems.}

In the ``covariant phase space" formalism for gRGFT,  alluded to in the main text, one key objective is to define a symplectic phase space for field theory that is invariant under the local symmetries (diffeos and gauge groups) -- this is the explicit intent of the original literature, see e.g. \cite{Witten1986,Zuckerman1986,Crnkovic1987, CrnkovicWitten1986}, also \cite{Gieres2021} for a review and discussion of related formalisms (Hamiltonian, multisymplectic, etc.). 

The role of the symplectic manifold, the phase space, is played by $\M_\S$, which  then  needs to be endowed with a symplectic 2-form. 
To obtain the latter, one first defines the \emph{presymplectic potential} as the integral of the presymplectic current over a codimension 1 submanifold $\Sigma \subset M$, with dim$M=n$: $\bs\theta_\Sigma \defeq \langle \bs\theta, \Sigma \rangle \in \Omega^1(\Phi, \RR)$. 
Then, one defines the $\bs d$-closed  2-form $\bs\Theta \defeq \bs{d\theta} \in \Omega^2\big(\Phi, \Omega^{n-1}(M) \big)$, and the corresponding $\bs\Theta_\Sigma =\langle \bs\Theta, \Sigma \rangle = \bs d\bs\theta_\Sigma \in \Omega^2(\Phi, \RR)$. 
Both $\bs\theta_\Sigma$ and $\bs\Theta_\Sigma$ are then considered restricted on the subbundle $\S$. 
The 2-form is a presymplectic form if $\ker \bs\Theta_\Sigma = \Gamma(V\S)$, i.e. if $\bs\Theta_\Sigma \in \Omega^2_\textbf{basic}(\S)$, in which case it induces (comes from) a form $\tilde{\bs\Theta}_\Sigma \in \Omega^2(\M_\S, \RR)$ -- s.t. ${\bs\Theta}_\Sigma = \pi^\star \tilde{\bs\Theta}_\Sigma$. 
The latter is $\bs d$-closed by naturality of $\bs d$, $[\bs d, \pi^\star]=0$, and s.t. $\ker \tilde{\bs\Theta}_\Sigma = 0$, so it is symplectic: 
one end-up with the covariant phase space $\big(\M_\S, \tilde{\bs\Theta}_\Sigma \big)$ for the field theory over the region $\Sigma$.

This program encounters obstruction the moment  ${\bs\Theta}_\Sigma$ is not basic on $\S$, which is typically the case in gRGFT: it fails to be horizontal, its verticality property involving terms at a codimension 2 boundary $\d \Sigma$, a.k.a. ``corner". 
One must then find ways to restore
horizontality, thus basicity, of ${\bs\Theta}_\Sigma$.
If the boundary is asymptotic (at infinity), it is enough to impose fall-off conditions on the fields $\phi$. 
If $\d\Sigma$ is at finite distance, boundary conditions are unnatural, and a typical strategy involves the \emph{ad hoc} introduction of so-called \emph{edge modes}; d.o.f. confined to $\d \Sigma$ whose gauge transformations are tuned to cancel the corner terms.\footnote{In general the verticality property of  $\bs\theta_\Sigma$ also involves ``bulk terms", integrated on $\Sigma$: see eq.  (5.65) in \cite{Francois2021} and eq.  (240) in \cite{Francois2023-a}.  
So, if the ``edge modes" strategy is used to restore its horizontality, these d.o.f. cannot be confined to the corner, but must also extend to the bulk.}
See e.g. \cite{DonnellyFreidel2016, Speranza2022, Geiller2017, Speranza2018, Geiller2018, Chandrasekaran_Speranza2021, Freidel-et-al2020-1, Freidel-et-al2020-2, Freidel-et-al2020-3, Ciambelli2023}. 
As shown in \cite{Francois2021, Francois-et-al2021} and  \cite{Francois2023-a}, this approach is  both  streamlined and conceptually clarified once understood to be but an instance of the Dressing Field Method (DFM):
 ``edge modes" are just \emph{ad hoc} dressing fields. But the DFM stresses that  they must 
be built from the existing field d.o.f. of the theory -- rather than than introduced by \emph{fiat} as is always done in this {literature} -- 
 allowing for a \emph{relational} 
understanding of the invariance  achieved via the DFM, as expressed in the main text and 
detailed in \cite{JTF-Ravera2024gRGFT}.
\enlargethispage{1\baselineskip}

A different use case of the covariant phase space formalism is the definition of conserved Noether charges associated to local symmetries, together with their Poisson bracket induced by $\bs\Theta_\Sigma$.
First a conserved Noether current $(n-1)$-form is defined from the (quasi-)invariance of the Lagrangian under the local symmetry: From the variational principle \eqref{dS-dL} and $\bs L_{\chi^v} L = \iota_{\chi^v} \bs dL = a(\chi; \phi)= d\beta(\chi; \phi)$ with $\chi$ the infinitesimal gauge parameter, one defines the Noether current as 
 $J(\chi; \phi)\defeq \iota_{\chi^v} \bs\theta - \beta(\chi; \phi)$\footnote{Or as any other member of its De Rham cohomology class, adding a $d$-exact term $d\gamma(\chi; \phi)$.} satisfying 
 $dJ(\chi; \phi)=-\iota_{\chi^v} \bs E$ and thus being $d$-closed on-shell. 
The Noether charge is then $Q_\Sigma(\chi; \phi) \defeq \langle J(\chi; \phi), \Sigma \rangle$.
Furthermore one finds, either by explicit computations or by deduction from its $d$-exactness on shell together with the 
 Poincaré lemma, that $J(\chi; \phi) = dq(\chi; \phi) + E(\chi; \phi)$. 
Meaning that  the Noether charges are  corner terms on-shell : $Q_\Sigma(\chi; \phi)_{|\S} =\langle q(\chi; \phi), \d \Sigma \rangle$.
See eq. (5.58) in 
\cite{Francois2021} and eq. (3.6) in \cite{Francois-et-al2021} for charges of internal gauge groups, and eq. (207) in \cite{Francois2023-a} for charges of diffeomorphisms. 

The Poisson bracket of charges is defined to be $\{ Q_\Sigma(\chi; \phi), Q_\Sigma(\chi'; \phi)\}\defeq \bs\Theta_\Sigma(\chi^v, {\chi'}^v)$, and  it is  found to be, on-shell, a central extension of the gauge Lie algebra:  $\{ Q_\Sigma(\chi; \phi), Q_\Sigma(\chi'; \phi)\} = Q_\Sigma([\chi, \chi']; \phi) + \mC(\lfloor X, Y\! \rfloor ;\phi )$, 
with $\mC(\lfloor X, Y\! \rfloor ;\phi )$ a 2-cocycle of the gauge Lie algebra whose expression depends on $L$ and $\bs\theta$.
See eq. (5.62) of \cite{Francois2021} for the internal gauge group case, and  eq. (224) of  \cite{Francois2023-a} for diffeomorphisms. 

With this background notions in place, let us now quickly work out the covariant phase space structure of~MFT. 
The first thing to notice is that in MFT there can be no Noether charges of $\Diff(I)$ on-shell, since there is no corner in dimension 1, so that the Poisson bracket will be trivial on-shell. 
Indeed, by 
\eqref{id-dL-E-theta} and \eqref{key-identity} above, we have  
the horizontality of $\bs E$ \eqref{vertic-E-theta}, meaning that the Noether current 0-form is 
 $J(X; \phi) \equiv 0$ -- as it obviously cannot be $d$-exact. 
 Correspondingly, for $\Sigma =\{\tau_0\} \in I$  a 0-dim (codimension 1) submanifold, on which  integration reduces to the evaluation operation: 
 $Q_\Sigma(X; \phi) \defeq \langle J(X; \phi), \Sigma \rangle=\int_\Sigma J(X; \phi) =\text{ev}_{\tau_0} J(X; \phi) =J\big(X; \phi(\tau_0)\big)\equiv 0$. 
 
Toward computing the Poisson bracket, and more importantly, later, the gauge transformation of $\bs\Theta_\Sigma$, we need its verticality property. We obtain it two ways: {\textit{first,}} from the infinitesimal equivariance and verticality of $\theta$ \eqref{equiv-E-theta}-\eqref{vertic-E-theta} {as well as from the variational principle} \eqref{dS-dL}, we have
\begin{align}
 \iota_{X^v} \bs\Theta 
 =  \iota_{X^v} \bs{d\theta}
 =  \bs L_{X^v} \bs\theta -  \bs d\iota_{X^v} \bs\theta
 = \mathfrak L_X \bs\theta - \bs d(\iota_X L)
 = \iota_X d\bs\theta + d\iota_X\bs\theta -\iota_X \bs dL
 = \iota_X d\bs\theta - \iota_X (\bs E + d\bs\theta)
 =- \iota_X \bs E,
\end{align}
where $\iota_X\bs\theta=0$ since $\bs\theta$ is a 0-form on $I$. 
{\textit{Second} and more simply,} one may use $\bs d^2 L =0=\bs{dE} + d\bs\Theta$, which yields, by the horizontality \eqref{vertic-E-theta} and infinitesimal equivariance \eqref{equiv-E-theta} of $\bs E$, 
\begin{align}
  \label{vert-Theta}d\iota_{X^v}\bs\Theta=\iota_{X^v}d\bs\Theta 
  = -\iota_{X^v} \bs{dE}
  = - \bs L_{X^v} \bs E - \bs d \iota_{X^v} \bs E
  = -\mathfrak L_X \bs E
  = -d \iota_X \bs E 
  \quad \Rightarrow \quad
  \iota_{X^v}\bs\Theta = -\iota_X \bs E,
\end{align}
where we use that $\bs E$ is a top form on $I$, so $d \bs E =0$, and we can conclude because $\bs\Theta$ and $\iota_X\bs E$ are 0-forms on $I$. 
From~this, and the horizontality of $\bs E$, it is immediate that
\begin{align}
 \label{vert-2-Theta} 
 \bs\Theta(X^v, Y^v)
 =\iota_{Y^v} \iota_{X^v} \bs \Theta
 = - \iota_X \iota_{Y^v} \bs E =0. 
\end{align}
The triviality of the Poisson bracket 
$ \{ Q_\Sigma(X; \phi), Q_\Sigma(Y; \phi)\}
 \defeq 
 \bs\Theta_\Sigma(X^v, Y^v)=0$ follows,
as expected. 

The verticality property \eqref{vertic-E-theta} of $\bs\theta$ shows that is fails to be basic, even on-shell, i.e. on $\S$.
It is clear given its
$\bs\Diff(I)$-transformation  \eqref{GT-E-theta}, from which one immediately deduce that of 
$\bs\theta_\Sigma$:
\begin{align}
    \label{GT-theta-Sigma}
(\bs\theta_\Sigma)^{\bs \psi}
=
\bs\theta_\Sigma + \langle \iota_{\bs{d\psi}\circ\bs\psi\-}L, \Sigma \rangle.
\end{align}
The presymplectic potential failing to be basic on $\S$, it does not induce a symplectic potential on $\M_\S$.
Surprisingly, this does not imply the same for the presymplectic form. 
Indeed, we compute geometrically, via \eqref{pushforward-X} and using \eqref{vert-Theta}-\eqref{vert-2-Theta}, the gauge-transformation of $\bs\Theta$ to be
\begin{align}
\bs\Theta^{\bs\psi}(\mathfrak{X}, \mathfrak Y) \defeq&\, \Xi^\star \bs\Theta(\mathfrak{X}, \mathfrak Y)
=
\bs\Theta\big( \Xi_\star\mathfrak{X} ,  \Xi_\star\mathfrak{Y} \big)
=
\bs\Theta \left( 
R_{\bs\psi\star} 
\big( \mathfrak{X} + 
\{ \bs{d\psi}\circ\bs\psi\-(\mathfrak X)\}^v \big),  
R_{\bs\psi\star} 
\big( \mathfrak{Y} + 
\{ \bs{d\psi}\circ\bs\psi\- (\mathfrak Y)\}^v \big)
\right) \notag\\
=&\, 
R^\star_{\bs\psi}\bs\Theta
\left( 
\mathfrak{X} + 
\{ \bs{d\psi}\circ\bs\psi\-(\mathfrak X)\}^v,
\mathfrak{X} + 
\{ \bs{d\psi}\circ\bs\psi\-(\mathfrak X)\}^v
\right) \notag\\
=&\,
\bs\psi^* \left(\,
\bs\Theta (\mathfrak X, \mathfrak Y)
+ 
 \bs\Theta \left(\mathfrak X, \{ \bs{d\psi}\circ\bs\psi\-(\mathfrak Y)\}^v \right)
+ 
 \bs\Theta \left(\bs{d\psi}\circ\bs\psi\-(\mathfrak X)\}^v, \mathfrak Y \right)  
+  \cancel{\bs\Theta
\left(\bs{d\psi}\circ\bs\psi\-(\mathfrak X)\}^v, \bs{d\psi}\circ\bs\psi\-(\mathfrak Y)\}^v\right)}\,  \right) \notag\\
=&\, 
\bs\psi^*  \left(\,
\bs\Theta (\mathfrak X, \mathfrak Y)
+ \iota_{\mathfrak X} \, \iota_{\bs{d\psi}\circ\bs\psi\-(\mathfrak Y)} \bs E
- \iota_{\mathfrak Y} \, \iota_{\bs{d\psi}\circ\bs\psi\-(\mathfrak X)} \bs E\,
\right). \notag 
\end{align}
Which finally yields 
\begin{align}
\label{GT-Theta-and-Sigma}
\bs\Theta^{\bs\psi} = \bs\psi^* \left( \bs\Theta - \iota_{\bs{d\psi}\circ\bs\psi\-} \bs E\right), 
\quad \text{ and } \quad 
(\bs\Theta_\Sigma)^{\bs\psi} &=   \bs\Theta_\Sigma - \langle \iota_{\bs{d\psi}\circ\bs\psi\-} \bs E , \Sigma\rangle.
\end{align}
Thus, $\bs\Theta_\Sigma \in \Omega^2_\text{basic}(\S)$ so $\bs\Theta_\Sigma =\pi^\star \tilde{\bs\Theta}_\Sigma$ with $\tilde{\bs\Theta}_\Sigma\in \Omega^2(\M_\S)$, and  $\big(\M_\S, \tilde{\bs\Theta}_\Sigma \big)$ is the  covariant phase space of~MFT.


As final remarks, consider the relation of the covariant phase space formalism to objects of the standard symplectic formalism in the case of the standard NR Lagrangian discussed in section \ref{Models of MFT}. 
We have that 
\begin{align}
\iota_{X^v} \bs\theta 
=
\theta \big(\iota_{X^v} \bs d\phi; \phi \big)
=
\theta \big(\mathfrak L_{X} \phi; \phi \big)
=
\theta \big(\iota_{X} d\phi; \phi \big)
=
\iota_X \b \theta,
\end{align}
where $\b\theta =\theta(d\phi; \phi)= p_\alpha d\phi^\alpha 
= \langle p_x, dx\rangle + p_t dt$ is the Tautological form, or Liouville-Poincaré 1-form, on the configuration space-time $\mC$, while  $\t\theta \defeq \langle p_x, dx\rangle$ is that of the configuration space $Q$ (a.k.a.   the standard symplectic potential). 
Since we have established that \eqref{theta-L-relation} holds, given the constraint \eqref{constraint-momenta}, we have that
the action functional can be written as $S=\int_i \b\theta$. 
We also have that the generalized Hamiltonian vanishes identically:
$\b H \defeq \b \theta - L = 0$.
A feature typical of general-relativistic theories (otherwise called ``parametrized theories"); GR in particular, where it leads to the ``problem of time" in its canonical analysis, a version of which we discussed  in section \ref{The relationality of MFT}.
The covariant symplectic 2-form is here $\bs\Theta = \bs{d\theta}= \bs d p_\alpha\, \bs d \phi^\alpha = \langle \bs d p_x, \bs d x\rangle + \bs d p_t \,\bs dt$, reminiscent of the standard symplectic 2-form $\t\Theta = d \t\theta = \langle d p_x,  d x \rangle$ on $T^*Q$. 

By \eqref{Dressing-form-bis-t}, or the DFM rule of thumb, we get that for the clock field as dressing field the dressed symplectic potential and 2-form are: 
$\bs\theta^\ups = \theta (\bs d\phi^\ups; \phi^\ups) = \langle p_{\b x}, \bs d \b x \rangle$ and 
$\bs\Theta^\ups = \bs d \bs \theta^\ups= \langle \bs d p_{\b x}, \bs d \b x \rangle$ on $\Phi^\ups \simeq \M \simeq \P(\mC) \rarrow \P(Q)$. 
One recognizes the close structural analogy with $\t \theta$ and $\t\Theta$ on $T^*Q$.

It is interesting to remark that the standard symplectic formalism for mechanics on $T^*Q$ (de-parametrized) was the motivation for the covariant phase space approach to GFT (and gRGFT) in the original literature \cite{Witten1986,Zuckerman1986,Crnkovic1987, CrnkovicWitten1986}. 
Applying it to (NR) MFT, understood as a model of gRGFT, and establishing the correspondence with the standard approach brings us full circle back to the origins.

\section{{List of symbols}}
\label{Principal mathematical notations}

\begin{center}
{\small{
\renewcommand{\arraystretch}{1.2}
\begin{longtable}{c||c|c|} 
  \textbf{Object}    &\textbf{Symbol}	\\[.2em] \hline\hline\\[-1em]
(Mechanical) Field Space &  $\Phi$ \\[.1em] \hline
Structure group / algebra of $\Phi$ &  $\Diff(I)$ / $\diff(I)$ \\[.1em] \hline
Structure group / algebra element &  $\psi/X$ \\[.1em] \hline
(Moduli) Base Space of $\Phi$ &  $\M$ \\[.1em] \hline
Point of $\Phi$ / $\M$ &  $\phi=\{t, x\}$ / $[\phi] $ \\[.1em] \hline
($1D$) Parameter space for $\phi$ &  $I$ \\[.1em] \hline
Automorphism group / algebra &  $\bs\Aut(\Phi)$ / $\bs\aut(\Phi)$ \\[.1em] \hline
Element of $\bs\Aut_{(v)}(\Phi)$ &  $\Xi$ \\[.1em] \hline
Vertical automorphism group / algebra &  $\bs\Aut_v(\Phi)$ / $\bs\aut_v(\Phi)$ \\[.1em] \hline
Gauge group / algebra &  $\bs\Diff(I)$ / $\bs\diff(I)$ \\[.1em] \hline
Element of $\bs\Diff(I)$ / $\bs\diff(I)$  &  $\bs\psi$ / $\bs X$ \\[.1em] \hline
Diffeomorphism group / algebra of $\M$ &  $\bs\Diff(
M)$ / $\bs\diff(\M)$ \\[.1em] \hline
Tangent space / Tangent vector fields on $\Phi$ &  $T\Phi$ /  $\Gamma(T\Phi)$ \\[.1em] \hline
Differential forms on $\Phi$ / $\M$ &  $\Omega^\bullet(\Phi)$ / $\Omega^\bullet(\M)$ \\[.1em] \hline
Right  action of $\Diff(I)$ on $\Phi$  &  $R_
\psi$  \\[.1em] \hline
Element of $\Gamma(T\Phi)$ / $\Omega^\bullet(\Phi)$ / $\Omega^\bullet(\M)$  &  $\mathfrak X$ / $\bs\alpha = \alpha\big(\! \w^\bullet \!\bs d\phi; \phi \big)$ / $\b{\bs\alpha}$ \\[.1em] \hline
De Rham differential on $I$ / on $\Phi$ &  $d$ / $\bs d$ \\[.1em] \hline
Pullback (pushforward) on $I$ / $\Phi$ by a map $f$ / $F$ &  $f^*$  ($f_* $) /  $ F^\star$ ($F_\star$) \\[.1em] \hline
Gauge transformation of $\bs\alpha$ (via vertical automorphism) &  $\bs\alpha^{\bs \psi}\defeq \Xi^\star \bs\alpha$ \\[.1em] \hline
Lie derivative on $I$ / on $\Phi$ &  $\mathfrak L_X$ / $\bs L_\mathfrak{X}$ \\[.1em] \hline
$\Diff(I)$  $1$-Cocycle on $\Phi$&  $C(\phi; \psi)$ \\[.1em] \hline
$\diff(I)$  $1$-Cocycle on $\Phi$&  $a(X; \phi)$ \\[.1em] \hline
Principal / cocyclic connection on $\Phi$ (curvature) &  $\bs\omega$ / $\bs\varpi$ ($\bs\Omega$ / $\b{\bs\Omega}$) \\[.1em] \hline
Space of regions of $I$ &  $\bs i(I)$ \\[.1em] \hline
Element of  $\bs i(I)$ /  1-form on $I$  &  $i$ / $\omega$  \\[.1em] \hline
Integration (pairing) on $I$ &  $\langle \omega, i \rangle=\int_i \omega $ \\[.1em] \hline
Associated bundle of regions &  $\mathcal R(I)
 \defeq   \Phi \times  \bs i(I) /\!\sim$ \\[.1em] \hline
Lagrangian 1-form / Action on $I$ / Path integral (PI) &  $L=L(\phi) $ / $S=
\int_i L$ / $Z=Z(\phi)$ \\[.1em] \hline
Field equations /  Symplectic potential &  $\bs E=E(\bs d\phi; \phi)$ / $\bs \theta$ \\[.1em] \hline
Space of solutions / Moduli space of solutions &  $\S$ / $\M_\S$ \\[.1em] \hline
Extended config. space /  Config. space &  $\mC$ / $Q$ \\[.1em] \hline
Curve in $\mC$ / $Q$ &  $\gamma$ / $\gamma_q$  \\[.1em] \hline
Dressing field / Dressed fields / Dressed regions &  $\bs\upupsilon$ / $\phi^{\bs\upupsilon}$ / $i^{\bs\upupsilon}$  \\[.1em] \hline
Dressing map / Dressed forms & $F_{\bs\upupsilon}$ / $\bs\alpha^{\bs\upupsilon}\defeq F^\star_{\bs\upupsilon} \bs \alpha$ \\[.1em] \hline
Dressed Lagrangian / field equations / Dressed PI  & $L^{\bs\upupsilon}$ / $\bs E^{\bs\upupsilon}$ / $Z^{\bs\upupsilon}$ \\[.1em] \hline
\end{longtable}
\captionof{table}{{Main mathematical notations.}}\label{Table01}}} 
\end{center}


{
\normalsize 
 \bibliography{Biblio13}
}

\end{document}